\documentclass[12pt]{article}
\usepackage{amsmath}
\usepackage{graphicx}

\usepackage{amsthm}
\usepackage{amsfonts}
\usepackage{mathrsfs}
\usepackage[colorlinks,allcolors=black]{hyperref}
\usepackage[linesnumbered,ruled,vlined]{algorithm2e}
\usepackage[protrusion=true,expansion=true]{microtype}
\usepackage[utf8]{inputenc}

\numberwithin{equation}{section}
\theoremstyle{plain}

\newtheorem{proposition}{Proposition}[section]
\newtheorem{remark}{Remark}[section]
\theoremstyle{definition}
\newtheorem{definition}{Definition}[section]
\newtheorem{example}{Example}[section]

\usepackage{booktabs}
\usepackage{longtable}
\usepackage{array}
\usepackage{multirow}
\usepackage{wrapfig}
\usepackage{float}
\usepackage{colortbl}
\usepackage{pdflscape}
\usepackage{tabu}
\usepackage{threeparttable}
\usepackage{threeparttablex}
\usepackage[normalem]{ulem}
\usepackage{makecell}
\usepackage{xcolor}
\usepackage{caption}

\usepackage{natbib}

\DeclareMathOperator{\diag}{diag}	

\DeclareMathOperator{\rank}{rank}
\DeclareMathOperator{\sign}{sign}

\DeclareMathOperator{\argmin}{arg\,min}

\newcommand \bv {\mathbf{v}}
\newcommand \bu {\mathbf{u}}
\newcommand \bl {\boldsymbol{\lambda}}

\usepackage[margin=1in]{geometry}

\begin{document}

\title{\bf Co-factor analysis of citation networks\thanks{The Version of Record of this manuscript has been published and is available in the Journal of Computational Graphics and Statistics at \url{http://www.tandfonline.com/10.1080/10618600.2024.2394464}.}}
\author{Alex Hayes\thanks{
    This project was supported by the NSF under Grant DMS-1916378; NSF under Grant DMS-1612456; and the ARO under Grant W911NF-15-1-0423.}\hspace{.2cm}\\
  Department of Statistics, University of Wisconsin-Madison\\
  and \\
  Karl Rohe \\
  Department of Statistics, University of Wisconsin-Madison}
\maketitle
\bigskip
\begin{abstract}
  One compelling use of citation networks is to characterize papers by their relationships to the surrounding literature. We propose a method to characterize papers by embedding them into two distinct ``co-factor'' spaces: one describing how papers send citations, and the other describing how papers receive citations. This approach presents several challenges. First, older documents cannot cite newer documents, and thus it is not clear that co-factors are even identifiable. We resolve this challenge by developing a co-factor model for asymmetric adjacency matrices with missing lower triangles and showing that identification is possible. We then frame estimation as a matrix completion problem and develop a specialized implementation of matrix completion because prior implementations are memory bound in our setting. Simulations show that our estimator has promising finite sample properties, and that naive approaches fail to recover latent co-factor structure. We leverage our estimator to investigate 255,780 papers published in statistics journals from 1898 to 2024, resulting in the most comprehensive topic model of the statistics literature to date. We find interpretable co-factors corresponding to many statistical subfields, including time series, variable selection, spatial methods, graphical models, GLM(M)s, causal inference, multiple testing, quantile regression, semiparametrics, dimension reduction, and several more.
\end{abstract}

\noindent%
\vfill

\newpage

\section{Introduction}
\label{sec:introduction}

Suppose we have a collection of written documents, and these documents cite each other. For example, the documents might be academic papers, judicial opinions, or patents, among other possibilities. One useful way to understand individual documents in the collection, and the collection as a whole, is to find documents that cite, and are cited, in similar ways. These documents are likely to be about the same subject, and can thus reveal information about important topics in the corpus.

We develop a network-based approach to understanding the structure in citation corpora, called \texttt{CitationImpute}. \texttt{CitationImpute} begins by representing a corpus as a network, where each document corresponds to a node, and citations between documents correspond to directed edges. Then, it uses a spectral factorization technique to embed each document into two distinct latent spaces, one characterizing how papers cite, and the other characterizing how papers get cited.

Unlike prior approaches to citation analysis, \texttt{CitationImpute} models citations from older documents to newer documents as structurally missing. As a consequence, our algorithm must estimate singular subspaces via matrix completion methods. Existing matrix completion methods are computationally prohibitive in this setting, so we develop a singular subspace estimator with reasonable time and space complexity.

After estimating singular subspaces, \texttt{CitationImpute} uses varimax rotation to identify latent factors in the network (as opposed to k-means, or k-medians clustering). This allows each document to have a weighted membership in each cluster. The overall procedure can be understood intuitively in the context of stochastic blockmodels, but is appropriate for a much broader class of low-rank network models.

We validate the new procedure with a simulation study, finding that the new estimator recovers latent factors under a partially observed stochastic blockmodel. Finally, we analyze 255,780 statistics papers and 2.2 million citations published in journals on statistics and probability, producing a comprehensive breakdown of topics in the statistics literature. We present the keywords most associated with these topics in Table \ref{tab:citation_impute_preclipped2-y-keywords-elly-50000-ellz-1e+05-k-30} (factors describing how papers get cited) and Table \ref{tab:citation_impute_preclipped2-z-keywords-elly-50000-ellz-1e+05-k-30} (factors describing how papers cite).

\texttt{CitationImpute} is related to several lines of extant work, most notably empirical investigations of the academic statistics literature. \cite{selby2020} and \cite{stigler1994} consider relationships between statistics papers and the larger academic literature, with \cite{selby2020} reviewing approaches to community detection in networks and suggesting a number of diagnostic techniques for assessing model fit. \cite{ji2022}, an expansion of \cite{ji2016}, considers a dataset with about a third as many papers as our own, and investigates undirected (and dynamic) networks of academic authors based on co-authorship and co-citation. \cite{ji2022} estimates researcher interests by embedding researchers into a three-dimensional latent space. In contrast, we model the topics of individual manuscripts, and co-embed manuscripts into much more detailed thirty-dimensional ``sending'' and ``receiving'' latent spaces. \cite{rohe2023} co-factor a directed network of journal-journal citation counts using varimax factor analysis, but aggregates citations over time and thus avoids the chronological missingness we consider here.

\begingroup\fontsize{8}{10}\selectfont

\begin{longtable}[t]{l>{\raggedright\arraybackslash}p{34em}l}
\caption{\label{tab:citation_impute_preclipped2-y-keywords-elly-50000-ellz-1e+05-k-30}Keywords for Y (incoming citation) factors}\\
\toprule
Factor Name & Top words & ID\\
\midrule
non-convex penalties & selection, variable, penalized, oracle, lasso, nonconcave & y01\\
feature screening & screening, dimensional, ultrahigh, feature, independence, high & y02\\
bayesian model selection & bayesian, models, complexity, disease, model, fit & y03\\
post-selection inference & high, dimensional, lasso, regression, confidence, dantzig & y04\\
survival analysis & survival, censored, hazards, proportional, cox, regression & y05\\
\addlinespace
information criteria & model, clustering, mixture, selection, dimension, mixtures & y06\\
causal inference & propensity, causal, score, observational, treatment, effects & y07\\
multiple testing & false, discovery, multiple, rate, testing, controlling & y08\\
graphical models & graphical, covariance, estimation, sparse, lasso, high & y09\\
bayesian non-parametrics & dirichlet, bayesian, nonparametric, mixture, mixtures, priors & y10\\
\addlinespace
supervised dimension reduction & dimension, reduction, regression, sliced, inverse, sufficient & y11\\
LASSO (optimization) & lasso, regularization, coordinate, descent, selection, via & y12\\
LASSO (classic) & lasso, selection, shrinkage, regression, via, longitudinal & y13\\
kriging & spatial, gaussian, datasets, covariance, large, temporal & y14\\
empirical likelihood & empirical, likelihood, confidence, ratio, intervals, regions & y15\\
\addlinespace
GLM(M)s & longitudinal, data, generalized, models, estimating, binary & y16\\
functional data & functional, regression, principal, data, linear, longitudinal & y17\\
skew normals & skew, normal, distributions, multivariate, distribution, t & y18\\
quantile regression & quantile, regression, quantiles, censored, median, estimation & y19\\
bayesian model selection & bayesian, selection, variable, bayes, priors, prior & y20\\
\addlinespace
missing data & missing, imputation, data, longitudinal, nonignorable, nonresponse & y21\\
adaptive clinical trials & trials, clinical, adaptive, sequential, group, multiple & y22\\
splines + random effects & models, mixed, splines, smoothing, longitudinal, regression & y23\\
multivariate analysis & covariance, matrices, high, dimensional, large, matrix & y24\\
MCMC & monte, carlo, markov, metropolis, chain, bayesian & y25\\
\addlinespace
single index models & coefficient, varying, models, index, single, partially & y26\\
causal semiparametrics & missing, semiparametric, regression, sampling, data, estimation & y27\\
individual/optimal treatment & treatment, regimes, individualized, learning, optimal, estimating & y28\\
RIDGE & ridge, regression, biased, linear, estimators, estimator & y29\\
cure models & cure, survival, censored, rate, mixture, hazards & y30\\
\bottomrule
\end{longtable}
\endgroup{}

Methodologically, \texttt{CitationImpute} is an extension of the varimax rotation technique studied in \cite{rohe2023}, and is closely related to co-clustering methods \citep{rohe2016, choi2014, choi2017}, as well as clustering methods for bipartite networks \citep{larremore2014, razaee2019, yen2020a}, some of which can be extended to handle missing data \citep{zhao2022d, peixoto2018}. While there is a large literature on network clustering with missing data, these techniques cannot be used for co-factoring and co-clustering. Nonetheless, some techniques similarly leverage nuclear norm penalized singular subspace estimation to handle missing edges \citep{chen2014, vinayak2014, li2020c}. There have also been some efforts to incorporate topic structure into preferential attachment models \citep{pollner2006, hajek2019}, bridging the gap between mixture modelling and more traditional bibliometric analysis \citep{price1976}.

Finally, our work is related to the general matrix completion literature, in particular nuclear norm penalization approaches for estimating partially observed matrices \citep{kim2013a, gu2014, klopp2014, cui2015, hosono2016, gu2017, zhang2019a, yang2022a, shamir2014, bhojanapalli2014,  cho2019, mazumder2010}. While this literature has recently made impressive inroads regarding the consistency of nuclear-norm regularization for spectral recovery in deterministic and non-uniform sampling settings \citep{foucart2021, zhu2022}, we are unaware of consistency results for the upper triangular observation pattern present in citation data, and thus validate our approach with simulations.

\subsection*{Notation}

Let $\bu_i (A), \bl_i (A), \bv_i (A)$ be functions that return the $i^{th}$ left singular vector, singular value, and right singular vector of a matrix $A$, respectively. Similarly, define $\bl_i^2 (A) = \left(\bl_i (A)\right)^2$. We use $\langle \cdot, \cdot \rangle$ to denote the Frobenius inner product and $\Vert \cdot \Vert_F$ the Frobenius norm. Let $A_{i \cdot}$ denote the $i^{th}$ row of a matrix $A$ and $A_{\cdot j}$ denote the $j^{th}$ column. For a partially observed matrix $A$, let $\Omega_A$ be the set $\{ (i, j) : A_{ij} \text{ is observed}\}$ and $\tilde \Omega_A$ be the set $\{ (i, j) : A_{ij} \text{ is observed and non-zero}\}$; when $A$ is clear from context we will omit the subscript $A$. By $Y_A$ ($Y$ when the context is clear) we denote the binary matrix such that $Y_{ij}$ is one when $(i, j) \in \Omega_A$ and zero otherwise. $\odot$ indicates elementwise multiplication between two matrices with the same dimensions. We use $P_{\Omega_A} (B) = B \odot Y_A$ to denote the projection of a matrix $B$ onto observed support of another matrix $A$, and $P_{\Omega_A}^\perp (B) = B \odot (1 - Y_A)$. Let $P_\ell (A)$ denote the ``clipping'' projection that sets the first $\ell$ columns and the last $\ell$ rows of $A$ all to zero. Finally, $g(n) = \mathcal O (f(n))$ means that $\lim_{n \to \infty} g(n) / f(n) \le M$ for some constant $M$. All proofs are deferred to the Appendix.

\section{Model}
\label{sec:operationalization}

\subsection{Co-factor model}

We use the co-factor model of \citet{rohe2023} as a model for latent similarities between documents. The co-factor model is a low-rank, distributionally agnostic generalization of the stochastic co-blockmodel \citep{holland1983,rohe2016}, and includes sub-models such as stochastic blockmodels, degree-corrected stochastic blockmodels \citep{karrer2011}, (degree-corrected) mixed membership stochastic blockmodels \citep{airoldi2008, jin2024}, latent dirichlet allocation \citep{blei2003}, and (generalized) random dot product graphs \citep{lyzinski2014}, many of which are closely related to topic models \citep{gerlach2018}.

In the co-factor model, each document $i$ possesses two co-factors. One co-factor, $Z_i \in \mathbb{R}^k$, controls outgoing citations, or the topics that a paper is likely to cite, and the other co-factor, $Y_i \in \mathbb{R}^k$, controls incoming citations, or the topics that a paper is likely to be cited by. The co-factor structure of the model operationalizes the fundamental difference between citing and being cited. Mathematically, co-factor models are generalizations of factor models, and there are compelling reasons to model full co-factor structure: co-factor structure is theoretically necessary to capture key features of real world network data \citep{chanpuriya2020}, an observation empirically verified by \citet{rohe2016} and \citet{qing2022}, among others.

\begin{example}
  Consider \citet{tibshirani1996}, which introduced \texttt{LASSO} regression. The \texttt{LASSO} paper builds upon a small body of statistical work on variable selection and resampling, but itself forms the basis for a large body of applied work, especially in genomics and biomedical settings. The directionality of citations is clear in the reference counts: \citet{tibshirani1996} cites twenty papers, but is cited by tens of thousands of papers. If we do not distinguish between papers cited and citing papers, we might fail to distinguish between the genomics literature (incoming co-topic) and the variable selection literature (outgoing co-topic), as well as differing propensities to cite and to be cited.
\end{example}

In the co-factor model, conditional on the latent factors, each edge $A_{ij}$ of the network is sampled independently from a distribution with expectation $\mathcal A \equiv \mathbb{E} (A \mid Z, B, Y) = Z B Y^T \in \mathbb R$ where $B \in \mathbb{R}^{k \times k}$ is a mixing matrix that controls how the outgoing and incoming latent factors interact. In the citation setting, $\mathcal A$ represents the similarities between documents in the latent topic space. $B$ is a weighting matrix that describes how likely it is that a document $i$ loading on outgoing factor $Z_{\cdot k}$ forms an edge to a document $j$ loading on incoming factor $Y_{\cdot \ell}$. As the $B$-mediated similarity between the outgoing topic of document $i$ and the incoming similarity of document $j$ increases, (i.e. $\mathcal A_{ij}$ gets larger), the probability of citation $i \to j$ goes up.

For the co-factor model to be identified, the co-factors $Z$ and $Y$ and the mixing matrix $B$ must satisfy several assumptions: the mixing matrix $B$ must be full rank, the rows of $Z$ and $Y$ must be independent and identically distributed (that is, $Z_{1 \cdot}, Z_{2 \cdot}, ..., Z_{n \cdot}$ must be i.i.d, and $Y_{1 \cdot}, Y_{2 \cdot}, ..., Y_{n \cdot}$ must be i.i.d.), and the distribution of the $Z_{i \cdot}$ and $Y_{i \cdot}$ must be leptokurtic (i.e., skewed). Skewness is the key assumption for $Z$ and $Y$ to be identified. When $Z$ and $Y$ come from leptokurtic distributions, the co-factors $Z$ and $Y$ are identified up to sign-flips and permutations of the column order.

The co-factor model is similar in form to mixed membership stochastic blockmodels, and generalizes the mixed membership stochastic blockmodel (see the supplement of \citet{rohe2023} for a precise characterization). Unlike mixed membership stochastic blockmodels, the rows of $Z$ and $Y$ do not need to normalized, and can take on negative values. In practice, the sign ambiguity of $Z$ and $Y$ can almost always be resolved by forcing the columns of $Z$ and $Y$ to be skew positive, in which case $Z$ and $Y$ typically consist of sparse, axis-aligned, positive values. The sparsity of $Z$ and $Y$ often enables substantive interpretations of the latent factors, as each node typically loads on a small number of factors.

\subsection{Chronological observation mechanism}

To specialize the co-factor model to the citation setting, we incorporate an observation mechanism.

\begin{definition}
  Given a corpus of documents $i = 1, ..., n$ published at times $T_1, ..., T_n$, the partially observed adjacency matrix is
  \begin{equation}
    \begin{aligned}
      A_{ij} =
      \begin{cases}
        1                 & \text{if $T_j \le T_i$ and $i$ cites $j$,}              \\
        0                 & \text{if $T_j \le T_i$ and $i$ does not cite $j$, and } \\
        \text{unobserved} & \text{if $T_j > T_i$.}
      \end{cases} \label{def:topical-similarity-adjacency-matrix}
    \end{aligned}
  \end{equation}
  For convenience, we re-index the documents in order of publishing times, forcing $T_1 \ge  ... \ge T_n$, such that $T_1$ is the most recent publishing time, and $T_n$ is the earliest publishing time. Using this indexing scheme, the observed portion of the network is nearly upper triangular, but elements can occur in the lower triangle when $T_i = T_j$.
\end{definition}

Under the citation observation mechanism, citations from older papers to newer papers are missing. This is because the lack of citations from older papers to newer papers should be uninformative about the outgoing co-factor of the older paper and the incoming co-factor of the newer paper.

If we presume that the older paper definitively cites the newer paper, or definitively does not cite the newer paper, this will force the corresponding co-factors closer together or farther apart in the latent topic space. \texttt{CitationImpute} thus treats citations forward-in-time as missing rather than precisely observed zeroes or ones. This allows the estimation procedure to spectrally infer co-factors without introducing chronological artifacts.

\begin{example}
  Consider \citet{hoerl1970}, which introduced \texttt{RIDGE} regression. Since the \texttt{RIDGE} paper was published long before the \texttt{LASSO} paper, \citet{hoerl1970} does not cite \citet{tibshirani1996}. But, since \texttt{RIDGE} regression and \texttt{LASSO} regression are closely related, it is plausible that the two papers are close to each other in outgoing topic space. The impossibility of citation forward-in-time is uninformative about the latent similarity between the two papers.
\end{example}

\begin{remark}
  The chronological observation mechanism is only relevant if citations are directed relationships. If there is no semantic information contained in the direction of a citation, we can impute the lower triangle of $A$ based on the upper triangle of $A$ by setting $A_{ij} = A_{ji}$ for all missing edges.
\end{remark}

\begin{remark}
  In some settings, such as the scientific literature, documents might build on each other, with later documents iterating on past work. In a co-factor model, one could argue that this should be modeled as dependence amongst the latent factors. We are not aware of any approaches to handle such dependence, but believe they are an interesting topic for future work.
\end{remark}

\subsection{Statistical identification of latent co-factors}

The chronological observation mechanism presents several challenges. First, it is unclear if the co-factors $Z$ and $Y$ are identified based on the information observed in the upper triangle of $A$. 

In Proposition \ref{prop:identifiability}, we show that outgoing community memberships $Z_{i \cdot}$ are identified for all but the very earliest documents, and that the incoming community memberships $Y_{i \cdot}$ are identified for all but the most recent documents. Some co-factors are unidentified because the most recent documents have not been around long enough to possibly be cited by papers from all topics and because oldest documents were written too early to possibly cite papers from all topics.

\begin{figure}[ht]
  \centering
  \includegraphics[width=0.5\textwidth]{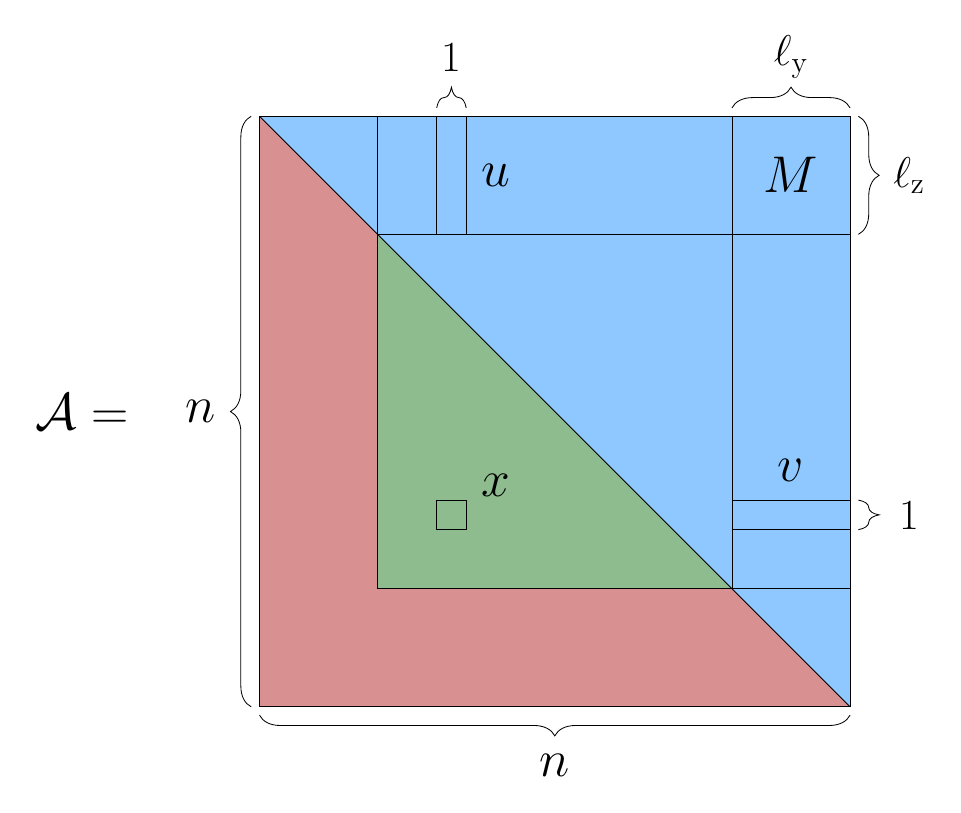}
  \caption{A decomposition of a conditional expectation matrix $\mathcal A$. Elements in the upper triangle are observed in the citation setting. We would like to recover elements of the lower triangle based on the information in the upper triangle. When $\rank(\mathcal A) = \rank(M)$, a portion of the lower triangle is identified, but the left-most rows and bottom-most columns cannot be recovered.}
  \label{fig:understanding-identification}
\end{figure}

More precisely, Proposition \ref{prop:identifiability} states that if the conditional expectation of a citation network $\mathcal A$ is rank $k$ and the $\ell_\text{z} \times \ell_\text{y}$ submatrix in the top right of $\mathcal A$ is rank $k$, it is possible to reconstruct all of $\mathcal A$ except for the elements in the last $\ell_\text{z}$ rows and the elements in the first $\ell_{y}$ columns. Observing a full rank matrix $M$ in the top right of $\mathcal A$ ensures that no information is hidden in the lower triangle (see Figure \ref{fig:understanding-identification}).

The statement of Proposition \ref{prop:identifiability} requires some additional notation. Let $\mathscr R_{n, k}$ be the set of rank $k$ matrices contained in $\mathbb{R}^{n \times n}$. Imagine that $\mathcal A, \mathcal B \in \mathscr R_{n, k}$ are the conditional expectations of two semi-parametric factor models. $P_U(\mathcal A)$ and $P_U(\mathcal B)$ are projections of $\mathcal A$ and $\mathcal B$ onto the space of upper triangular matrices. $P_U(\mathcal A)$ and $P_U(\mathcal B)$ represent the conditional expectations of the observed portion of $\mathcal A$ and $\mathcal B$, respectively.

\begin{proposition}
  \label{prop:identifiability}
  Let $\mathcal A, \mathcal B \in \mathscr R_{n, k}$. If $P_U(\mathcal A) = P_U(\mathcal B)$ and there exist $\ell_\text{z}, \ell_\text{y} \in \{k, ..., n/2\}$ such that $M = \mathcal A_{[1:\ell_\text{z}, (n - \ell_\text{y}):n]}$ has rank $k$, then $\mathcal A_{ij} = \mathcal B_{ij}$ for all $i, j \in \mathbb Z$ satisfying $1 < i \le (n - \ell_\text{z})$ and $\ell_\text{y} < j \le n$.
\end{proposition}

\begin{remark}
  Proposition \ref{prop:identifiability} can be understood constructively as approximating $\mathcal A$ from $P_U(A)$ using the Nystr\"om method \citep{drineas2005, gittens2016}. Because $\mathcal A$ is rank $k$, the Nystr\"om method has zero approximation error.
\end{remark}

When the conditions of Proposition \ref{prop:identifiability} are violated, it is possible that no elements in the lower triangle of $\mathcal A$ are recoverable. That is, if $M$ is rank $k$ only for $\ell_\text{z}, \ell_\text{y} > n / 2$, there is information hidden in the lower triangle of $\mathcal A$ that is lost upon projecting onto the upper triangle. A concrete example where recovery is impossible is given by the following matrix $\mathcal A$, which has rank two. Let $J_{n}$ denote an $n \times n$ matrix of ones and suppose
\begin{equation*}
  \mathcal A
  =
  \begin{bmatrix}
    a \, J_{n/2} & a \, J_{n/2} \\
    b \, J_{n/2} & a \, J_{n/2}
  \end{bmatrix}.
\end{equation*}
\noindent Every element in the upper triangle of $\mathcal A$ is $a$, and thus there is no way of estimating $b$ when the lower triangle of $\mathcal A$ is missing. $\mathcal A$ corresponds to a two-block stochastic blockmodel where the first $n/2$ documents are in one block and the last $n/2$ documents are in a separate block. Because these blocks of documents do not overlap in time (here represented by node order) and they have asymmetric citation probabilities, all information about asymmetric citation probabilities is lost. However, if $n$ is large and the rows and columns of $\mathcal A$ are permuted according to the same random permutation, then there are $\ell_\text{z}, \ell_\text{y} \ll n/2$ that satisfies the conditions of Proposition \ref{prop:identifiability} with high probability.

For balanced stochastic co-blockmodels, such as the one used in the simulation study, it is sufficient to take $\ell_\text{z} = \ell_\text{y} = 2 k \log k$ to achieve identifiability with high probability. This demonstrates that identification with $\ell_\text{z}, \ell_\text{y} \ll n/2$ is reasonable in blockmodel-like settings.

\begin{proposition}
  \label{prop:coupon}
  Suppose $\mathcal A$ characterizes the expected adjacency matrix of the simulation test-bed model (Definition \ref{def:cite-co-sbm}). Let $\ell_\text{z} = \ell_\text{y} = 2 k \log k$  and let $M$ be as defined in Proposition \ref{prop:identifiability}. Then $\mathbb P(\rank(M) = k) = 1 - 2 n^{-1}$.
\end{proposition}

\section{Spectral estimation}
\label{sec:algorithm}

Spectral clustering typically proceeds in three steps. First, the network is represented as a matrix, often the adjacency matrix, but sometimes normalized or regularized versions of the graph Laplacian. Second, the leading singular vectors of this matrix are estimated, which associates each node in the graph with a point in Euclidean space. Lastly, these node embeddings are analyzed using standard methods for Euclidean data. While this estimation strategy may seem ad hoc, spectral estimators performs statistical inference under network models that are identified by their singular subspaces, a large class of models that includes stochastic blockmodels and many generalizations thereof \citep{vonluxburg2007, ji2016, jin2015, lei2015, rohe2016, lyzinski2017, athreya2015, athreya2018, lyzinski2014, priebe2019}. 

\subsection{The algorithm}
\label{subsec:the-algorithm}

\texttt{CitationImpute} adapts the standard spectral estimation pipeline to the citation setting. The main difference is that we cannot estimate singular subspaces of the adjacency matrix using a singular value decomposition, due to missing data. Instead, \texttt{CitationImpute} uses the \texttt{AdaptiveImpute} algorithm of \citet{cho2019}, which is a self-tuning variant of the \texttt{softImpute} algorithm of \citet{mazumder2010}.

\texttt{CitationImpute} accepts as input a network adjacency matrix $A \in \mathbb{R}^{n \times n}$ where the lower triangle is assumed to be mostly missing, a desired number of co-factors $k \in \{2, ..., n\}$, and clipping parameters $\ell_\text{z}, \ell_\text{y} \in \{k, ..., n/2\}$. The algorithm then proceeds as follows.

\begin{enumerate}
  \item Set all elements in the first $\ell_\text{y}$ columns of $A$ and last $\ell_\text{z}$ rows of $A$ to zero. This means that edges corresponding to the unidentified rows of $Z$ and $Y$ are ignored during estimation; see Proposition \ref{prop:identifiability} for details.
  \item Estimate the singular vectors and singular values of $A \approx \widehat U \widehat D \widehat V^T$ using \texttt{AdaptiveImpute} \citep{cho2019}. In Section \ref{subsec:the-computational-problem} we describe why a naive implementation is computationally infeasible, and in Section \ref{subsec:fast-adaptive-impute} we outline our computational contributions and a practical implementation of \texttt{AdaptiveImpute} for upper triangular data.
  \item Compute the varimax rotations of $\widehat U$ and $\widehat V$ and construct rotated singular vector matrices $\widehat Z, \widehat B$ and $\widehat Y$, respectively. We briefly review varimax rotation in Section \ref{subsec:varimax-details}.
\end{enumerate}

\texttt{CitationImpute} has several hyperparameters: the number of desired co-factors $k$, and the clipping parameters $\ell_\text{z}$ and $\ell_\text{y}$. In simulations, we find that $\ell_\text{z} = \ell_\text{y} = n / 10$ are good default values, but recommend applying domain knowledge as appropriate, and conducting a sensitivity analysis (see Section \ref{subsec:application:methods} for an example).

\subsection{The computational problem}
\label{subsec:the-computational-problem}

One contribution of this paper is a collection of algebraic identities (Propositions \ref{prop:multiplication-identity} and \ref{prop:alpha-identity}) that allow for an efficient implementation of \texttt{AdaptiveImpute} on citation matrices with hundreds of thousands of documents.

To understand why these identities are useful, we must disambiguate between two senses of sparsity. A matrix is \emph{sparse} if most of its elements are zero. These matrices can be represented very efficiently on a computer by recording only the small number of non-zero elements and their indices. On the other hand, a matrix is \emph{sparsely observed} if only a few of its entries are observed, regardless of the value of those entries. These two notations of sparsity are often conflated, and sparsely observed matrices are often represented as sparse matrices, where implicit zeroes are considered missing, and the observed zeroes must be explicitly tracked.

In the citation setting, the data matrix $A$, as defined in $\eqref{def:topical-similarity-adjacency-matrix}$, is densely observed; at least half of the entries are defined by the data.  However, in the portion of the network that is observed, the data is sparse, i.e., mostly zero-valued. Thus, the usual conflation of sparse and sparsely observed matrices leads to issues: there are $n \, (n - 1) / 2$ elements in the upper triangle of $A$ that must be explicitly tracked even if they are zero. Using this representation, even moderately sized corpora cannot be held in memory on commodity hardware. Beyond memory considerations, adding approximately $n \, (n - 1) / 2$ explicit zeroes to a sparse matrix slows down matrix operations like matrix-vector multiplication.

This makes matrix completion algorithms infeasible in both time and space when using the naive sparse representation of $A$. Both \texttt{AdaptiveImpute} and \texttt{softImpute} rely on iterated singular value decompositions of a running low-rank approximation $\tilde A^{(t)}$ to $A$. In the typical setting where the number of nodes is $n$, the rank of the decomposition is $k$, and $n \gg k$, naively taking a singular value decomposition of $\tilde A^{(t)}$ has time complexity per iteration $\mathcal O \left( n^2 \, k \right)$. This high computational complexity constrains researchers to inference on networks with at most thousands of nodes.

We are able to reduce the both the time and space complexity of the matrix completion problem. The solution requires leveraging the fact that $A$ is sparse, even if it is not sparsely observed. In particular, there is no need to explicitly track zeroes in the upper triangle of $A$, and $A$ may be represented as a sparse matrix that records only non-zero elements of $A$ and zeroes in the lower triangle of $A$.  Using this representation, with some algebraic tricks, all the operations necessary for \texttt{AdaptiveImpute} are computationally feasible. In brief, by representing $\tilde A^{(t)}$ as the sum of four carefully constructed matrices, we can reduce the naive time complexity from $\mathcal O(n^2 \, k)$ down to $\mathcal O( |\tilde \Omega| \, k + n \, k^2)$, where $|\tilde \Omega|$ is the number of observed non-zero elements of $A$. In real world datasets $|\tilde \Omega|$ represents the number of citations between documents, and empirical evidence suggests that each document in a citation network cites a fixed number of other documents, regardless of the overall size of the corpus. That is, $|\tilde \Omega|$ is $\mathcal O(n)$. Thus the effective per-iteration runtime reduces from $\mathcal O (n^2 \, k)$ to $\mathcal O (n \, k^2)$.

\subsection{\texttt{AdaptiveImpute}}
\label{subsec:fast-adaptive-impute}

The \texttt{AdaptiveImpute} algorithm is similar to \texttt{softImpute} \citep{hastie2015}, with two key differences.  First, \texttt{AdaptiveImpute} initializes with a debiased singular value decomposition. Second, on each iteration, \texttt{AdaptiveImpute} adaptively varies the \texttt{softImpute} thresholding parameter. This procedure is defined in Algorithm \ref{alg:adaptive-impute}, which is identical to the algorithm as defined in \citep{cho2019} but with some minor notation changes and the introduction of a maximum number of iterations $T$.

\begin{algorithm}
  \DontPrintSemicolon
  \KwIn{partially observed matrix $A \in \mathbb{R}^{n \times n}$, rank $k \in \{2, ..., n\}$, convergence tolerance $\varepsilon > 0$, and maximum allowable iterations $T \in \mathbb{Z}^+$.}
  
  $Z^{(1)} \gets \texttt{AdaptiveInitialize}(A, k)$ \;
  
  \Repeat{$\| Z^{(t+1)} - Z^{(t)} \|^2_F / \| Z^{(t+1)} \|^2_F < \varepsilon$ \text{ or } $t \ge T$}{
    $\tilde A^{(t)} \gets P_\Omega (A) + P_\Omega^\perp \left( Z^{(t)} \right)$ \;
    $\widehat V_i^{(t)} \gets \bv_i \left( \tilde A^{(t)} \right)$ for $i = 1, ..., k$ \;
    $\widehat U_i^{(t)} \gets \bu_i \left( \tilde A^{(t)} \right)$ for $i = 1, ..., k$ \;
    $\tilde \alpha^{(t)} \gets \displaystyle \frac{1}{n - k} \sum_{i=k+1}^n \bl_i^2 \left( \tilde A^{(t)} \right)$ \;
    $\hat \lambda_i^{(t)} \gets \sqrt{\bl_i^2 \left( \tilde A^{(t)} \right) - \tilde \alpha^{(t)}}$ for $i = 1, ..., k$ \;
    $Z^{(t+1)} \gets \sum_{i=1}^k \hat \lambda_i^{(t)} \widehat U_i^{(t)} \widehat V_i^{(t)^T}$ \;
    $t \gets t + 1$ \;
  }
  \Return{$\hat \lambda_i^{(t)}, \widehat U_i^{(t)}, \widehat V_i^{(t)}$ for $i = 1, ..., k$}\;
  \caption{\textsc{AdaptiveImpute}}
  \label{alg:adaptive-impute}
\end{algorithm}

The initializer is given by running Algorithm \ref{alg:adaptive-initialize}, which we defer to the appendix. If we compute $Z^{(1)}$ by taking a rank $k$ singular value decomposition of $P_\Omega(A)$ and fix $\alpha^{(t)} = \lambda$ for all $t$ (note that $\tilde \alpha^{(t)}$ is the data adaptive thresholding parameter), \texttt{AdaptiveImpute} reduces to \texttt{softImpute}. This implies that a naive implementation of \texttt{AdaptiveImpute} inherits the per-iteration time complexity of \texttt{softImpute}, which is $\mathcal O( |\Omega| \, k + n \, k^2)$, plus the cost of evaluating $\tilde \alpha^{(t)}$.

\subsubsection{Feasible implementation}

In practice, the runtime for each iteration of \texttt{AdaptiveImpute} and \texttt{softImpute} is dominated by the singular value decomposition, which is computed using an algorithm such as the implicitly restarted Lanczos bidiagonalization algorithm. The time complexity of this decomposition depends fundamentally on an underlying bidiagonalization subroutine (Algorithm \ref{alg:lanczos_bidiagonalization}), and the time complexity of the bidiagonalization subroutine in turn depends on cost of left and right matrix-multiplication of $\tilde A^{(t)}$ with an appropriately sized vector \citep{baglama2005}.

When $A$ is sparsely observed, $\tilde A^{(t)}$ can be expressed as a sparse matrix plus a low-rank matrix
\begin{equation*} \begin{aligned}
    \tilde A^{(t)}
     & = \underbrace{P_\Omega (A - Z^{(t)})}_\text{sparse} + \underbrace{Z^{(t)}}_\text{low-rank},
     & \text{sparsely observed setting}
  \end{aligned} \end{equation*}
\noindent and matrix-vector multiplication has time complexity $\mathcal O(|\Omega| \, k)$ for the sparse part and $\mathcal O(n \, k^2)$ for the low-rank part. In the citation setting, naively re-using this decomposition in the bidiagonalization subroutine is inefficient since $|\Omega| \approx n^2 / 2$.

However, a similar trick can improve the time complexity of multiplication with $\tilde A^{(t)}$: we can drop observed zeroes from consideration if we partition $\tilde A^{(t)}$ carefully. Since $P_{\tilde \Omega} (A) = P_\Omega (A)$, we can compute only on $\tilde \Omega$. Let $U = \{(i, j) : i < j\}$ denote the indices of the upper triangle of $A$ and $L$ denote the indices of the observed elements of $A$ on the lower triangle, such that $\Omega = U \cup L$. Then
\begin{equation*} \begin{aligned}
    \tilde A^{(t)}
     & = P_\Omega(A) + P_\Omega^\perp (Z^{(t)})                                           & \text{citation setting} \\
     & = P_\Omega(A) - P_\Omega (Z^{(t)}) + P_\Omega (Z^{(t)}) + P_\Omega^\perp (Z^{(t)})                           \\
     & = P_{\tilde \Omega} (A) - P_\Omega(Z^{(t)}) + Z^{(t)}                                                        \\
     & = \underbrace{P_{\tilde \Omega} (A)}_\text{sparse} -
    \underbrace{P_{L}(Z^{(t)})}_\text{sparse} -
    \underbrace{P_U (Z^{(t)})}_\text{low-rank until projection} +
    \underbrace{Z^{(t)}}_\text{low-rank}.
  \end{aligned} \end{equation*}
Efficient implementation strategies for matrix-vector multiplications  with the sparse and low-rank terms are well known. This leaves the $P_U (Z^{(t)})$ term, which is low-rank until it is projected onto the upper triangle. There one can use the same implementation strategy as for the low-rank component, but summing over fewer indices.

\begin{proposition}
  \label{prop:multiplication-identity}
  Let $Z^{(t)} \in \mathbb{R}^{n \times n}$ be a rank $k$ matrix with singular value decomposition $Z^{(t)} = U D V^T$ and let $x \in \mathbb{R}^n.$ Then
  \begin{equation*} \begin{aligned}
      \left[ P_U (Z^{(t)}) \, x \right]_i = \langle U_{i \cdot}, \tilde W_i \rangle,
    \end{aligned} \end{equation*}
  where $\tilde W_{ki} = \sum_{j=i+1}^n W_{kj}$ and $W_{\cdot j} = (DV^T)_{\cdot j} \cdot x_j$.
\end{proposition}

We defer the proof to the appendix. Proposition \ref{prop:multiplication-identity} is a straightforward result that suggests a computational scheme for evaluating the term $P_U (Z^{(t)}) \, x$. In particular, it suggests constructing $W$, then $\tilde W$, and then obtaining elements of $P_U (Z^{(t)}) \, x$ element by element. This procedure requires $\mathcal O(n \, k^2)$ flops as opposed to the $\mathcal O(n^2 \, k)$ flops of a naive implementation. The left-multiplication case is analogous.

The last requirement to implement \texttt{AdaptiveImpute} is a similarly efficient calculation of $\alpha^{(t)}$.

\begin{proposition}
  \label{prop:alpha-identity}
  
  Let $\tilde A^{(t)}, Z^{(t)}$ and $\alpha^{(t)}$ be as defined in Algorithm \ref{alg:adaptive-impute}. Recall that $Z^{(t)}$ is a low-rank matrix of the form $U D V^T$ with $U, V \in \mathbb{R}^{n \times k}$ orthonormal and $D \in \mathbb{R}^{k \times k}$ diagonal. Then
  \begin{equation*} \begin{aligned}
      \alpha^{(t)}
      = \frac{1}{n - k} \left[
        \left \Vert P_{\tilde \Omega} (A) \right \Vert_F^2 +
        \left \Vert Z^{(t)} \right \Vert_F^2
        - \left \Vert P_L \left( Z^{(t)} \right) \right \Vert_F^2
        - \left \Vert P_U \left( Z^{(t)} \right) \right \Vert_F^2
        - \sum_{i = 1}^k \bl_i^2 \left( \tilde A^{(t)} \right)
        \right].
    \end{aligned} \end{equation*}
  Additionally, define $U^{r q} \in \mathbb{R}^n$ and $V^{r q \triangle} \in \mathbb R^n$ such that
  \begin{equation*} \begin{aligned}
      U^{r q}_i = U_{i r} \, U_{i q}, \qquad \text{ and } \qquad
      V^{r q \triangle}_i = \sum_{j=i+1}^n (DV)^T_{r j} (DV)^T_{q j} \qquad \forall i = 1, ..., n.
    \end{aligned} \end{equation*}
  Then
  \begin{equation*} \begin{aligned}
      \left \Vert P_U \left( Z^{(t)} \right) \right \Vert_F^2
      = \sum_{r=1}^k \sum_{q=1}^k \left \langle U^{r q}, V^{r q \triangle} \right \rangle.
    \end{aligned} \end{equation*}
\end{proposition}

To understand the computational complexity of this expression we proceed term by term. First, consider the $\sum_{i = 1}^k \bl_i^2 (\tilde A^{(t)})$ term. Each iteration of \texttt{AdaptiveImpute} computes a truncated singular value decomposition of $\tilde A^{(t)}$ of rank $k$ before computing $\alpha^{(t)}$, so evaluating this term is a trivial $\mathcal O(k)$ summation since $\bl_i (\tilde A^{(t)})$ is available for $i = 1, ..., k$. Next, observe that $\Vert P_{\tilde \Omega} (A) \Vert_F^2$ and $ \Vert P_L ( Z^{(t)} ) \Vert_F^2$ are collectively $\mathcal O(\vert \tilde \Omega \vert \, k)$. This leaves the terms $\Vert Z^{(t)} \Vert_F^2$ and $\Vert P_U (Z^{(t)}) \Vert_F^2$, both of which can require $\mathcal O(n \, k^2)$ flops. As in Proposition \ref{prop:multiplication-identity}, the idea is that evaluating $\Vert P_U (Z^{(t)}) \Vert_F^2$ is essentially the same evaluating $\Vert Z^{(t)} \Vert_F^2$ case, modulo some care while indexing. The time complexity to compute $\alpha^{(t)}$ is then $\mathcal O( \vert \tilde \Omega \vert \, k + n \, k^2)$ flops. Using this scheme to evaluate $\alpha^{(t)}$, the overall time complexity of each iteration of \texttt{AdaptiveImpute} is $\mathcal O( \vert \tilde \Omega \vert \, k + n \, k^2)$.

\begin{figure}[ht]
  \centering
  \includegraphics[width=\textwidth]{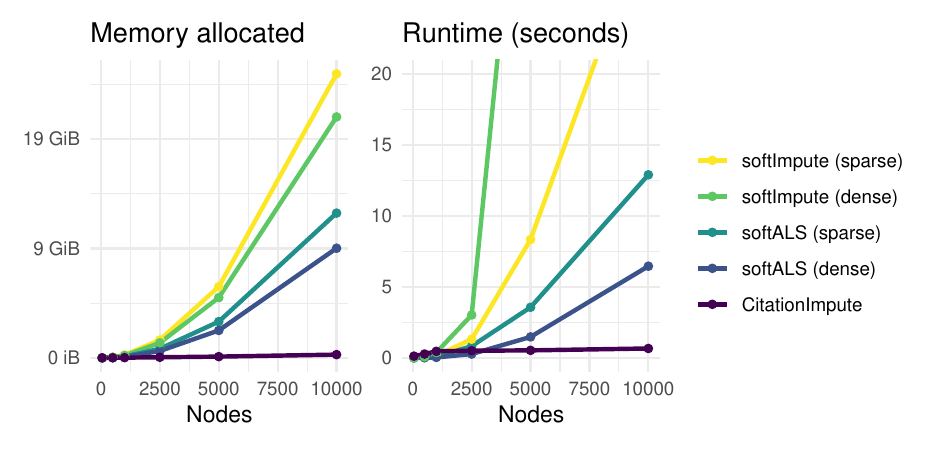}
  \caption{Comparison of memory and time complexity of \texttt{CitationImpute} with some existing options for low-rank matrix completion when applied our simulation test-bed model (Definition \ref{def:cite-co-sbm}). Some run-times are truncated at 20 seconds in the right panel. Each estimator is iterative; we compare time and memory use for five iterations. Existing implementations for both sparse and dense data representations are memory bound and do not scale to networks with more than several thousand nodes. Our implementation, although un-optimized, uses less memory and is faster for large networks.}
  \label{fig:memory-use}
\end{figure}

\subsection{Varimax rotation}
\label{subsec:varimax-details}

After obtaining an estimated singular value decomposition $A \approx \widehat U \widehat D \widehat V^T$ from \texttt{AdaptiveImpute}, \texttt{CitationImpute} varimax rotates the estimates to obtain latent factors for each node. Given an $n \times k$ matrix orthonormal matrix $U$, varimax rotation finds a $k \times k$ orthogonal matrix $R$ that maximizes
\begin{equation*}
  v(R, U) = \sum_{\ell=1}^{k} \frac{1}{n} \sum_{i=1}^{n} \left(
  [U R]_{i \ell}^{4} - \left( \frac{1}{n} \sum_{j=1}^{n}[U R]_{j \ell}^{2} \right)^{2}
  \right)
\end{equation*}
\noindent over the set of $k \times k$ orthonormal matrices. In particular, it compute $\widehat R_U$ that maximizes $v(\cdot, \widehat U)$ and $\widehat R_V$ that maximizes $v(\cdot, \widehat V)$ where $\widehat R_U$ and $\widehat R_V$ are $k \times k$ orthonormal matrices. Calculating these rotation matrices is a routine operation available in many statistical packages. After the rotation matrices $\widehat R_U$ and $\widehat R_V$ have been found, the latent factors are estimated as
\begin{align}
  \widehat{Z}=\sqrt{n} \widehat{U} R_{\widehat{U}},
  \quad \widehat{Y}=\sqrt{n} \widehat{V} R_{\widehat{V}},
  \quad \text { and } \quad \widehat{B}=R_{\widehat{U}}^{T} \widehat{D} R_{\widehat{V}} / n. \label{eq:rotation}
\end{align}
\citet{rohe2023} show that, for $\widehat U, \widehat V$ obtained from the singular value decomposition in the fully observed case, varimax rotated estimates $\widehat Z, \widehat B$ and $\widehat Y$ are consistent for population terms $Z, B$ and $Y$.

\section{Simulation study}
\label{sec:simulation-study}

To assess the performance of \texttt{CitationImpute}, we perform a simulation study using a co-stochastic blockmodel, a sub-model of the co-factor model. In the simulation study, \texttt{CitationImpute} recovers singular subspaces of $\mathcal A$ and the latent factors $Z$ and $Y$ at the same rate as an oracle estimator that has access to all of $A$. The simulations also show that naive imputation of missing data leads to inconsistent estimates.

For the simulations, we use a Poisson degree-corrected stochastic co-blockmodel subject to lower triangular missingness.

\begin{definition}[Degree-corrected stochastic co-blockmodel]
  The \emph{degree-corrected stochastic co-blockmodel} is random graph model on $n$ nodes. Each node $i$ is assigned an incoming community $z(i) \in \{1, ..., k\}$ and an outgoing community $y(i) \in \{1, ..., k\}$ according to parameters $\pi^\text{in} \in [0, 1]^k$ and $\pi^\text{out} \in [0, 1]^k$, such that $\mathbb P(z(i) = j) = \pi^\text{in}_j$ and $\mathbb P(y(i) = j) = \pi^\text{out}_j$ for $j \in \{1, ..., k\}$. Each node $i$ is also assigned a propensity $\theta^\text{out}_i \in \mathbb{R}_+$ to send edges, and a propensity $\theta^\text{in} \in \mathbb{R}_+$ to receive edges. Conditional on community memberships and edge formation propensities, integer-valued edges occur independently according to a Poisson distribution with expectation.
  \begin{equation*}
    \mathbb E(A_{ij} \mid z(i), y(j)) = \theta^\text{out}_i \, B_{z(i), y(j)} \, \theta^\text{in}_j.
  \end{equation*}
  \noindent where $B \in [0, 1]^{k \times k}$ is a rank $k$ mixing matrix denoting propensities of edge formation between communities. $B$ can be rescaled by a constant to enforce that the expected density of edges in the network is $\rho$.
\end{definition}

The idea behind the simulation model is to mimic the behavior we expect in citation networks, where papers in a given field will primarily cite papers from that same field (strong diagonal structure in $B$), but will intermittently cite papers from other fields (some active elements of $B$ on the off-diagonal). This is motivated by the observation that the topics that \citet{tibshirani1996} cites and the topics that cite \citet{tibshirani1996} are distinct.

\begin{definition}[simulation model]
  \label{def:cite-co-sbm}
  The simulation model is a degree-corrected stochastic co-blockmodel with $n$ nodes, $k$ co-communities, and expected density $\rho = 0.15$. Let $\pi^\text{in}_j = \pi^\text{out}_j = 1 / k$ for $j = 1, ..., k$, such that the co-communities are balanced. Let be $\theta^\text{in}$ and $\theta^\text{out}$ be generated by sampling $n$ independent realizations from an exponential distribution with mean eight, and then adding one to each realization, inducing some degree-heterogeneity. The diagonal elements of $B$ are set to $B_\text{within} = 0.8$. $k$ elements of the off-diagonal to $B_\text{between} = (B_\text{within} / 3 - (k - 2) \, B_\text{inactive})$ (in particular, the off-diagonal values in the first row of $B$, and the last element of the second column of $B$). The remaining elements of the off diagonal to $B_\text{inactive} = 0.01$. This ensures that $B$ is rank $k$ and that there is strong assortative structure in the network. In the simulations, we use $k \in \{3, 6, 9\}$, with corresponding values of $B_\text{between}= 0.257, 0.227, 0.197$.
\end{definition}

We compare the \texttt{CitationImpute} to an oracle estimator with access to the full data $A$, and also two imputation estimators. In total, we compare four estimators:

\begin{enumerate}
  \item \texttt{CitationImpute}, with $\ell_\text{z} = \ell_\text{y} = n / 10$,
  \item singular value decomposition applied after imputing all missing data as zeros (call this the \emph{zero-imputed} estimator),
  \item singular value decomposition applied after imputing all missing data by symmetrizing the observed data (call this the \emph{symmetrized} estimator), and
  \item oracle singular value decomposition applied to a fully observed similarity data (call this the \emph{fully observed} estimator).
\end{enumerate}

For the last three estimators, after estimating singular subspace, the singular vectors are varimax rotated according to \eqref{eq:rotation} to obtain co-factor estimates.

To measure how well various estimators recover the singular subspaces of $\mathcal A$, we compute the $\sin \Theta$ distance between the subspaces spanned by $U$ and $\widehat U$ \citep{vu2013, bhatia1997}, for identified rows only. Given two orthonormal bases $U \in \mathbb{R}^{n \times k}$ and $\widehat U \in \mathbb{R}^{n \times k}$, the singular values $\sigma_1, ..., \sigma_k$ of $U^T \widehat U$ are the cosines of the principal angles $\cos \theta_1, ..., \cos \theta_k$ between the span of $U$ and the span of $\widehat U$. Define $\sin \Theta(U, \widehat U)$ to be a diagonal matrix containing the sine of the principle angles of $U^T \widehat U$. Then the $\sin \Theta$ distance between the subspaces spanned by $U$ and $\widehat U$ is given by
\begin{equation*} \begin{aligned}
    d(U, \widehat U) = \Vert \sin \Theta (U, \widehat U) \Vert_F.
  \end{aligned} \end{equation*}
We aggregate error across identified rows of the estimates $\widehat U$ and $\widehat V$ and report a single metric
\begin{equation*} \begin{aligned}
    \mathcal L_\text{subspace} (U, \widehat U, V, \widehat V)
    = \Vert \sin \Theta (U, \widehat U) \Vert_F + \Vert \sin \Theta (V, \widehat V) \Vert_F.
  \end{aligned} \end{equation*}
To measure how well the estimators recover the latent factors $Z$ and $Y$, we report root mean squared error on individual elements of identified rows $\widehat Z$ and $\widehat Y$. Since varimax estimates $\widehat Z$ and $\widehat Y$ are only determined up to sign-flips and column reordering, this requires an alignment step to match $\widehat Z$ with $Z$, and $\widehat Y$ with $Y$. Let $\mathscr P(k)$ be the set of $k \times k$ orthogonal matrices whose entries $P_{ij}$ are elements of $\{-1, 0, 1\}$. Define
\begin{align}
  P_Z & = \argmin_{P \in \mathscr P(k)} \Vert Z - \widehat Z P \Vert_F \label{eq:P_Z}  \\
  P_Y & = \argmin_{P \in \mathscr P(k)} \Vert Y - \widehat Y P \Vert_F. \label{eq:P_Y}
\end{align}
We find $P_Z$ and $P_Y$ by using the Hungarian algorithm to match columns of the estimates $\widehat Z, \widehat Y$ to the corresponding population values $Z, Y$. Then the elementwise factor root mean squared error is
\begin{equation*} \begin{aligned}
    \mathcal L_\text{factor} (Z, \widehat Z, Y, \widehat Y)
    = \sqrt{\frac{1}{n \,  k} \left( \Vert Z - \widehat Z P_Z \Vert_F^2 + \Vert Y - \widehat Y P_Y \Vert_F^2 \right) }.
  \end{aligned} \end{equation*}

\begin{figure}
  \centering
  \includegraphics[width=0.99\textwidth]{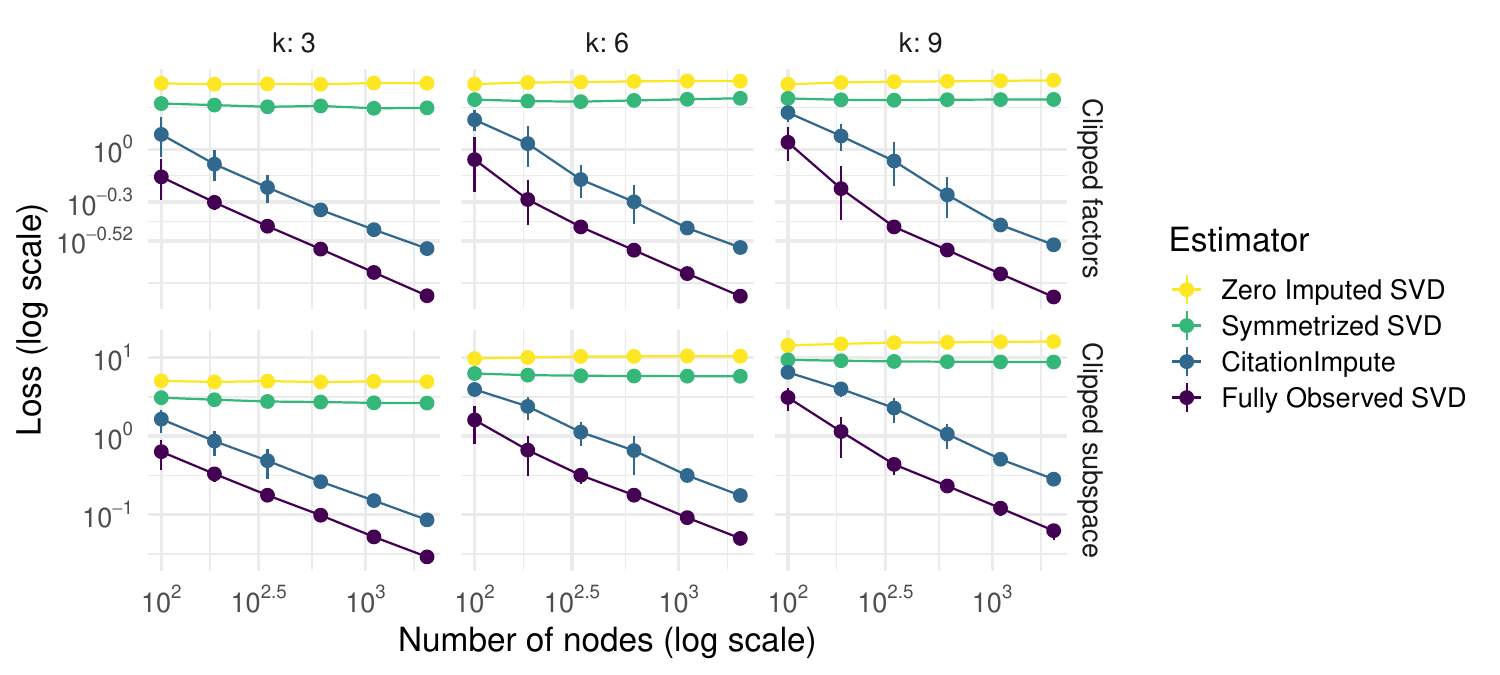}
  \caption{Average estimation error as a function of sample size, on $\log-\log$ scale. The top row of panels visualizes estimation error of the factors $Z$ and $Y$, excluding clipped factors. The bottom row of panels visualizes estimation error of the singular subspaces, again excluding clipped rows of $U$ and $V$. Each column of panels represents a simulation model with a different number of latent communities. Within each panel, each line corresponds to the loss of a single estimator. Average loss plus and minus one standard deviation are shown as a dotplot; in most cases the standard deviations are too small to see.}
  \label{fig:consistency}
\end{figure}

To perform the simulation, we evaluate the subspace loss and the factor loss 200 times for every estimator, every $k \in \{3, 6, 9\}$, and every $n \in \{100, 182, 331, 603, 1099, 2000\}$. In Figure \ref{fig:consistency}, we report the average subspace loss and the average factor loss for these combinations. Estimation error for \texttt{CitationImpute} decreases at approximately $\sqrt n$-rates, suggesting that \texttt{CitationImpute} is a consistent estimator of the singular subspaces of $\mathcal A$ and also of the latent factors $Z$ and $Y$. The rate for \texttt{CitationImpute} parallels that of the oracle estimator with access to all of $A$, although it unsurprisingly advantageous to observe the full data.

In contrast, the symmetric imputation strategy and the zero-imputation strategies are not reliable ways to estimate singular subspaces or latent factors. Estimation error for both imputation strategies is constant as a function of $n$, suggesting that estimators based on naive imputation approaches are inconsistent. The symmetric imputation strategy is always better than treating the unobserved entries as zeroes, which makes sense as the model has some underlying symmetry. Some additional simulation results investigating the imputation estimators are available in Appendix \ref{app:additional-sims}.

\section{Analysis of the statistics literature}
\label{sec:application}

We next leveraged \texttt{CitationImpute} to analyze of the academic statistics literature.

\subsection{Data}

We used proprietary Web of Science data that we obtained through an institutional agreement with Clarivate Analytics. The complete Web of Science corpus contains hundreds of millions of documents, which amount to nearly a terabyte of data. We considered only papers published in a subset of 125 journals focused on probability and statistics (see Appendix \ref{app:data-analysis-supplement} for a list of the journals). The node-induced subgraph formed by considering only these papers and the citations between them had 281,883 nodes, 2,224,775 edges, and $24,051$ weakly connected components (a weakly connected component in a subgraph where there is a path between every pair of nodes, ignoring the direction of edges). Most of the 24,051 weakly connected components were singletons. The largest weakly connected component contained 255,780 nodes and 2,222,363 edges. From this point onward, when we refer to the ``citation network'' or ``citation graph'' we are referring exclusively to this largest connected component. For each document we additionally knew the authors, publication date, and the abstract text, although some of this information was missing.

Papers in the citation network were published between 1898 and 2024. The number of citations received from other papers in the largest connected component (i.e. in-degree) ranged from 0 to 4,759 and the number of citations sent to other papers in the largest connected component (i.e. out-degree) ranged from 0 to 603. There are several articles in the sample that cite hundreds of other papers; these articles are typically bibliographies or reviews. A small number of papers mutually cited each other.

\subsection{Methods}
\label{subsec:application:methods}

First we constructed the partially observed adjacency matrix of the citation graph. We ordered nodes chronologically, and then clipped data using $\ell_\text{z} = 100,000$ and $\ell_{y} = 50,000$. This amounted to discarding outgoing citations for papers published before 2004, and incoming citations for papers published after 2018. These clipping parameters were primarily selected on the basis of domain knowledge -- we supposed that the topics in the modern statistics literature were present by 2004. We did not estimate incoming co-factors for papers published after 2018, because it can take several years to publish an academic paper, and we believed papers in 2018 are the latest papers that reasonably had the chance to be discovered, cited, and included in the Web of Science dataset.  

We then ran \texttt{AdaptiveImpute} to obtain a low-rank decomposition $A \approx \widehat U \widehat D \widehat V^T$. Here we report the results for a rank $k = 30$ decomposition. After computing a low-rank decomposition $A \approx \widehat U \widehat D \widehat V^T$, we performed varimax rotation of $\widehat U$ and $\widehat V$ to obtain a final low-rank decomposition $A \approx \widehat Z \widehat B \widehat Y^T$, as described in Section \ref{sec:algorithm}. The rows of $\widehat Z$ and the rows of $\widehat Y$ thus correspond to document-level latent co-factors \citep{rohe2023, rohe2016}. The rows of $\widehat Z$ contained outgoing-citation factors, and the rows of $\widehat Y$ contained incoming-citation factors. Both $\widehat Z$ and $\widehat Y$ were relatively sparse. To interpret the co-factors $\widehat{Y}$ and $\widehat{Z}$, we took several approaches.

First, we found keywords most associated with each factor by examining the words in paper titles following the ``best features'' approach of \citet{zhang2021d} and \citet{chen2021a}. We constructed a document-term matrix from the manuscript title. Letting $X \in \mathbb Z^{255,780 \times 11298}$, $X_{i \ell}$ indicates the number of times word $\ell$ appears in manuscript title $i$. We restricted our analysis to words that appeared in at least five manuscript titles. Then, for each factor $j$, define the sets $in(j) = \{i : \widehat Y_{ij} \ge 0\}$ and $out(j) = \{i : \widehat Y_{ij} < 0\}$. Then the importance of word $\ell$ to factor $j$ is 
\begin{equation*}
  \texttt{bff}(j, \ell) = \sqrt{
    \frac{\sum_{i \in in(j)} \widehat Y_{ij} X_{i \ell}}{\sum_{j\in in(i)} \widehat Y_{ij}}} -
  \sqrt{\frac{\sum_{i \in out(j)}X_{i \ell}}{|out(j)|}},
\end{equation*}
\noindent and in Tables \ref{tab:citation_impute_preclipped2-y-keywords-elly-50000-ellz-1e+05-k-30} and \ref{tab:citation_impute_preclipped2-z-keywords-elly-50000-ellz-1e+05-k-30} we report the six words most important to each factor. To complement this keyword analysis, we found the papers with the largest loadings for each dimension of $\widehat{Y}$ and $\widehat{Z}$, which we refer to as hub papers (see Table \ref{tab:citation_impute_preclipped2-y-hubs-elly-50000-ellz-1e+05-k-30} and \ref{tab:citation_impute_preclipped2-z-hubs-elly-50000-ellz-1e+05-k-30} in the Appendix).

\subsection{Results}

The incoming and outgoing co-factors were interpretable and associated with meaningful statistical sub-fields. We found co-factors corresponding to statistical sub-field such as GLMM(s), GEE, multiple testing, feature selection, post-selection inference, survival analysis, MCMC, causal inference, clinical trial design, experimental design, functional data, multivariate analysis, graphical models, semiparametrics, kriging, model selection (both Bayesian and frequentist).

\begin{table}
\centering
\caption{\label{tab:citation_impute_preclipped2-z-keywords-elly-50000-ellz-1e+05-k-30}Keywords for Z (outgoing citation) factors}
\centering
\resizebox{\ifdim\width>\linewidth\linewidth\else\width\fi}{!}{
\fontsize{8}{10}\selectfont
\begin{tabular}[t]{l>{\raggedright\arraybackslash}p{34em}l}
\toprule
Factor Name & Top words & ID\\
\midrule
non-convex penalties & selection, variable, dimensional, high, penalized, lasso & z01\\
experimental design & screening, dimensional, high, ultrahigh, feature, supersaturated & z02\\
bayesian spatial stats & bayesian, models, spatial, model, longitudinal, hierarchical & z03\\
post-selection inference & high, dimensional, lasso, recurrent, selection, regression & z04\\
survival analysis & survival, hazards, censored, cox, data, proportional & z05\\
\addlinespace
mixture models & selection, clustering, model, mixture, models, mixtures & z06\\
causal inference & propensity, causal, score, treatment, missing, observational & z07\\
multiple testing & false, discovery, testing, multiple, rate, microarray & z08\\
graphical models & graphical, high, dimensional, models, sparse, estimation & z09\\
bayesian non-parametrics & bayesian, dirichlet, nonparametric, mixture, clustering, process & z10\\
\addlinespace
supervised dimension reduction & dimension, reduction, sufficient, index, inverse, sliced & z11\\
times series & garch, volatility, series, models, time, change & z12\\
sparse multivariate analysis & selection, lasso, high, sparse, variable, dimensional & z13\\
kriging & spatial, spatio, temporal, gaussian, fields, bayesian & z14\\
empirical likelihood & empirical, likelihood, inference, missing, partially, jackknife & z15\\
\addlinespace
GEE & longitudinal, data, generalized, binary, estimating, clustered & z16\\
functional data & functional, data, regression, longitudinal, principal, linear & z17\\
skew normals & skew, normal, distributions, multivariate, distribution, t & z18\\
quantile regression & quantile, regression, quantiles, censored, composite, expectile & z19\\
bayesian model selection & bayesian, selection, variable, priors, prior, model & z20\\
\addlinespace
missing data & missing, imputation, data, longitudinal, with, nonignorable & z21\\
adaptive clinical trials & adaptive, trials, clinical, sequential, designs, group & z22\\
splines + random effects & models, mixed, splines, penalized, regression, additive & z23\\
multivariate analysis & high, dimensional, covariance, matrices, matrix, factor & z24\\
MCMC & bayesian, carlo, monte, mcmc, metropolis, chain & z25\\
\addlinespace
single index models & varying, coefficient, models, index, single, partially & z26\\
joint longitudinal/survival models & longitudinal, mixed, models, data, joint, effects & z27\\
causal inference reviews & causal, treatment, effects, propensity, instrumental, effect & z28\\
RIDGE & ridge, regression, estimator, liu, linear, estimators & z29\\
cure models & cure, censored, survival, model, rate, data & z30\\
\bottomrule
\end{tabular}}
\end{table}

One particularly interesting feature of the $\widehat Y$ co-factors was the presence of numerous incoming dimensions related to penalized regression. These factors covered the \texttt{LASSO} proper (y13), optimization methods for $L_1$ penalization (y14), non-convex penalties (y01), post-selection inference (y04), feature screening (y02), graphical models (y09) and RIDGE regression (y29). Several other incoming $\widehat{Y}$ co-factors were interesting because they corresponded to more niche statistical subfields. For example, we found incoming factors corresponding to empirical likelihood (y15), supervised dimension reduction (y11), and skew normals (y18). We suspect these co-factors emerged due to strong assortative structure in the sub-field: that is, a tendency to cite heavily within the factor while citing limited papers outside the factor. The tendency for spectral methods to find assortative clusters is widely known within spectral clustering literature, and it makes sense that they would pick up smaller but self-contained topics.

\begin{figure}[ht]
  \begin{minipage}{0.49\textwidth}
    \centering
    \includegraphics[width=\textwidth]{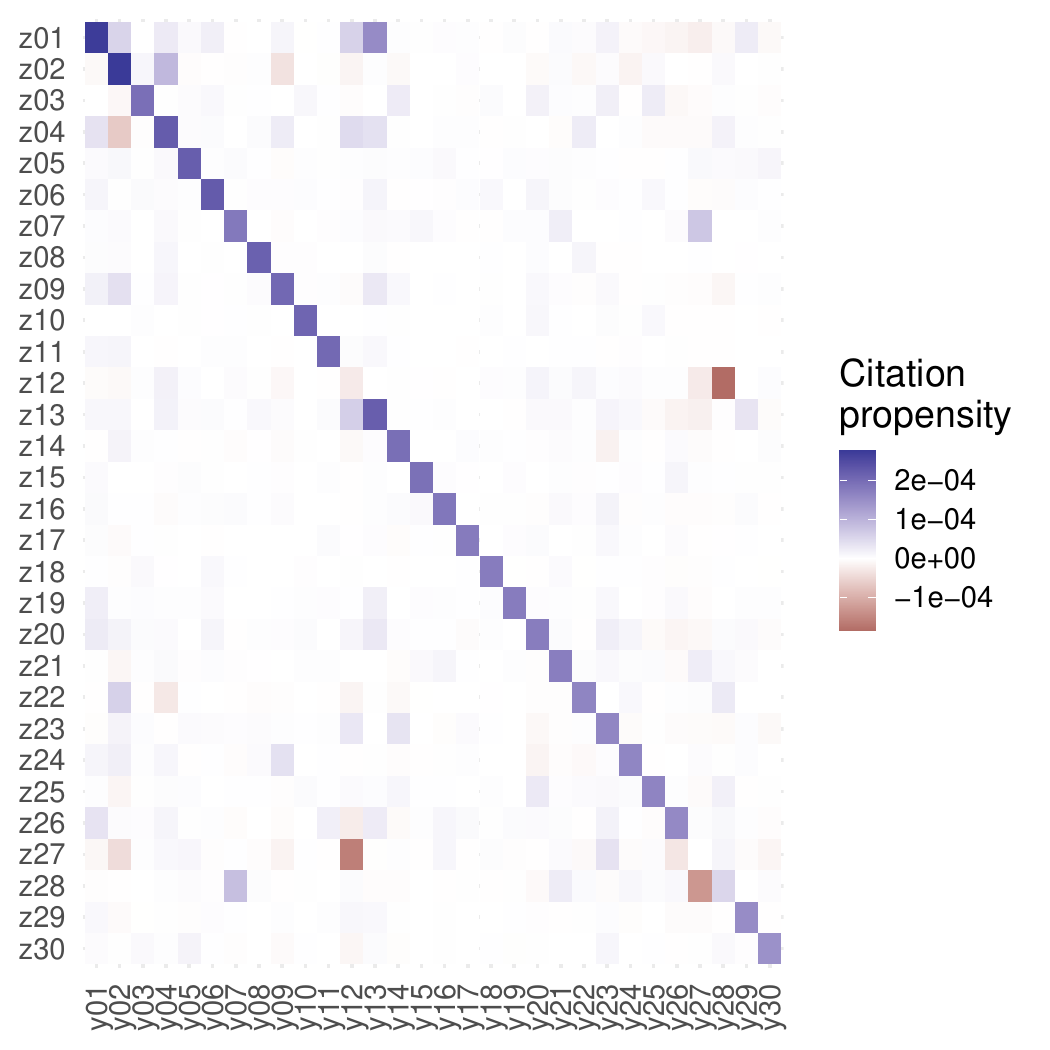}
  \end{minipage}\hfill
  \begin{minipage}{0.49\textwidth}
    \centering
    \includegraphics[width=\textwidth]{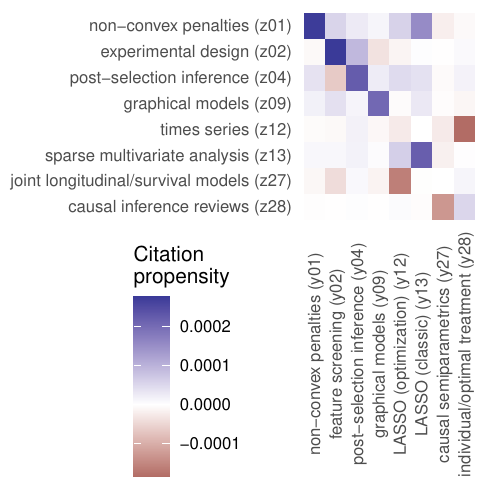}
  \end{minipage}
  \caption{Left: The varimax estimate $\widehat{B}$. Each entry $\widehat B_{ij}$ denotes the estimated citation propensity from papers loading on $i^{th}$ outgoing co-factor $Z_{\cdot i}$ to the $j^{th}$ incoming co-factor $Y_{\cdot j}$. Right: A labelled sub-matrix of $\widehat B$ considering the co-factors exhibiting off-diagonal structure.}
  \label{fig:hat-B}
\end{figure}

Most of the incoming co-factors $\widehat{Y}$ correspond closely with an outgoing co-factor $\widehat{Z}$ on the same topic. For instance, there is an incoming survival analysis co-factor (y05) and also an outgoing survival analysis co-factor (z05). The hubs for the incoming co-factor are highly cited methods papers such as \cite{cox1972} and \cite{andersen1982}. The hubs for the outgoing co-factor are review papers that cite many of these works while reviewing few citations themselves, such as \cite{guo2014} and \cite{kalbfleisch2023}.  To investigate correspondences between $\widehat Y$ and $\widehat Z$ factors, we plotted the mixing matrix $\widehat B$ in the left panel of Figure \ref{fig:hat-B}.

We found several $\widehat Z$ factors that did not correspond closely with any incoming $\widehat Y$ factor. For instance, z12 is a co-factor describing propensity to cite papers on times series analysis, while y12 is a co-factor related to the LASSO and optimization. Similarly, z27, a co-factor about joint longitudinal models, and z28, a causal inference co-factor, did not exhibit topical correspondence with y27, on causal semiparametrics, and y28, on individualized treatment rules. We visualized the relationships between the unmatched factors in the right panel of Figure \ref{fig:hat-B}, where it is clear that some co-factors are not in one-to-one correspondence with one another.

One question was how to interpret co-factors exhibiting one-to-one incoming-to-outgoing correspondence. For example, what was the difference between the outgoing survival analysis factor (z10) and the incoming survival analysis factor (y10)? To answer this question, we looked at the hub papers for each co-factor. For the survival analysis factor, for example, the top incoming hub was \citet{cox1972}, which introduced the proportional hazards model, and the top outgoing hub was \citet{guo2014}, a survey of semiparametric models in survival analysis. Incoming $\widehat{Y}$ hub papers were typically highly cited, important papers in each sub-field. In contrast, the outgoing $\widehat Z$ hub papers were typically review articles, retrospectives, tutorials, and papers with good literature reviews that summarized the past literature. Put differently, statistical papers tended to either: (1) perform important synthesis of past work but be cited very little, (2) cite a limited number of papers while receiving many citations, or (3) cite and be cited very little.

This distinct behavior from the $\widehat Y$ paper hubs and $\widehat Z$ paper hubs is evidence of co-factor structure in the statistics literature, and more broadly, evidence that papers do indeed cite and get cited in fundamentally different ways.

\subsection{Sensitivity to choice of rank and clipping parameters}

We repeated our analysis for $k \in \{5, 10, 20, 30, 40\}$, holding $\ell_\text{z} = 100,000$ and $\ell_\text{y} = 50,000$ fixed. Factor keywords, hubs, and mixing matrices for these analyses can be found in the supplemental material. We obtained qualitatively consistent results across values of $k$, and found that the co-factors can be coherently interpreted at all values of $k$ that we explored. In practice, increasing $k$ revealed additional, finer-grain factor structure. We chose to analyze $k=30$ co-factors because those co-factors revealed rich structure in the statistics literature while remaining interpretable and digestible.

We additionally explored $\ell_\text{z} = \ell_{y} \in \{1, 25000, 50000, 70000\}$, holding $k = 30$. Factors keywords, hubs, and mixing matrices for these analyses can also be found in the supplemental material. We found that the $Y$ keywords, the $Y$ factor hubs, and the $Z$ factor hubs remained fairly stable across choices of $\ell_\text{z}$ and $\ell_\text{y}$. However, $Z$ keywords and factor identities were more varied, and mixing matrices $\widehat B$ also exhibited substantial variation. At low clipping parameters, $\widehat{B}$ exhibited substantial off-diagonal structure. As the clipping parameters increased, $\widehat{B}$ became more and more diagonal. The results for  $\ell_\text{z} = 100,000$ and $\ell_\text{y} = 50,000$ had the least off-diagonal structure in $\widehat B$. 

Altogether, the sensitivity analysis for the clipping parameters indicated that $\widehat B$, and to a lesser extent, the outgoing co-factors $Z$, were somewhat unstable across hyperparameter values. Ultimately, our choice of $\ell_\text{z} = 100,000$ and $\ell_\text{y} = 50,000$ was based on domain knowledge: we assumed that it would take until 2004 for all outgoing co-topics to appear in the statistical literature, and that papers published after 2018 would not have the chance to be cited by papers from each incoming co-topic, due to the lengthy academic publication process. Regardless, since we do not definitely know how to select $\ell_\text{z}$ and $\ell_\text{y}$, results should be treated as somewhat tentative. 

\subsection{How the past would cite the future}
\label{sec:citation-forward-in-time}

One of the interesting features of our missing data framework is that it allows us to impute latent similarities from older documents to newer documents, or, with conceptual abuse, citations forward in time. In particular, if a paper $i$ was published before paper $j$, we can estimate the latent similarity from paper $i$ to paper $j$ via the real-valued imputation $\widehat A_{ij} \approx \widehat Z_{i \cdot} \, \widehat B \, \widehat {Y_{j \cdot}}^T$. We suggest interpreting these imputed similarities as you would interpret probability estimates from a linear probability model; as in the linear probability model, we have no guarantee that $\widehat A_{ij} \in [0, 1]$, such that $\widehat A_{ij}$ represents a valid probability of ``citation''. However, we can still think of $\widehat A_{ij}$ as indicative of probability of citation, had citation been possible.

In particular, for each paper, we calculated all of these imputed similarities from prior papers. Summing over these imputations, we obtained an estimate of the number of times papers from the past would have cited papers from the future on the basis of topical similarity, were they so able. We computed these estimates for each of the papers in our citation network and report the 15 papers with the highest imputed in-degree in Table \ref{tab:citation_impute_preclipped2-forward-citations} and the 15 papers with highest imputed out-degree in Table \ref{tab:citation_impute_preclipped2-backward-citations}. Most of the papers with high imputed in-degree are related to feature screening, the graphical \texttt{LASSO}, or some form of high dimensional regression. Most of the papers with high imputed out-degree are review articles published in Biometrika.

\begingroup\fontsize{7.6}{9.6}\selectfont

\begin{longtable}[t]{>{\raggedright\arraybackslash}p{46em}rr}
\caption{\label{tab:citation_impute_preclipped2-forward-citations}Imputed incoming citations (identified edges only)}\\
\toprule
Title & Imputed & Cited by\\
\midrule
On asymptotically optimal confidence regions and tests for high-dimensional models (2014) & 1632 & 360\\
Confidence intervals for low dimensional parameters in high dimensional linear models (2014) & 1564 & 350\\
Sure independence screening for ultrahigh dimensional feature space (2008) & 1387 & 905\\
Estimating individualized treatment rules using outcome weighted learning (2012) & 1215 & 280\\
Regularization paths for generalized linear models via coordinate descent (2010) & 1135 & 1124\\
\addlinespace
Feature screening via distance correlation learning (2012) & 1094 & 327\\
A robust method for estimating optimal treatment regimes (2012) & 1014 & 210\\
Model-free feature screening for ultrahigh-dimensional data (2011) & 871 & 246\\
Sure independence screening in generalized linear models with np-dimensionality (2010) & 847 & 305\\
Double/debiased machine learning for treatment and structural parameters (2018) & 831 & 222\\
\addlinespace
Nonparametric independence screening in sparse ultra-high-dimensional additive models (2011) & 812 & 262\\
Exact post-selection inference, with application to the lasso (2016) & 757 & 188\\
Performance guarantees for individualized treatment rules (2011) & 753 & 219\\
Simultaneous analysis of lasso and dantzig selector (2009) & 728 & 617\\
Sparse inverse covariance estimation with the graphical lasso (2008) & 717 & 754\\
\bottomrule
\end{longtable}
\endgroup{}

\begingroup\fontsize{7.6}{9.6}\selectfont

\begin{longtable}[t]{>{\raggedright\arraybackslash}p{46em}rr}
\caption{\label{tab:citation_impute_preclipped2-backward-citations}Imputed outgoing citations (identified edges only)}\\
\toprule
Title & Imputed & Cites\\
\midrule
Bayesian statistics in medicine: a 25 year review (2006) & 754 & 511\\
Joint modeling of longitudinal and time-to-event data: an overview (2004) & 172 & 36\\
Joint longitudinal-survival-cure models and their application to prostate cancer (2004) & 165 & 34\\
Methodological issues with adaptation of clinical trial design (2006) & 155 & 41\\
Adaptive statistical analysis following sample size modification based on interim review of effect size (2005) & 143 & 26\\
\addlinespace
Semiparametric regression during 2003-2007 (2009) & 137 & 219\\
A 25-year review of sequential methodology in clinical studies (2007) & 135 & 85\\
Group sequential and adaptive designs - a review of basic concepts and points of discussion (2008) & 134 & 76\\
Maximum likelihood estimation in semiparametric regression models with censored data (2007) & 131 & 53\\
Adaptive seamless designs: selection and prospective testing of hypotheses (2007) & 128 & 61\\
\addlinespace
A regulatory view on adaptive/flexible clinical trial design (2006) & 128 & 30\\
An overview of statistical approaches for adaptive designs and design modifications (2006) & 124 & 35\\
An investigation of two-stage tests (2006) & 122 & 30\\
Efficient group sequential designs when there are several effect sizes under consideration (2006) & 122 & 28\\
Joint modeling of longitudinal and survival data via a common frailty (2004) & 118 & 21\\
\bottomrule
\end{longtable}
\endgroup{}

\section{Discussion}
\label{sec:discussion}

We proposed a new method to co-factor documents in citation networks. The method is motivated by the observation that factors should be based on similarity measurements, and citations are only partially observed similarity measurements. Factoring a partially observed network complicated standard spectral clustering procedures and required use of matrix completion methods to estimate singular subspaces of the graph adjacency matrix. Here we found computational difficulties due to the precise observation pattern of citation data, which we resolved via a careful new implementation of the \texttt{AdaptiveImpute} algorithm. Because of dependence in the observation mechanism in the citation setting, existing theoretical results for \texttt{AdaptiveImpute}, and nuclear norm minimization more generally, were not applicable, and we validated our approach to matrix completion via a simulation study.

Our work suggests several avenues for methodological and theoretical exploration. Methodologically, it may be interesting to propose computationally efficient estimation procedures for other matrix completion methods in the upper triangular observation setting, or more generally in settings where sparse data is densely observed. Methods designed for independent but general sampling distributions, such as weighted nuclear norm minimization, may perform particularly well in the citation setting. Alternatively, further computational improvements would allow for larger scale bibliometric exploration of scientific citation networks. Current bibliometric databases contain hundreds of millions of papers and billions of references, more data than our method can handle. While our analysis of the statistics literature is one of the most extensive to date, incorporating additional papers could illuminate the relationships between statistical methodology and scientific practice at large. Another open question is how to extend our approach to the tensor, or multi-layer, citation network case, which would be appropriate for data like U.S. Court Opinions, where there are several distinct and explicitly labelled types of citation that documents may use when referencing each other. Finally, it may be of significant practical use to develop a better theoretical understanding of how matrix completion methods perform in settings with dependent observation mechanisms.

\section*{Acknowledgements}

We thank Steve Meyer at UW-Madison Libraries for assistance with the Web of Science dataset; Keith Levin, Vivak Patel and several anonymous reviewers for feedback on this manuscript; Yunyi Shen for a code contribution; and Alexander Tahk, Ben Bolker, Mark Padgham, Noam Ross, Max Kuhn, Dan Simpson, Sam Power, Patrick Girardet, and Cannon Lewis for generative discussions throughout the course of the project.

\section{Replication package}

We implemented a proof-of-concept implementation of the \texttt{AdaptiveImpute} estimator specialized to the citation setting in the \texttt{fastadi} R package, which is available on CRAN and at \url{https://github.com/RoheLab/fastadi}.

Code to reproduce the simulations and performance comparison is available at \url{https://github.com/alexpghayes/citation-cofactoring-replication/}. Due to licensing agreements, we cannot publish the Web of Science data. Nonetheless, the replication package contains the code we used to analyze the Web of Science data.

\bibliographystyle{chicago.bst}
\bibliography{citation-impute,software}

\begin{thebibliography}{}

\bibitem[\protect\citeauthoryear{Airoldi, Blei, Fienberg, and Xing}{Airoldi et~al.}{2008}]{airoldi2008}
Airoldi, E.~M., D.~M. Blei, S.~E. Fienberg, and E.~P. Xing (2008).
\newblock Mixed {{Membership Stochastic Blockmodels}}.
\newblock {\em Journal of Machine Learning Research\/}~{\em 9}, 1981--2014.

\bibitem[\protect\citeauthoryear{Andersen and Gill}{Andersen and Gill}{1982}]{andersen1982}
Andersen, P.~K. and R.~D. Gill (1982).
\newblock Cox's {{Regression Model}} for {{Counting Processes}}: {{A Large Sample Study}}.
\newblock {\em The Annals of Statistics\/}~{\em 10\/}(4), 1100--1120.

\bibitem[\protect\citeauthoryear{Athreya, Fishkind, Tang, Priebe, Park, Vogelstein, Levin, Lyzinski, Qin, and Sussman}{Athreya et~al.}{2018}]{athreya2018}
Athreya, A., D.~E. Fishkind, M.~Tang, C.~E. Priebe, Y.~Park, J.~T. Vogelstein, K.~Levin, V.~Lyzinski, Y.~Qin, and D.~L. Sussman (2018).
\newblock Statistical {{Inference}} on {{Random Dot Product Graphs}}: A {{Survey}}.
\newblock {\em Journal of Machine Learning Research\/}~{\em 18}, 1--92.

\bibitem[\protect\citeauthoryear{Athreya, Priebe, Tang, Lyzinski, Marchette, and Sussman}{Athreya et~al.}{2015}]{athreya2015}
Athreya, A., C.~E. Priebe, M.~Tang, V.~Lyzinski, D.~J. Marchette, and D.~L. Sussman (2015).
\newblock A {{Limit Theorem}} for {{Scaled Eigenvectors}} of {{Random Dot Product Graphs}}.
\newblock {\em Sankhya A: The Indian Journal of Statistics\/}~{\em 78\/}(1), 1--18.

\bibitem[\protect\citeauthoryear{Baglama and Reichel}{Baglama and Reichel}{2005}]{baglama2005}
Baglama, J. and L.~Reichel (2005).
\newblock Augmented {{Implicitly Restarted Lanczos Bidiagonalization Methods}}.
\newblock {\em SIAM Journal on Scientific Computing\/}~{\em 27\/}(1), 19--42.

\bibitem[\protect\citeauthoryear{Bhatia}{Bhatia}{1997}]{bhatia1997}
Bhatia, R. (1997).
\newblock {\em Matrix Analysis}.
\newblock Springer.

\bibitem[\protect\citeauthoryear{Bhojanapalli and Jain}{Bhojanapalli and Jain}{2014}]{bhojanapalli2014}
Bhojanapalli, S. and P.~Jain (2014).
\newblock Universal {{Matrix Completion}}.
\newblock In {\em Proceedings of the 31st {{International Conference}} on {{Machine Learning}}}.

\bibitem[\protect\citeauthoryear{Blei, Ng, and Jordan}{Blei et~al.}{2003}]{blei2003}
Blei, D.~M., A.~Y. Ng, and M.~I. Jordan (2003).
\newblock Latent {{Dirichlet Allocation}}.
\newblock {\em Journal of Machine Learning Research\/}~{\em 3}, 993--1022.

\bibitem[\protect\citeauthoryear{Chanpuriya, Tsourakakis, Musco, and Sotiropoulos}{Chanpuriya et~al.}{2020}]{chanpuriya2020}
Chanpuriya, S., C.~E. Tsourakakis, C.~Musco, and K.~Sotiropoulos (2020).
\newblock Node {{Embeddings}} and {{Exact Low-Rank Representations}} of {{Complex Networks}}.
\newblock In {\em 34th {{Conference}} on {{Neural Information Processing Systems}}}, Vancouver, Canada.

\bibitem[\protect\citeauthoryear{Chen}{Chen}{2021}]{chen2021a}
Chen, F. (2021).
\newblock {\em Spectral {{Methods}} for {{Social Media Data Analysis}}}.
\newblock Ph.\ D. thesis, University of Wisconsin-Madison.

\bibitem[\protect\citeauthoryear{Chen, Jalali, Sanghavi, and Xu}{Chen et~al.}{2014}]{chen2014}
Chen, Y., A.~Jalali, S.~Sanghavi, and H.~Xu (2014).
\newblock Clustering {{Partially Observed Graphs}} via {{Convex Optimization}}.
\newblock {\em Journal of Machine Learning Research\/}~{\em 15}, 2213--2238.

\bibitem[\protect\citeauthoryear{Cho, Kim, and Rohe}{Cho et~al.}{2019}]{cho2019}
Cho, J., D.~Kim, and K.~Rohe (2019).
\newblock Intelligent {{Initialization}} and {{Adaptive Thresholding}} for {{Iterative Matrix Completion}}: {{Some Statistical}} and {{Algorithmic Theory}} for {{{\emph{Adaptive-Impute}}}}.
\newblock {\em Journal of Computational and Graphical Statistics\/}~{\em 28\/}(2), 323--333.

\bibitem[\protect\citeauthoryear{Choi}{Choi}{2017}]{choi2017}
Choi, D. (2017).
\newblock Co-clustering of nonsmooth graphons.
\newblock {\em The Annals of Statistics\/}~{\em 45\/}(No. 4), 1488--1515.

\bibitem[\protect\citeauthoryear{Choi and Wolfe}{Choi and Wolfe}{2014}]{choi2014}
Choi, D. and P.~J. Wolfe (2014).
\newblock Co-clustering separately exchangeable network data.
\newblock {\em The Annals of Statistics\/}~{\em 42\/}(1), 29--63.

\bibitem[\protect\citeauthoryear{Cox}{Cox}{1972}]{cox1972}
Cox, D.~R. (1972).
\newblock Regression {{Models}} and {{Life-Tables}}.
\newblock {\em Journal of the Royal Statistical Society: Series B (Methodological)\/}~{\em 34\/}(2), 187--202.

\bibitem[\protect\citeauthoryear{Cui, Zhang, Wang, Zhang, Li, and Zuo}{Cui et~al.}{2015}]{cui2015}
Cui, Z., D.~Zhang, K.~Wang, H.~Zhang, N.~Li, and W.~Zuo (2015).
\newblock Weighted {{Nuclear Norm Minimization Based Tongue Specular Reflection Removal}}.
\newblock {\em Mathematical Problems in Engineering\/}~{\em 2015}, 1--15.

\bibitem[\protect\citeauthoryear{Drineas and Mahoney}{Drineas and Mahoney}{2005}]{drineas2005}
Drineas, P. and M.~W. Mahoney (2005).
\newblock On the {{Nystrom Method}} for {{Approximating}} a {{Gram Matrix}} for {{Improved Kernel-Based Learning}}.
\newblock {\em Journal of Machine Learning Research\/}~{\em 6}, 2153--2175.

\bibitem[\protect\citeauthoryear{Foucart, Needell, Pathak, Plan, and Wootters}{Foucart et~al.}{2021}]{foucart2021}
Foucart, S., D.~Needell, R.~Pathak, Y.~Plan, and M.~Wootters (2021).
\newblock Weighted {{Matrix Completion From Non-Random}}, {{Non-Uniform Sampling Patterns}}.
\newblock {\em IEEE Transactions on Information Theory\/}~{\em 67\/}(2), 1264--1290.

\bibitem[\protect\citeauthoryear{Gerlach, Peixoto, and Altmann}{Gerlach et~al.}{2018}]{gerlach2018}
Gerlach, M., T.~P. Peixoto, and E.~G. Altmann (2018).
\newblock A network approach to topic models.
\newblock {\em Science Advances\/}~{\em 4\/}(7), 1--11.

\bibitem[\protect\citeauthoryear{Gittens and Mahoney}{Gittens and Mahoney}{2016}]{gittens2016}
Gittens, A. and M.~W. Mahoney (2016).
\newblock Revisiting the {{Nystrom Method}} for {{Improved Large-scale Machine Learning}}.
\newblock {\em Journal of Machine Learning Research\/}~{\em 17}, 1--65.

\bibitem[\protect\citeauthoryear{Gu, Xie, Meng, Zuo, Feng, and Zhang}{Gu et~al.}{2017}]{gu2017}
Gu, S., Q.~Xie, D.~Meng, W.~Zuo, X.~Feng, and L.~Zhang (2017).
\newblock Weighted {{Nuclear Norm Minimization}} and {{Its Applications}} to {{Low Level Vision}}.
\newblock {\em International Journal of Computer Vision\/}~{\em 121\/}(2), 183--208.

\bibitem[\protect\citeauthoryear{Gu, Zhang, Zuo, and Feng}{Gu et~al.}{2014}]{gu2014}
Gu, S., L.~Zhang, W.~Zuo, and X.~Feng (2014).
\newblock Weighted {{Nuclear Norm Minimization}} with {{Application}} to {{Image Denoising}}.
\newblock In {\em 2014 {{IEEE Conference}} on {{Computer Vision}} and {{Pattern Recognition}}}, Columbus, OH, USA, pp.\  2862--2869. IEEE.

\bibitem[\protect\citeauthoryear{Guo and Zeng}{Guo and Zeng}{2014}]{guo2014}
Guo, S. and D.~Zeng (2014).
\newblock An overview of semiparametric models in survival analysis.
\newblock {\em Journal of Statistical Planning and Inference\/}~{\em 151--152}, 1--16.

\bibitem[\protect\citeauthoryear{Hajek and Sankagiri}{Hajek and Sankagiri}{2019}]{hajek2019}
Hajek, B. and S.~Sankagiri (2019).
\newblock Community {{Recovery}} in a {{Preferential Attachment Graph}}.
\newblock {\em IEEE Transactions on Information Theory\/}~{\em 65\/}(11), 6853--6874.

\bibitem[\protect\citeauthoryear{Hastie, Mazumder, Lee, and Zadeh}{Hastie et~al.}{2015}]{hastie2015}
Hastie, T., R.~Mazumder, J.~D. Lee, and R.~Zadeh (2015).
\newblock Matrix {{Completion}} and {{Low-Rank SVD}} via {{Fast Alternating Least Squares}}.
\newblock {\em Journal of Machine Learning Research\/}~{\em 16}, 3367--3402.

\bibitem[\protect\citeauthoryear{Hoerl and Kennard}{Hoerl and Kennard}{1970}]{hoerl1970}
Hoerl, A.~E. and R.~W. Kennard (1970).
\newblock Ridge {{Regression}}: {{Biased Estimation}} for {{Nonorthogonal Problems}}.
\newblock {\em Technometrics\/}~{\em 12\/}(1), 55--67.

\bibitem[\protect\citeauthoryear{Holland, Laskey, and Leinhardt}{Holland et~al.}{1983}]{holland1983}
Holland, P.~W., K.~B. Laskey, and S.~Leinhardt (1983).
\newblock Stochastic blockmodels: First steps.
\newblock {\em Social Networks\/}~{\em 5\/}(2), 109--137.

\bibitem[\protect\citeauthoryear{Hosono, Ono, and Miyata}{Hosono et~al.}{2016}]{hosono2016}
Hosono, K., S.~Ono, and T.~Miyata (2016).
\newblock Weighted tensor nuclear norm minimization for color image denoising.
\newblock In {\em 2016 {{IEEE International Conference}} on {{Image Processing}} ({{ICIP}})}, Phoenix, AZ, USA, pp.\  3081--3085. IEEE.

\bibitem[\protect\citeauthoryear{Ji and Jin}{Ji and Jin}{2016}]{ji2016}
Ji, P. and J.~Jin (2016).
\newblock Coauthorship and citation networks for statisticians.
\newblock {\em The Annals of Applied Statistics\/}~{\em 10\/}(4), 1779--1812.

\bibitem[\protect\citeauthoryear{Ji, Jin, Ke, and Li}{Ji et~al.}{2022}]{ji2022}
Ji, P., J.~Jin, Z.~T. Ke, and W.~Li (2022).
\newblock Co-citation and {{Co-authorship Networks}} of {{Statisticians}}.
\newblock {\em Journal of Business \& Economic Statistics\/}~{\em 40\/}(2), 469--485.

\bibitem[\protect\citeauthoryear{Jin}{Jin}{2015}]{jin2015}
Jin, J. (2015).
\newblock Fast community detection by {{SCORE}}.
\newblock {\em The Annals of Statistics\/}~{\em 43\/}(1), 57--89.

\bibitem[\protect\citeauthoryear{Jin, Ke, and Luo}{Jin et~al.}{2024}]{jin2024}
Jin, J., Z.~T. Ke, and S.~Luo (2024).
\newblock Mixed membership estimation for social networks.
\newblock {\em Journal of Econometrics\/}~{\em 239\/}(2), 105369.

\bibitem[\protect\citeauthoryear{Kalbfleisch and Schaubel}{Kalbfleisch and Schaubel}{2023}]{kalbfleisch2023}
Kalbfleisch, J.~D. and D.~E. Schaubel (2023).
\newblock Fifty {{Years}} of the {{Cox Model}}.
\newblock {\em Annual Review of Statistics and Its Application\/}~{\em 10\/}(Volume 10, 2023), 1--23.

\bibitem[\protect\citeauthoryear{Karrer and Newman}{Karrer and Newman}{2011}]{karrer2011}
Karrer, B. and M.~E.~J. Newman (2011).
\newblock Stochastic blockmodels and community structure in networks.
\newblock {\em Physical Review E\/}~{\em 83\/}(1), 016107.

\bibitem[\protect\citeauthoryear{Kim and Choi}{Kim and Choi}{2013}]{kim2013a}
Kim, Y.-D. and S.~Choi (2013).
\newblock Variational {{Bayesian View}} of {{Weighted Trace Norm Regularization}} for {{Matrix Factorization}}.
\newblock {\em IEEE Signal Processing Letters\/}~{\em 20\/}(3), 261--264.

\bibitem[\protect\citeauthoryear{Klopp}{Klopp}{2014}]{klopp2014}
Klopp, O. (2014).
\newblock Noisy low-rank matrix completion with general sampling distribution.
\newblock {\em Bernoulli\/}~{\em 20\/}(1), 282--303.

\bibitem[\protect\citeauthoryear{Larremore, Clauset, and Jacobs}{Larremore et~al.}{2014}]{larremore2014}
Larremore, D.~B., A.~Clauset, and A.~Z. Jacobs (2014).
\newblock Efficiently inferring community structure in bipartite networks.
\newblock {\em Physical Review E\/}~{\em 90\/}(1), 012805.

\bibitem[\protect\citeauthoryear{Lei and Rinaldo}{Lei and Rinaldo}{2015}]{lei2015}
Lei, J. and A.~Rinaldo (2015).
\newblock Consistency of spectral clustering in stochastic block models.
\newblock {\em The Annals of Statistics\/}~{\em 43\/}(1), 215--237.

\bibitem[\protect\citeauthoryear{Li, Levina, and Zhu}{Li et~al.}{2020}]{li2020c}
Li, T., E.~Levina, and J.~Zhu (2020).
\newblock Network cross-validation by edge sampling.
\newblock {\em Biometrika\/}~{\em 107\/}(2), 257--276.

\bibitem[\protect\citeauthoryear{Lyzinski, Sussman, Tang, Athreya, and Priebe}{Lyzinski et~al.}{2014}]{lyzinski2014}
Lyzinski, V., D.~L. Sussman, M.~Tang, A.~Athreya, and C.~E. Priebe (2014).
\newblock Perfect clustering for stochastic blockmodel graphs via adjacency spectral embedding.
\newblock {\em Electronic Journal of Statistics\/}~{\em 8\/}(2), 2905--2922.

\bibitem[\protect\citeauthoryear{Lyzinski, Tang, Athreya, Park, and Priebe}{Lyzinski et~al.}{2017}]{lyzinski2017}
Lyzinski, V., M.~Tang, A.~Athreya, Y.~Park, and C.~E. Priebe (2017).
\newblock Community {{Detection}} and {{Classification}} in {{Hierarchical Stochastic Blockmodels}}.
\newblock {\em IEEE Transactions on Network Science and Engineering\/}~{\em 4\/}(1), 13--26.

\bibitem[\protect\citeauthoryear{Mazumder, Hastie, and Tibshirani}{Mazumder et~al.}{2010}]{mazumder2010}
Mazumder, R., T.~Hastie, and R.~Tibshirani (2010).
\newblock Spectral {{Regularization Algorithms}} for {{Learning Large Incomplete Matrices}}.
\newblock {\em Journal of Machine Learning Research\/}~{\em 11}, 2287--2322.

\bibitem[\protect\citeauthoryear{Mitzenmacher and Upfal}{Mitzenmacher and Upfal}{2017}]{mitzenmacher2017}
Mitzenmacher, M. and E.~Upfal (2017).
\newblock {\em Probability and Computing\/} (Second edition ed.).
\newblock Cambridge, United Kingdom ; New York, NY, USA: Cambridge University Press.

\bibitem[\protect\citeauthoryear{Peixoto}{Peixoto}{2018}]{peixoto2018}
Peixoto, T.~P. (2018).
\newblock Reconstructing {{Networks}} with {{Unknown}} and {{Heterogeneous Errors}}.
\newblock {\em Physical Review X\/}~{\em 8\/}(4), 041011.

\bibitem[\protect\citeauthoryear{Pollner, Palla, and Vicsek}{Pollner et~al.}{2006}]{pollner2006}
Pollner, P., G.~Palla, and T.~Vicsek (2006).
\newblock Preferential attachment of communities: {{The}} same principle, but a higher level.
\newblock {\em Europhysics Letters (EPL)\/}~{\em 73\/}(3), 478--484.

\bibitem[\protect\citeauthoryear{Price}{Price}{1976}]{price1976}
Price, D. D.~S. (1976).
\newblock A general theory of bibliometric and other cumulative advantage processes.
\newblock {\em Journal of the American Society for Information Science\/}~{\em 27\/}(5), 292--306.

\bibitem[\protect\citeauthoryear{Priebe, Park, Vogelstein, Conroy, Lyzinski, Tang, Athreya, Cape, and Bridgeford}{Priebe et~al.}{2019}]{priebe2019}
Priebe, C.~E., Y.~Park, J.~T. Vogelstein, J.~M. Conroy, V.~Lyzinski, M.~Tang, A.~Athreya, J.~Cape, and E.~Bridgeford (2019).
\newblock On a two-truths phenomenon in spectral graph clustering.
\newblock {\em Proceedings of the National Academy of Sciences\/}~{\em 116\/}(13), 5995--6000.

\bibitem[\protect\citeauthoryear{Qing and Wang}{Qing and Wang}{2022}]{qing2022}
Qing, H. and J.~Wang (2022).
\newblock Directed mixed membership stochastic blockmodel.
\newblock {\em arXiv:2101.02307\/}.

\bibitem[\protect\citeauthoryear{Razaee, Amini, and Li}{Razaee et~al.}{2019}]{razaee2019}
Razaee, Z.~S., A.~A. Amini, and J.~J. Li (2019).
\newblock Matched {{Bipartite Block Model}} with {{Covariates}}.
\newblock {\em Journal of Machine Learning Research\/}~{\em 20}, 1--44.

\bibitem[\protect\citeauthoryear{Rohe, Qin, and Yu}{Rohe et~al.}{2016}]{rohe2016}
Rohe, K., T.~Qin, and B.~Yu (2016).
\newblock Co-clustering directed graphs to discover asymmetries and directional communities.
\newblock {\em Proceedings of the National Academy of Sciences\/}~{\em 113\/}(45), 12679--12684.

\bibitem[\protect\citeauthoryear{Rohe and Zeng}{Rohe and Zeng}{2023}]{rohe2023}
Rohe, K. and M.~Zeng (2023).
\newblock Vintage factor analysis with {{Varimax}} performs statistical inference.
\newblock {\em Journal of the Royal Statistical Society Series B: Statistical Methodology\/}~{\em 85\/}(4), 1037--1060.

\bibitem[\protect\citeauthoryear{Selby}{Selby}{2020}]{selby2020}
Selby, D.~A. (2020).
\newblock {\em Statistical Modelling of Citation Networks, Research Influence and Journal Prestige}.
\newblock Ph.\ D. thesis, University of Warwick.

\bibitem[\protect\citeauthoryear{Shamir and {Shalev-Shwartz}}{Shamir and {Shalev-Shwartz}}{2014}]{shamir2014}
Shamir, O. and S.~{Shalev-Shwartz} (2014).
\newblock Matrix {{Completion}} with the {{Trace Norm}}: {{Learning}}, {{Bounding}}, and {{Transducing}}.
\newblock {\em Journal of Machine Learning Research\/}~{\em 15}, 3401--3423.

\bibitem[\protect\citeauthoryear{Stigler}{Stigler}{1994}]{stigler1994}
Stigler, S.~M. (1994).
\newblock Citation {{Patterns}} in the {{Journals}} of {{Statistics}} and {{Probability}}.
\newblock {\em Statistical Science\/}~{\em 9\/}(1).

\bibitem[\protect\citeauthoryear{Tian}{Tian}{2004}]{tian2004}
Tian, Y. (2004).
\newblock More on maximal and minimal ranks of {{Schur}} complements with applications.
\newblock {\em Applied Mathematics and Computation\/}~{\em 152\/}(3), 675--692.

\bibitem[\protect\citeauthoryear{Tibshirani}{Tibshirani}{1996}]{tibshirani1996}
Tibshirani, R. (1996).
\newblock Regression {{Shrinkage}} and {{Selection Via}} the {{Lasso}}.
\newblock {\em Journal of the Royal Statistical Society: Series B (Methodological)\/}~{\em 58\/}(1), 267--288.

\bibitem[\protect\citeauthoryear{Vinayak, Oymak, and Hassibi}{Vinayak et~al.}{2014}]{vinayak2014}
Vinayak, R.~K., S.~Oymak, and B.~Hassibi (2014).
\newblock Graph {{Clustering With Missing Data}} : {{Convex Algorithms}} and {{Analysis}}.
\newblock In {\em Advances in {{Neural Information Processing Systems}}}.

\bibitem[\protect\citeauthoryear{{von Luxburg}}{{von Luxburg}}{2007}]{vonluxburg2007}
{von Luxburg}, U. (2007).
\newblock A tutorial on spectral clustering.
\newblock {\em Statistics and Computing\/}~{\em 17\/}(4), 395--416.

\bibitem[\protect\citeauthoryear{Vu and Lei}{Vu and Lei}{2013}]{vu2013}
Vu, V.~Q. and J.~Lei (2013).
\newblock Minimax sparse principal subspace estimation in high dimensions.
\newblock {\em The Annals of Statistics\/}~{\em 41\/}(6), 2905--2947.

\bibitem[\protect\citeauthoryear{Yang, Li, and Wang}{Yang et~al.}{2022}]{yang2022a}
Yang, M., Y.~Li, and J.~Wang (2022).
\newblock Feature and {{Nuclear Norm Minimization}} for {{Matrix Completion}}.
\newblock {\em IEEE Transactions on Knowledge and Data Engineering\/}~{\em 34\/}(5), 2190--2199.

\bibitem[\protect\citeauthoryear{Yen and Larremore}{Yen and Larremore}{2020}]{yen2020a}
Yen, T.-C. and D.~B. Larremore (2020).
\newblock Community detection in bipartite networks with stochastic block models.
\newblock {\em Physical Review E\/}~{\em 102\/}(3), 032309.

\bibitem[\protect\citeauthoryear{Zhang and Ng}{Zhang and Ng}{2019}]{zhang2019a}
Zhang, X. and M.~K. Ng (2019).
\newblock A {{Corrected Tensor Nuclear Norm Minimization Method}} for {{Noisy Low-Rank Tensor Completion}}.
\newblock {\em SIAM Journal on Imaging Sciences\/}~{\em 12\/}(2), 1231--1273.

\bibitem[\protect\citeauthoryear{Zhang, Chen, and Rohe}{Zhang et~al.}{2021}]{zhang2021d}
Zhang, Y., F.~Chen, and K.~Rohe (2021).
\newblock Social {{Media Public Opinion}} as {{Flocks}} in a {{Murmuration}}: {{Conceptualizing}} and {{Measuring Opinion Expression}} on {{Social Media}}.
\newblock {\em Journal of Computer-Mediated Communication\/}~{\em 27\/}(1), zmab021.

\bibitem[\protect\citeauthoryear{Zhao, Sun, Chen, and Chiu}{Zhao et~al.}{2022}]{zhao2022d}
Zhao, J., M.~Sun, F.~Chen, and P.~Chiu (2022).
\newblock Understanding {{Missing Links}} in {{Bipartite Networks With MissBiN}}.
\newblock {\em IEEE Transactions on Visualization and Computer Graphics\/}~{\em 28\/}(6), 2457--2469.

\bibitem[\protect\citeauthoryear{Zhu, Wang, and Samworth}{Zhu et~al.}{2022}]{zhu2022}
Zhu, Z., T.~Wang, and R.~J. Samworth (2022).
\newblock High-{{Dimensional Principal Component Analysis}} with {{Heterogeneous Missingness}}.
\newblock {\em Journal of the Royal Statistical Society Series B: Statistical Methodology\/}~{\em 84\/}(5), 2000--2031.

\end{thebibliography}

\appendix

\clearpage

\section{\texttt{AdaptiveInitialize} and \texttt{Lanczos Bidiagonalization}}

For convenience, we report the algorithmic details of the \texttt{Lanczos Bidiagonalization} and \texttt{AdaptiveInitialize} sub-routines.

\begin{algorithm}
  \linespread{1.3}\selectfont
  \DontPrintSemicolon
  \KwIn{matrix $A \in \mathbb{R}^{\ell \times n}$ or functions for evaluating matrix-vector products with the matrices $A$ and $A^T$, initial vector of unit length $p_1 \in \mathbb{R}^n$, number of bidiagonalization steps $m$}
  $P_1 \gets p_1$; $q_1 \gets A p_1$\;
  $\alpha_1 \gets \|q_1\|$, $q_1 \gets q_1 / \alpha_1$, $Q_1 \gets q_1$\;
  \For{$j = 1$ to $m$}{
    $r_j \gets A^T q_1 - \alpha_j p_j$\;
    \If{$j < m$}{
      $\beta_j \gets \|r_j\|$, $p_{j+1} \gets r_j / \beta_j$, $P_{j+1} \gets [P_j, p_{j+1}]$\;
      $q_{j+1} \gets Ap_{j+1} - \beta_j q_j$\;
      $\alpha_{j+1} \gets \|q_{j+1}\|$, $q_{j+1} \gets q_{j+1} / \alpha_{j+1}$, $Q_{j+1} \gets [Q_j, q_{j+1}]$\;
      $B_{j, j+1} \gets \beta_j$\;
    }
    $B_{jj} \gets \alpha_j$\;
  }
  \Return{$P_m, Q_m, B_m, r_m$}
  \caption{\textsc{Lanczos Bidiagonalization}}
  \label{alg:lanczos_bidiagonalization}
\end{algorithm}

\texttt{AdaptiveInitialize} is computes an initial estimate for a low-rank decomposition, and was originally reported in \citet{cho2019}. We have made minor notational changes for greater clarity.

\begin{algorithm}
  \DontPrintSemicolon
  \KwIn{partially observed matrix $A \in \mathbb{R}^{n \times n}$, desired rank $k \in \{2, ..., n \}$}
  $\hat p \gets \vert \Omega_A \vert \, / n^2 $ \;
  $\Sigma_V \gets A^T A - (1 - \hat p) \diag \left( A^T A \right)$ \;
  $\Sigma_U \gets A A^T - (1 - \hat p) \diag \left( A A^T \right)$ \;
  $\widehat V_i \gets \bv_i \left( \Sigma_V \right)$ for $i = 1, ..., k$ \;
  $\widehat U_i \gets \bu_i \left( \Sigma_U \right)$ for $i = 1, ..., k$ \;
  $\tilde \alpha \gets \displaystyle \frac{1}{n - k} \sum_{i=k+1}^n \bl_i \left( \Sigma_V \right)$ \;
  $\hat \lambda_i \gets \displaystyle \frac{1}{\hat p} \displaystyle \sqrt{\bl_i \left( \Sigma_V \right) - \tilde \alpha}$ for $i = 1, ..., k$ \;
  $\hat s_i \gets \sign \left(\langle \widehat V_i, \bv_i (A) \rangle \right) \cdot \sign \left( \langle \widehat U_i, \bu_i (A) \rangle \right)$ for $i = 1, ..., k$ \;
  \Return{$\hat s_i, \hat \lambda_i, \widehat U_i, \widehat V_i$ for $i = 1, ..., k$}\;
  \caption{\textsc{AdaptiveInitialize}}
  \label{alg:adaptive-initialize}
\end{algorithm}

Note that the left and right singular vectors estimates are initialized separately in \texttt{AdaptiveInitialize}, and $\hat s_i \in \{-1, 1\}$ can be used to ensure sign consistency between the singular vector pairs.

\section{Additional simulation results}
\label{app:additional-sims}

\begin{figure}
  \centering
  \includegraphics[width=0.99\textwidth]{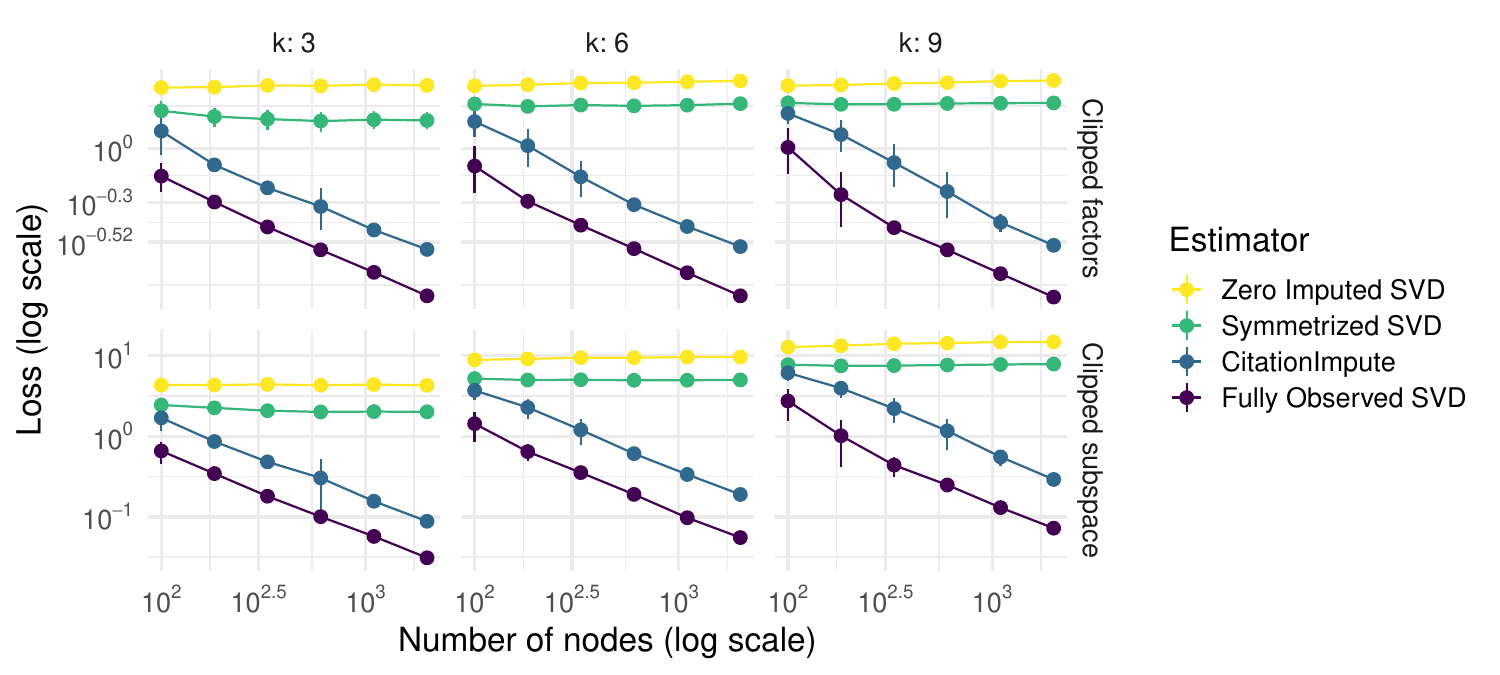}
  \caption{Average estimation error as a function of sample size, on $\log-\log$ scale. The top row of panels visualizes estimation error of the factors $Z$ and $Y$, excluding clipped factors. The bottom row of panels visualizes estimation error of the singular subspaces, again excluding clipped rows of $U$ and $V$. Each column of panels represents a simulation model with a different number of latent communities. Within each panel, each line corresponds to the loss of a single estimator. Average loss plus and minus one standard deviation are shown as a dotplot; in most cases the standard deviations are too small to see.}
  \label{fig:consistency5}
\end{figure}

Here, we investigate what happens if one of the naive imputation approaches is correct. We consider two symmetric variants of the simulation model. In the first variant, we symmetrize $\mathbb E[\mathcal A \mid Z, B, Y]$ by using the asymmetric model from Section \ref{sec:simulation-study} and setting $B = (B + B^T) / 2$ and $\theta^\text{in} = \theta^\text{out}$. This yields a directed stochastic blockmodel that is symmetric \emph{in expectation}. We then repeat the simulation study of Section \ref{sec:simulation-study} and report the results in Figure \ref{fig:consistency5}. In this case, the symmetric imputation strategy is insufficient to estimate $Z$ and $Y$, despite the symmetry in the conditional expectation of $\mathcal A$.

\begin{figure}
  \centering
  \includegraphics[width=0.99\textwidth]{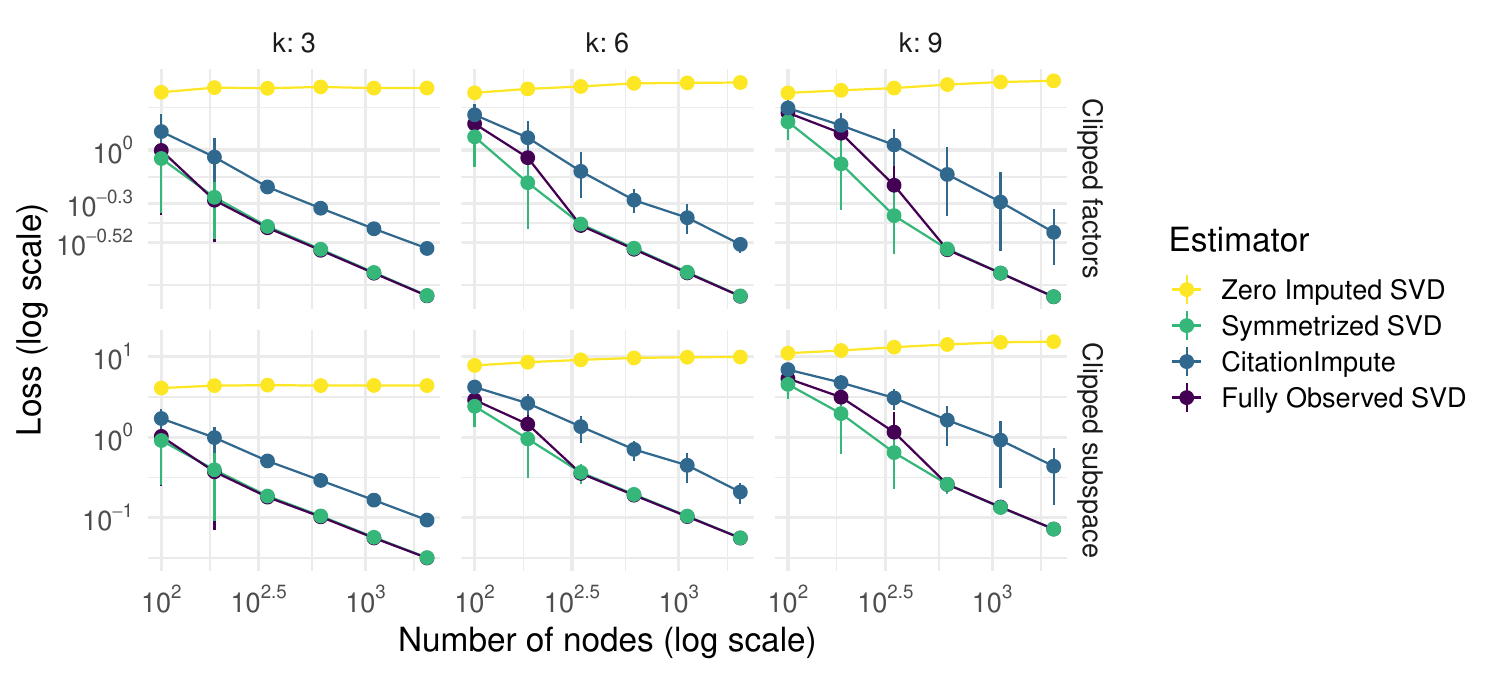}
  \caption{Average estimation error as a function of sample size, on $\log-\log$ scale. The top row of panels visualizes estimation error of the factors $Z$ and $Y$, excluding clipped factors. The bottom row of panels visualizes estimation error of the singular subspaces, again excluding clipped rows of $U$ and $V$. Each column of panels represents a simulation model with a different number of latent communities. Within each panel, each line corresponds to the loss of a single estimator. Average loss plus and minus one standard deviation are shown as a dotplot; in most cases the standard deviations are too small to see.}
  \label{fig:consistency6}
\end{figure}

Then, we consider a genuinely symmetric generative model, by sampling the upper half of a directed stochastic co-blockmodel, and then setting $A_{ij} = A{ji}$ for all $i, j \in [n]$. Results for this model are visualized in Figure \ref{fig:consistency6}. For truly symmetric $A$, the symmetrized SVD estimator achieves the same performance as the oracle estimator with access to all of $A$. \texttt{CitationImpute} continues to achieve the same rate as the oracle and symmetrized estimators in this setting, although with some performance penalty.

\section{Proofs}
\label{app:proofs}

\subsection{Proof of Proposition \ref{prop:identifiability}}

\begin{proof}

  Let $P^{-1}_U (\mathcal A)$ be the pre-image of $\mathcal A$ under $P_U$, which must contain at least one element, and let $\mathcal C$ be an arbitrary element of $P^{-1}_U (\mathcal A)$. Consider a decomposition of $\mathcal C$ as in Figure \ref{fig:understanding-identification}. let $i, j \in \mathbb Z$ satisfying $1 < i \le (n - \ell_\text{z})$ and $\ell_\text{y} < j \le n$. That is, $i$ and $j$ index an element of $\mathcal C$ that is sent to zero by $P_U$ but that we wish to recover. Put $x \equiv \mathcal C_{ij}$. By hypothesis $M = \mathcal C_{[1:\ell_\text{z}, (n - \ell_\text{y}):n]}$. Let $M^{-}$ be an arbitrary generalized inverse of $M$. From equation (1.5) of \citet{tian2004} it follows that
  \begin{align*}
    \rank \left(x - v M^{-} u \right)
    \le \rank \left(
    \begin{bmatrix}
        u & M  \\
        x & v
      \end{bmatrix}
    \right) - \rank \left( M \right)
    = k - k = 0,
  \end{align*}
  \noindent and thus that $x = v M^{-} u$. That is, any pre-image of $\mathcal A$ is uniquely specified at indexes$ i, j \in \mathbb Z$ satisfying $1 < i \le (n - \ell_\text{z})$ and $\ell_\text{y} < j \le n$, as desired.
\end{proof}

\subsection{Proof of Proposition \ref{prop:coupon}}

\begin{proof}
  In order to show that $M$ has rank $k$, it is sufficient to show that the first $\ell_\text{z}$ nodes are collectively members of all $k$ outgoing blocks, and the last $\ell_\text{y}$ nodes are collectively members of all $k$ incoming blocks.
  
  The problem thus reduces to the well-known coupon collector's problem. The probability of sampling $k$ out of $k$ distinct and equiprobable items given a sample of size $2 k \log k$ is $1 - n^{-1}$ \cite[p125-126]{mitzenmacher2017}. We first use this bound for the incoming $Y$ blocks and then again for the outgoing $Z$ blocks, and combine them with a union bound to complete the proof.
\end{proof}

\subsection{Proof of Proposition \ref{prop:multiplication-identity}}

\begin{proof}
  \begin{align*}
    \left( P_U (Z_t) \, x \right)_i
     & = \sum_{j=1}^n Z_{ij} \cdot \mathbf 1 (i < j) \cdot x_j                                         \\
     & = \sum_{j=1}^n \left( \sum_{r=1}^k U_{ir} (DV^T)_{rj} \right) \cdot x_j \cdot \mathbf 1 (i < j) \\
     & = \sum_{r=1}^k U_{ir} \sum_{j=1}^n (DV^T)_{rj} \cdot x_j \cdot \mathbf 1 (i < j)                \\
     & = \sum_{r=1}^k U_{ir} \sum_{j=1}^n W_{rj} \cdot \mathbf 1 (i < j)                               \\
     & = \sum_{r=1}^k U_{ir} \sum_{j=i+1}^n W_{rj}                                                     \\
     & = \sum_{r=1}^k U_{ir} \tilde W_{ri}                                                             \\
     & = \langle U_{i \cdot}, \tilde W_i \rangle
  \end{align*}
\end{proof}

\subsection{Proof of Proposition \ref{prop:alpha-identity}}

\begin{proof}

  First observe that
  \begin{align*}
    \alpha^{(t)}
     & = \frac{1}{n - k} \sum_{i=k+1}^n \bl_i^2 \left( \tilde A^{(t)} \right)                                                                       \\
     & = \frac{1}{n - k} \left[ \sum_{i = 1}^n \bl_i^2 \left( \tilde A^{(t)} \right) - \sum_{i = 1}^k \bl_i^2 \left( \tilde A^{(t)} \right) \right] \\
     & = \frac{1}{n - k} \left[ \left \Vert \tilde A^{(t)} \right \Vert_F^2 - \sum_{i = 1}^k \bl_i^2 \left( \tilde A^{(t)} \right) \right].
  \end{align*}
  Further
  \begin{align*}
    \left \Vert \tilde A^{(t)} \right \Vert_F^2
     & = \left \Vert P_{\Omega} (A) + P_\Omega^\perp \left( Z^{(t)} \right) \right \Vert_F^2 \\
     & = \left \Vert P_{\tilde \Omega} (A) \right \Vert_F^2 +
    \left \Vert P_\Omega^\perp \left( Z^{(t)} \right) \right \Vert_F^2                       \\
     & = \left \Vert P_{\tilde \Omega} (A) \right \Vert_F^2 +
    \left \Vert Z^{(t)} \right \Vert_F^2
    - \left \Vert P_\Omega \left( Z^{(t)} \right) \right \Vert_F^2                           \\
     & = \left \Vert P_{\tilde \Omega} (A) \right \Vert_F^2 +
    \left \Vert Z^{(t)} \right \Vert_F^2
    - \left \Vert P_L \left( Z^{(t)} \right) \right \Vert_F^2
    - \left \Vert P_U \left( Z^{(t)} \right) \right \Vert_F^2.
  \end{align*}
  Finally,
  \begin{align*}
    \left \Vert P_U \left( Z^{(t)} \right) \right \Vert_F^2
     & = \sum_{i=1}^n \sum_{j=1}^n \left \langle U_{i \cdot}, DV^T_{\cdot j} \right \rangle^2 \,
    \mathbf 1 (i < j)                                                                                                                    \\
     & = \sum_{i=1}^n \sum_{j=1}^n
    \left( \sum_{r=1}^k U_{i r} (DV)^T_{r j} \right)^2 \mathbf 1 (i < j)                                                                 \\
     & = \sum_{i=1}^n \sum_{j=1}^n \left( \sum_{r=1}^k \sum_{q=1}^k U_{i r} (DV)^T_{r j} U_{i q} (DV)^T_{q j} \right) \mathbf 1 (i < j)  \\
     & = \sum_{r=1}^k \sum_{q=1}^k \left( \sum_{i=1}^n \sum_{j=1}^n U_{i r} (DV)^T_{r j} U_{i q} (DV)^T_{q j} \right) \mathbf 1 (i < j)  \\
     & = \sum_{r=1}^k \sum_{q=1}^k \left( \sum_{i=1}^n U_{i r} U_{i q} \sum_{j=1}^n  (DV)^T_{r j} (DV)^T_{q j} \right) \mathbf 1 (i < j) \\
     & = \sum_{r=1}^k \sum_{q=1}^k \left( \sum_{i=1}^n U_{i r} U_{i q} \sum_{j=i+1}^n (DV)^T_{r j} (DV)^T_{q j} \right)                  \\
     & = \sum_{r=1}^k \sum_{q=1}^k \left( \sum_{i=1}^n U_{i r} U_{i q} \sum_{j=i+1}^n (DV)^T_{r j} (DV)^T_{q j} \right)                  \\
     & = \sum_{r=1}^k \sum_{q=1}^k \left( \sum_{i=1}^n U^{r q}_i V^{r q \triangle}_i \right)                                             \\
     & = \sum_{r=1}^k \sum_{q=1}^k \left \langle U^{r q}, V^{r q \triangle} \right \rangle,
  \end{align*}
  
  \noindent and putting all three of these pieces together completes the proof.
\end{proof}

\section{Supplement to data analysis}
\label{app:data-analysis-supplement}

This section reports the journals included in the data analysis and the factor hubs referenced in the main body of the paper.

\begingroup\fontsize{8}{10}\selectfont

\begin{longtable}[t]{rl}
\caption{\label{tab:considered-journals}Journals considered in data analysis}\\
\toprule
\# & Journal Title\\
\midrule
\endfirsthead
\caption[]{Journals considered in data analysis \textit{(continued)}}\\
\toprule
\# & Journal Title\\
\midrule
\endhead

\endfoot
\bottomrule
\endlastfoot
1 & Journal Of Statistical Software\\
2 & Annual Review Of Statistics And Its Application\\
3 & Econometrica\\
4 & Journal Of The American Statistical Association\\
5 & Journal Of The Royal Statistical Society Series B-Statistical Methodology\\
\addlinespace
6 & Statistical Science\\
7 & R Journal\\
8 & Fuzzy Sets And Systems\\
\end{longtable}
\endgroup{}

\begingroup\fontsize{8}{10}\selectfont

\begin{longtable}[t]{l>{\raggedright\arraybackslash}p{35em}rr}
\caption{\label{tab:citation_impute_preclipped2-y-hubs-elly-50000-ellz-1e+05-k-30}Y (incoming citation) factor hubs}\\
\toprule
ID & Title & Cited by & Cites\\
\midrule
\endfirsthead
\caption[]{Y (incoming citation) factor hubs \textit{(continued)}}\\
\toprule
ID & Title & Cited by & Cites\\
\midrule
\endhead

\endfoot
\bottomrule
\endlastfoot
y01 & Variable Selection Via Nonconcave Penalized Likelihood And Its Oracle Properties & 2804 & 14\\
y01 & The Adaptive Lasso And Its Oracle Properties & 2052 & 14\\
y01 & Nearly Unbiased Variable Selection Under Minimax Concave Penalty & 987 & 36\\
y01 & One-Step Sparse Estimates In Nonconcave Penalized Likelihood Models & 481 & 27\\
y01 & Tuning Parameter Selectors For The Smoothly Clipped Absolute Deviation Method & 407 & 17\\
\addlinespace
y02 & Sure Independence Screening For Ultrahigh Dimensional Feature Space & 905 & 33\\
y02 & Feature Screening Via Distance Correlation Learning & 327 & 22\\
y02 & Sure Independence Screening In Generalized Linear Models With Np-Dimensionality & 305 & 24\\
\end{longtable}
\endgroup{}

\begingroup\fontsize{8}{10}\selectfont

\begin{longtable}[t]{l>{\raggedright\arraybackslash}p{35em}rr}
\caption{\label{tab:citation_impute_preclipped2-z-hubs-elly-50000-ellz-1e+05-k-30}Z (outgoing citation) factor hubs}\\
\toprule
ID & Title & Cited by & Cites\\
\midrule
\endfirsthead
\caption[]{Z (outgoing citation) factor hubs \textit{(continued)}}\\
\toprule
ID & Title & Cited by & Cites\\
\midrule
\endhead

\endfoot
\bottomrule
\endlastfoot
z01 & Scad-Penalized Regression In Additive Partially Linear Proportional Hazards Models With An Ultra-High-Dimensional Linear Part & 3 & 37\\
z01 & Scad-Penalized Least Absolute Deviation Regression In High-Dimensional Models & 5 & 29\\
z01 & Regularization Parameter Selections Via Generalized Information Criterion & 102 & 27\\
z01 & Partially Linear Structure Selection In Cox Models With Varying Coefficients & 11 & 43\\
z01 & Variable Selection And Estimation Using A Continuous Approximation To The  $L$ $_{0}$ Penalty & 1 & 28\\
\addlinespace
z02 & Bayesian Statistics In Medicine: A 25 Year Review & 23 & 511\\
z02 & Construction Of Optimal Multi-Level Supersaturated Designs & 47 & 32\\
z02 & Methodological Issues With Adaptation Of Clinical Trial Design & 12 & 41\\
\end{longtable}
\endgroup{}

\newpage
\section{Sensitivity to rank}

This section contains supplementary results for estimates using varying rank $k$.

\subsection{Y keywords as rank varies}

\begingroup\fontsize{8}{10}\selectfont

\begin{longtable}[t]{l>{\raggedright\arraybackslash}p{34em}l}
\caption{\label{tab:citation_impute_preclipped2-y-keywords-elly-50000-ellz-1e+05-k-5}Keywords for Y (incoming citation) factors - k = 5, $\ell_\text{z}$ = 1e+05, $\ell_\text{y}$ = 50000}\\
\toprule
Factor Name & Top words & ID\\
\midrule
y1 & selection, lasso, variable, regression, via, dimensional & y1\\
y2 & dimension, models, skew, mixture, model, selection & y2\\
y3 & survival, data, regression, censored, semiparametric, models & y3\\
y4 & false, discovery, multiple, testing, rate, high & y4\\
y5 & bayesian, dirichlet, models, spatial, priors, mixture & y5\\
\bottomrule
\end{longtable}
\endgroup{}

\begingroup\fontsize{8}{10}\selectfont

\begin{longtable}[t]{l>{\raggedright\arraybackslash}p{34em}l}
\caption{\label{tab:citation_impute_preclipped2-y-keywords-elly-50000-ellz-1e+05-k-10}Keywords for Y (incoming citation) factors - k = 10, $\ell_\text{z}$ = 1e+05, $\ell_\text{y}$ = 50000}\\
\toprule
Factor Name & Top words & ID\\
\midrule
y01 & selection, lasso, variable, regression, via, penalized & y01\\
y02 & bayesian, models, spatial, mixed, longitudinal, monte & y02\\
y03 & covariance, graphical, sparse, high, dimensional, lasso & y03\\
y04 & survival, censored, hazards, proportional, data, regression & y04\\
y05 & mixture, model, clustering, skew, selection, mixtures & y05\\
\addlinespace
y06 & false, discovery, multiple, testing, rate, controlling & y06\\
y07 & propensity, causal, score, treatment, missing, effects & y07\\
y08 & dimension, reduction, regression, sliced, inverse, sufficient & y08\\
y09 & bayesian, dirichlet, nonparametric, mixture, mixtures, priors & y09\\
y10 & likelihood, empirical, models, regression, longitudinal, quantile & y10\\
\bottomrule
\end{longtable}
\endgroup{}

\begingroup\fontsize{8}{10}\selectfont

\begin{longtable}[t]{l>{\raggedright\arraybackslash}p{34em}l}
\caption{\label{tab:citation_impute_preclipped2-y-keywords-elly-50000-ellz-1e+05-k-20}Keywords for Y (incoming citation) factors - k = 20, $\ell_\text{z}$ = 1e+05, $\ell_\text{y}$ = 50000}\\
\toprule
Factor Name & Top words & ID\\
\midrule
y01 & selection, variable, penalized, oracle, nonconcave, lasso & y01\\
y02 & screening, dimensional, ultrahigh, feature, independence, high & y02\\
y03 & bayesian, models, complexity, disease, longitudinal, model & y03\\
y04 & high, dimensional, lasso, selection, models, sparse & y04\\
y05 & survival, censored, hazards, proportional, cox, data & y05\\
\addlinespace
y06 & model, clustering, mixture, dimension, selection, mixtures & y06\\
y07 & false, discovery, multiple, testing, rate, controlling & y07\\
y08 & propensity, causal, treatment, score, observational, effects & y08\\
y09 & dirichlet, bayesian, nonparametric, mixture, mixtures, priors & y09\\
y10 & covariance, graphical, sparse, estimation, high, dimensional & y10\\
\addlinespace
y11 & dimension, reduction, regression, sliced, inverse, sufficient & y11\\
y12 & longitudinal, data, models, generalized, linear, estimating & y12\\
y13 & spatial, gaussian, models, datasets, covariance, large & y13\\
y14 & empirical, likelihood, confidence, ratio, semiparametric, partially & y14\\
y15 & lasso, selection, regression, via, shrinkage, regularization & y15\\
\addlinespace
y16 & functional, regression, longitudinal, linear, principal, data & y16\\
y17 & skew, normal, distributions, multivariate, distribution, t & y17\\
y18 & quantile, regression, quantiles, estimation, censored, models & y18\\
y19 & missing, data, imputation, longitudinal, nonignorable, models & y19\\
y20 & bayesian, selection, monte, carlo, variable, bayes & y20\\
\bottomrule
\end{longtable}
\endgroup{}

\begingroup\fontsize{8}{10}\selectfont

\begin{longtable}[t]{l>{\raggedright\arraybackslash}p{34em}l}
\caption{\label{tab:citation_impute_preclipped2-y-keywords-elly-50000-ellz-1e+05-k-40}Keywords for Y (incoming citation) factors - k = 40, $\ell_\text{z}$ = 1e+05, $\ell_\text{y}$ = 50000}\\
\toprule
Factor Name & Top words & ID\\
\midrule
y01 & lasso, selection, variable, oracle, elastic, net & y01\\
y02 & screening, dimensional, ultrahigh, feature, independence, high & y02\\
y03 & coordinate, descent, regularization, paths, selection, lasso & y03\\
y04 & bayesian, models, complexity, disease, fit, model & y04\\
y05 & high, dimensional, lasso, supersaturated, selection, confidence & y05\\
\addlinespace
y06 & survival, censored, hazards, proportional, regression, cox & y06\\
\bottomrule
\end{longtable}
\endgroup{}

\subsection{Z keywords as rank varies}

\begin{table}
\centering
\caption{\label{tab:citation_impute_preclipped2-z-keywords-elly-50000-ellz-1e+05-k-5}Keywords for Z (outgoing citation) factors - k = 5, $\ell_\text{z}$ = 1e+05, $\ell_\text{y}$ = 50000}
\centering
\resizebox{\ifdim\width>\linewidth\linewidth\else\width\fi}{!}{
\fontsize{8}{10}\selectfont
\begin{tabular}[t]{l>{\raggedright\arraybackslash}p{34em}l}
\toprule
Factor Name & Top words & ID\\
\midrule
z1 & selection, variable, high, dimensional, lasso, regression & z1\\
z2 & models, model, selection, bayesian, clustering, data & z2\\
z3 & data, survival, semiparametric, hazards, longitudinal, censored & z3\\
z4 & false, discovery, high, multiple, testing, dimensional & z4\\
z5 & bayesian, models, spatial, modeling, dirichlet, hierarchical & z5\\
\bottomrule
\end{tabular}}
\end{table}

\begin{table}
\centering
\caption{\label{tab:citation_impute_preclipped2-z-keywords-elly-50000-ellz-1e+05-k-10}Keywords for Z (outgoing citation) factors - k = 10, $\ell_\text{z}$ = 1e+05, $\ell_\text{y}$ = 50000}
\centering
\resizebox{\ifdim\width>\linewidth\linewidth\else\width\fi}{!}{
\fontsize{8}{10}\selectfont
\begin{tabular}[t]{l>{\raggedright\arraybackslash}p{34em}l}
\toprule
Factor Name & Top words & ID\\
\midrule
z01 & selection, variable, high, dimensional, regression, lasso & z01\\
z02 & bayesian, models, spatial, data, longitudinal, modeling & z02\\
z03 & high, graphical, dimensional, sparse, covariance, matrix & z03\\
z04 & survival, hazards, data, censored, cox, model & z04\\
z05 & selection, model, clustering, mixture, models, skew & z05\\
\addlinespace
z06 & false, discovery, testing, multiple, rate, high & z06\\
z07 & causal, missing, propensity, treatment, score, effects & z07\\
z08 & dimension, reduction, sufficient, index, single, regression & z08\\
z09 & bayesian, dirichlet, nonparametric, mixture, clustering, semiparametric & z09\\
z10 & empirical, likelihood, quantile, models, longitudinal, data & z10\\
\bottomrule
\end{tabular}}
\end{table}

\begin{table}
\centering
\caption{\label{tab:citation_impute_preclipped2-z-keywords-elly-50000-ellz-1e+05-k-20}Keywords for Z (outgoing citation) factors - k = 20, $\ell_\text{z}$ = 1e+05, $\ell_\text{y}$ = 50000}
\centering
\resizebox{\ifdim\width>\linewidth\linewidth\else\width\fi}{!}{
\fontsize{8}{10}\selectfont
\begin{tabular}[t]{l>{\raggedright\arraybackslash}p{34em}l}
\toprule
Factor Name & Top words & ID\\
\midrule
z01 & selection, variable, dimensional, high, penalized, regression & z01\\
z02 & screening, adaptive, dimensional, clinical, trials, ultrahigh & z02\\
z03 & bayesian, models, spatial, model, data, longitudinal & z03\\
z04 & high, dimensional, selection, lasso, models, sparse & z04\\
z05 & survival, hazards, censored, data, cox, model & z05\\
\addlinespace
z06 & selection, clustering, model, mixture, models, mixtures & z06\\
z07 & false, discovery, testing, multiple, rate, microarray & z07\\
z08 & causal, treatment, propensity, score, effects, observational & z08\\
z09 & bayesian, dirichlet, nonparametric, mixture, clustering, semiparametric & z09\\
z10 & graphical, high, covariance, dimensional, sparse, matrix & z10\\
\addlinespace
z11 & dimension, reduction, sufficient, index, single, inverse & z11\\
z12 & longitudinal, data, models, generalized, mixed, binary & z12\\
z13 & spatial, models, temporal, spatio, bayesian, gaussian & z13\\
z14 & empirical, likelihood, inference, partially, missing, models & z14\\
z15 & selection, lasso, regression, sparse, high, variable & z15\\
\addlinespace
z16 & functional, regression, data, longitudinal, varying, linear & z16\\
z17 & skew, normal, distributions, multivariate, t, distribution & z17\\
z18 & quantile, regression, models, varying, index, quantiles & z18\\
z19 & missing, data, longitudinal, imputation, with, nonignorable & z19\\
z20 & bayesian, selection, variable, models, priors, model & z20\\
\bottomrule
\end{tabular}}
\end{table}

\begin{table}
\centering
\caption{\label{tab:citation_impute_preclipped2-z-keywords-elly-50000-ellz-1e+05-k-40}Keywords for Z (outgoing citation) factors - k = 40, $\ell_\text{z}$ = 1e+05, $\ell_\text{y}$ = 50000}
\centering
\resizebox{\ifdim\width>\linewidth\linewidth\else\width\fi}{!}{
\fontsize{8}{10}\selectfont
\begin{tabular}[t]{l>{\raggedright\arraybackslash}p{34em}l}
\toprule
Factor Name & Top words & ID\\
\midrule
z01 & selection, lasso, variable, high, dimensional, sparse & z01\\
z02 & screening, dimensional, high, ultrahigh, feature, variable & z02\\
z03 & adaptive, clinical, trials, sequential, designs, group & z03\\
z04 & bayesian, models, spatial, model, data, hierarchical & z04\\
z05 & high, dimensional, supersaturated, designs, selection, regression & z05\\
\addlinespace
z06 & survival, hazards, censored, cox, data, proportional & z06\\
z07 & selection, clustering, model, mixture, models, mixtures & z07\\
\bottomrule
\end{tabular}}
\end{table}

\subsection{Mixing matrix as rank varies}

\begin{figure}[H]
  \centering
  \includegraphics[width=0.8\textwidth]{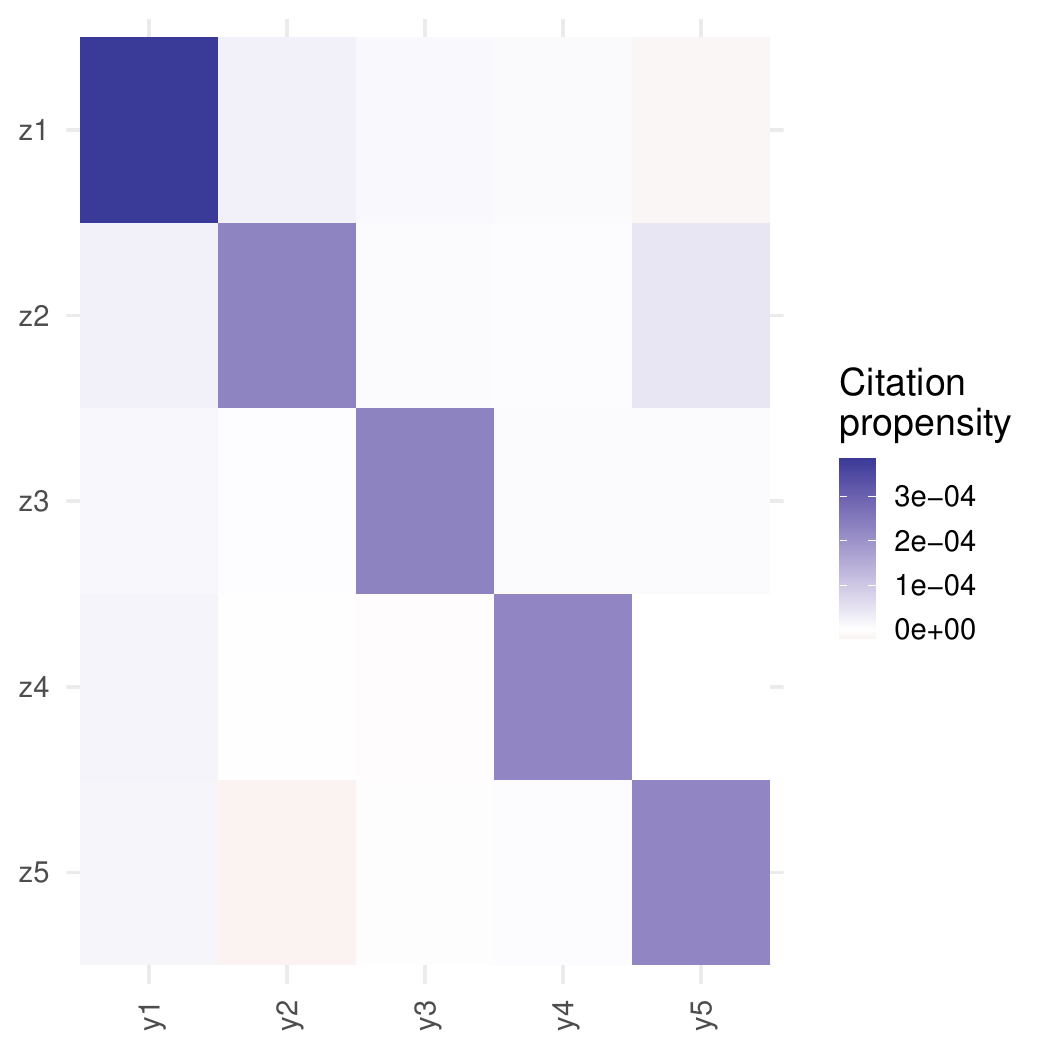}
  \caption{$\widehat B$ when $k = 5$}
\end{figure}

\begin{figure}[H]
  \centering
  \includegraphics[width=0.8\textwidth]{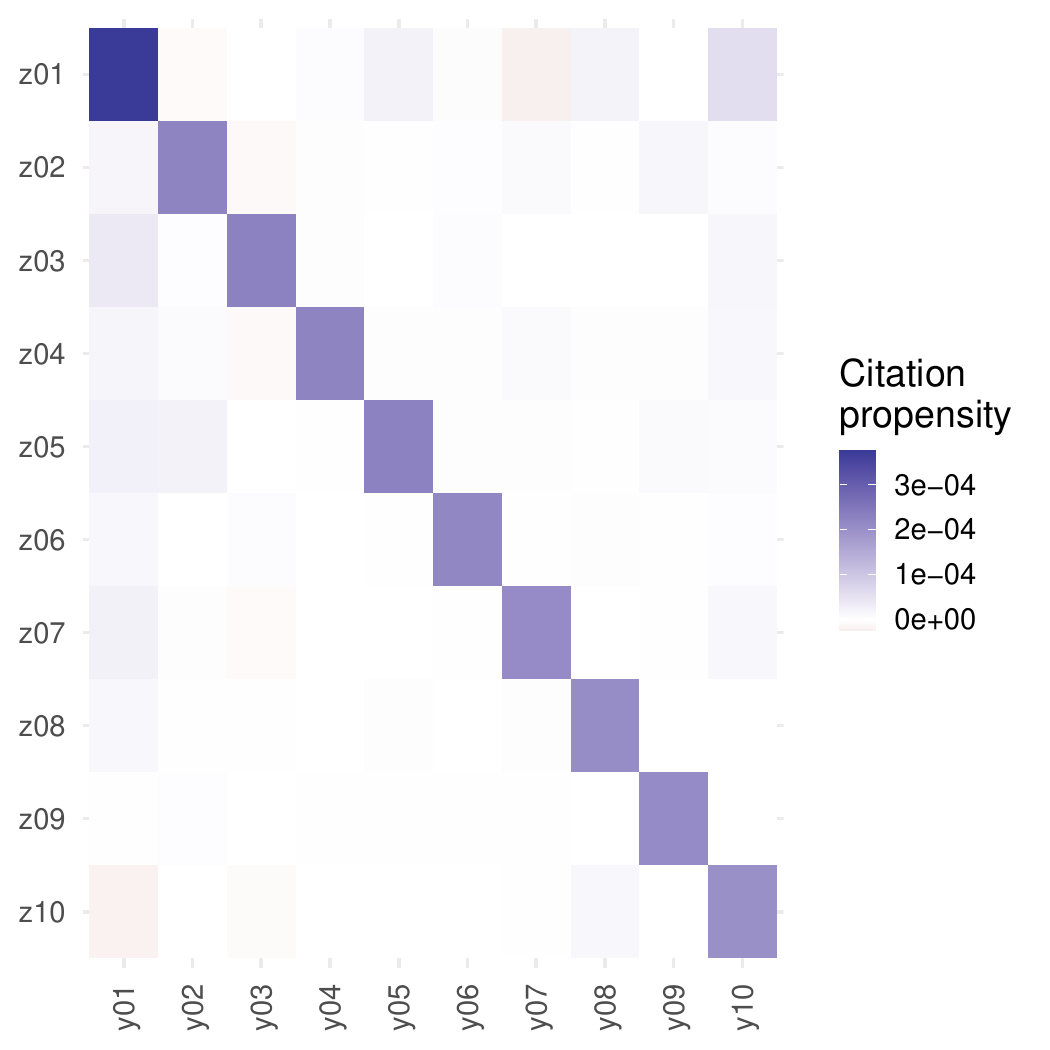}
  \caption{$\widehat B$ when $k = 10$}
\end{figure}

\begin{figure}[H]
  \centering
  \includegraphics[width=0.8\textwidth]{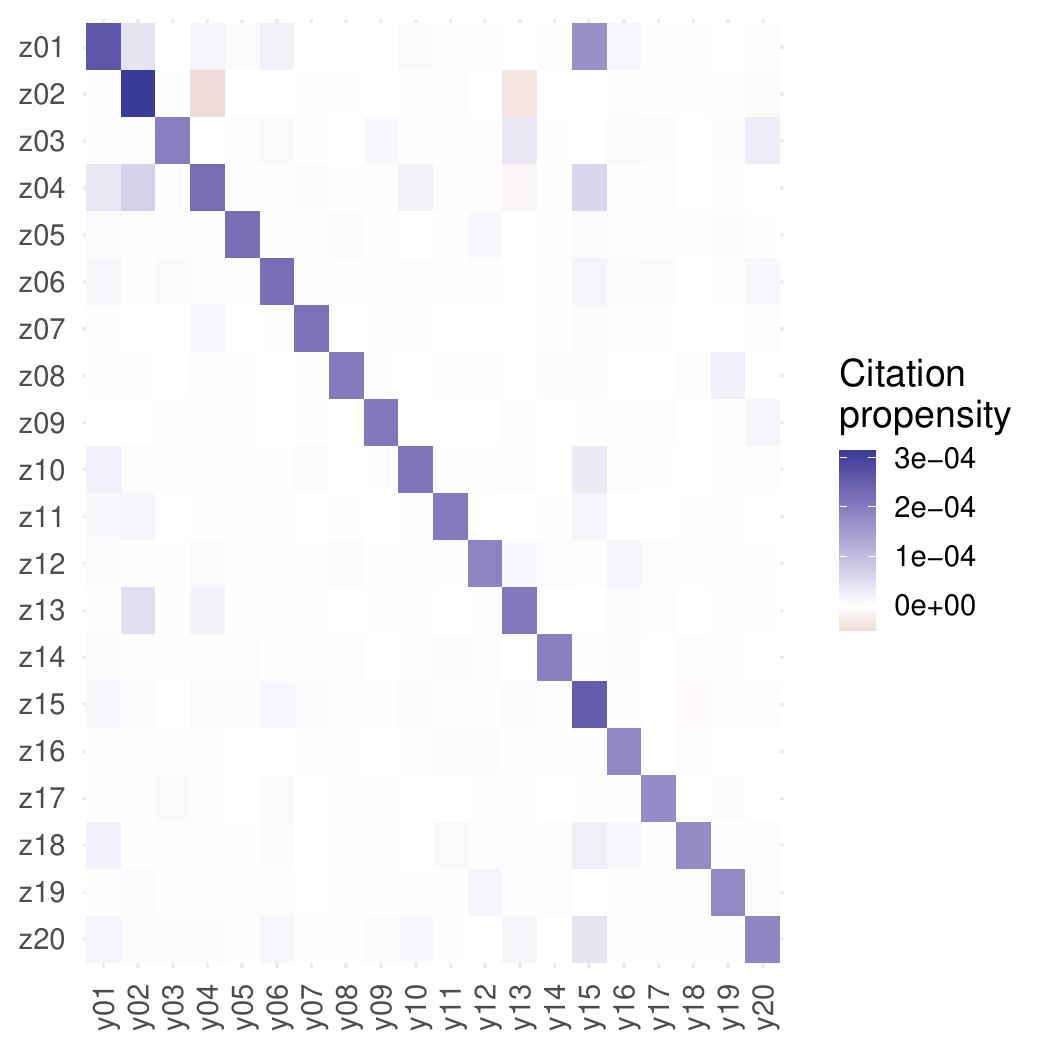}
  \caption{$\widehat B$ when $k = 20$}
\end{figure}

\begin{figure}[H]
  \centering
  \includegraphics[width=0.8\textwidth]{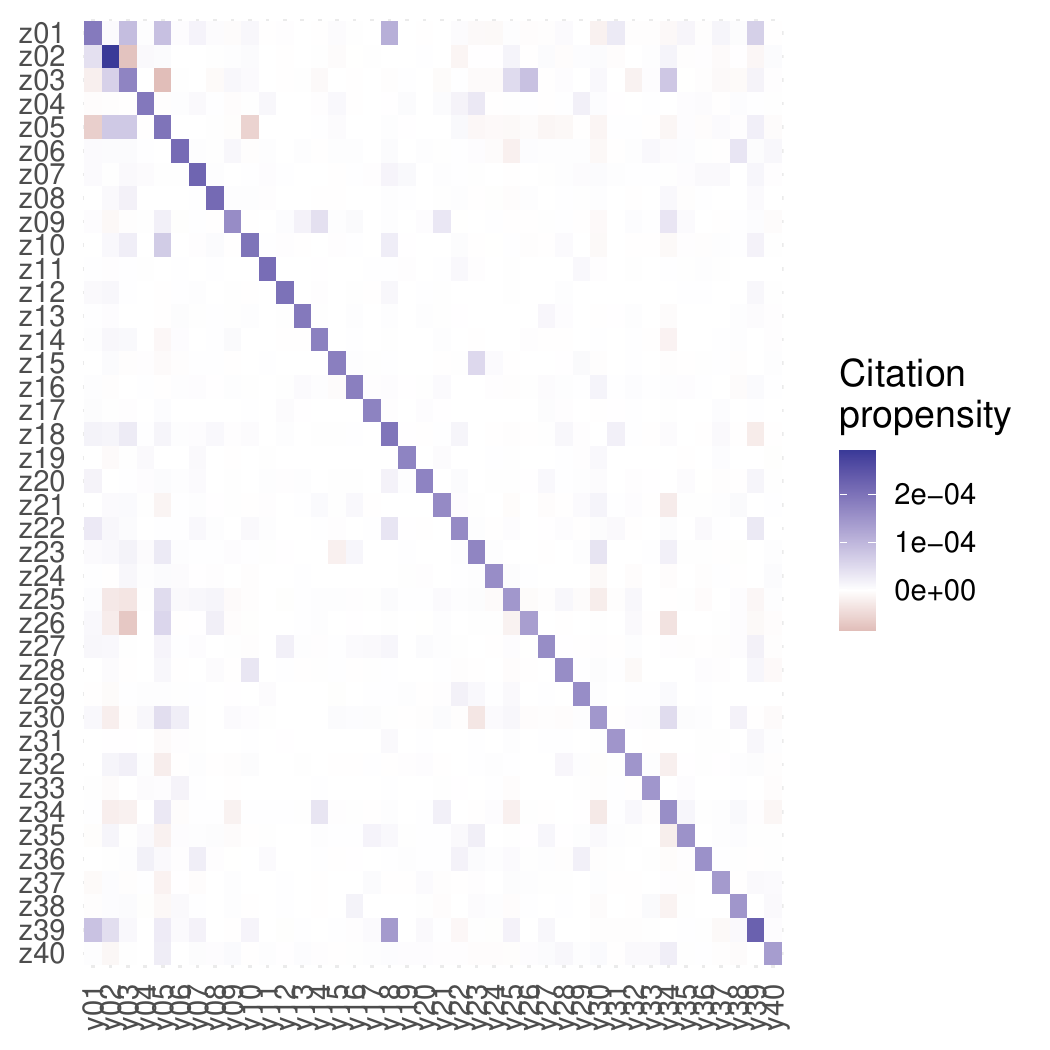}
  \caption{$\widehat B$ when $k = 40$}
\end{figure}

\subsection{Y hubs as rank varies}

\begingroup\fontsize{8}{10}\selectfont

\begin{longtable}[t]{l>{\raggedright\arraybackslash}p{35em}rr}
\caption{\label{tab:citation_impute_preclipped2-y-hubs-elly-50000-ellz-1e+05-k-5}Y (incoming citation) factor hubs - k = 5, $\ell_\text{z}$ = 1e+05, $\ell_\text{y}$ = 50000}\\
\toprule
ID & Title & Cited by & Cites\\
\midrule
\endfirsthead
\caption[]{Y (incoming citation) factor hubs - k = 5, $\ell_\text{z}$ = 1e+05, $\ell_\text{y}$ = 50000 \textit{(continued)}}\\
\toprule
ID & Title & Cited by & Cites\\
\midrule
\endhead

\endfoot
\bottomrule
\endlastfoot
y1 & Regression Shrinkage And Selection Via The Lasso & 4759 & 8\\
y1 & Variable Selection Via Nonconcave Penalized Likelihood And Its Oracle Properties & 2804 & 14\\
y1 & The Adaptive Lasso And Its Oracle Properties & 2052 & 14\\
\addlinespace
y2 & Estimating Dimension Of A Model & 3727 & 3\\
y2 & Bayes Factors & 1553 & 56\\
y2 & Bayesian Measures Of Model Complexity And Fit & 2107 & 38\\
\addlinespace
y3 & Regression Models And Life-Tables & 4087 & 21\\
y3 & Cox Regression-Model For Counting-Processes - A Large Sample Study & 1218 & 13\\
y3 & Longitudinal Data-Analysis Using Generalized Linear-Models & 2624 & 11\\
\addlinespace
y4 & Controlling The False Discovery Rate - A Practical And Powerful Approach To Multiple Testing & 2025 & 9\\
y4 & The Control Of The False Discovery Rate In Multiple Testing Under Dependency & 644 & 21\\
y4 & A Direct Approach To False Discovery Rates & 555 & 5\\
\addlinespace
y5 & Bayesian Measures Of Model Complexity And Fit & 2107 & 38\\
y5 & Bayesian Analysis Of Some Nonparametric Problems & 1373 & 4\\
y5 & A Constructive Definition Of Dirichlet Priors & 711 & 6\\*
\end{longtable}
\endgroup{}

\begingroup\fontsize{8}{10}\selectfont

\begin{longtable}[t]{l>{\raggedright\arraybackslash}p{35em}rr}
\caption{\label{tab:citation_impute_preclipped2-y-hubs-elly-50000-ellz-1e+05-k-10}Y (incoming citation) factor hubs - k = 10, $\ell_\text{z}$ = 1e+05, $\ell_\text{y}$ = 50000}\\
\toprule
ID & Title & Cited by & Cites\\
\midrule
\endfirsthead
\caption[]{Y (incoming citation) factor hubs - k = 10, $\ell_\text{z}$ = 1e+05, $\ell_\text{y}$ = 50000 \textit{(continued)}}\\
\toprule
ID & Title & Cited by & Cites\\
\midrule
\endhead

\endfoot
\bottomrule
\endlastfoot
y01 & Regression Shrinkage And Selection Via The Lasso & 4759 & 8\\
y01 & Variable Selection Via Nonconcave Penalized Likelihood And Its Oracle Properties & 2804 & 14\\
y01 & The Adaptive Lasso And Its Oracle Properties & 2052 & 14\\
\addlinespace
y02 & Bayesian Measures Of Model Complexity And Fit & 2107 & 38\\
y02 & Approximate Bayesian Inference For Latent Gaussian Models By Using Integrated Nested Laplace Approximations & 812 & 109\\
y02 & Prior Distributions For Variance Parameters In Hierarchical Models(Comment On An Article By Browne And Draper) & 906 & 28\\
\addlinespace
y03 & Sparse Inverse Covariance Estimation With The Graphical Lasso & 754 & 2\\
y03 & High-Dimensional Graphs And Variable Selection With The Lasso & 885 & 12\\
\end{longtable}
\endgroup{}

\begingroup\fontsize{8}{10}\selectfont

\begin{longtable}[t]{l>{\raggedright\arraybackslash}p{35em}rr}
\caption{\label{tab:citation_impute_preclipped2-y-hubs-elly-50000-ellz-1e+05-k-20}Y (incoming citation) factor hubs - k = 20, $\ell_\text{z}$ = 1e+05, $\ell_\text{y}$ = 50000}\\
\toprule
ID & Title & Cited by & Cites\\
\midrule
\endfirsthead
\caption[]{Y (incoming citation) factor hubs - k = 20, $\ell_\text{z}$ = 1e+05, $\ell_\text{y}$ = 50000 \textit{(continued)}}\\
\toprule
ID & Title & Cited by & Cites\\
\midrule
\endhead

\endfoot
\bottomrule
\endlastfoot
y01 & Variable Selection Via Nonconcave Penalized Likelihood And Its Oracle Properties & 2804 & 14\\
y01 & The Adaptive Lasso And Its Oracle Properties & 2052 & 14\\
y01 & Nearly Unbiased Variable Selection Under Minimax Concave Penalty & 987 & 36\\
\addlinespace
y02 & Sure Independence Screening For Ultrahigh Dimensional Feature Space & 905 & 33\\
y02 & Feature Screening Via Distance Correlation Learning & 327 & 22\\
y02 & Sure Independence Screening In Generalized Linear Models With Np-Dimensionality & 305 & 24\\
\addlinespace
y03 & Bayesian Measures Of Model Complexity And Fit & 2107 & 38\\
y03 & Prior Distributions For Variance Parameters In Hierarchical Models(Comment On An Article By Browne And Draper) & 906 & 28\\
\end{longtable}
\endgroup{}

\begingroup\fontsize{8}{10}\selectfont

\begin{longtable}[t]{l>{\raggedright\arraybackslash}p{35em}rr}
\caption{\label{tab:citation_impute_preclipped2-y-hubs-elly-50000-ellz-1e+05-k-40}Y (incoming citation) factor hubs - k = 40, $\ell_\text{z}$ = 1e+05, $\ell_\text{y}$ = 50000}\\
\toprule
ID & Title & Cited by & Cites\\
\midrule
\endfirsthead
\caption[]{Y (incoming citation) factor hubs - k = 40, $\ell_\text{z}$ = 1e+05, $\ell_\text{y}$ = 50000 \textit{(continued)}}\\
\toprule
ID & Title & Cited by & Cites\\
\midrule
\endhead

\endfoot
\bottomrule
\endlastfoot
y01 & The Adaptive Lasso And Its Oracle Properties & 2052 & 14\\
y01 & Regularization And Variable Selection Via The Elastic Net & 1584 & 11\\
y01 & Least Angle Regression & 1259 & 10\\
\addlinespace
y02 & Sure Independence Screening For Ultrahigh Dimensional Feature Space & 905 & 33\\
y02 & Feature Screening Via Distance Correlation Learning & 327 & 22\\
y02 & Sure Independence Screening In Generalized Linear Models With Np-Dimensionality & 305 & 24\\
\addlinespace
y03 & Regularization Paths For Generalized Linear Models Via Coordinate Descent & 1124 & 15\\
y03 & Nearly Unbiased Variable Selection Under Minimax Concave Penalty & 987 & 36\\
\end{longtable}
\endgroup{}

\subsection{Z hubs as rank varies}

\begingroup\fontsize{8}{10}\selectfont

\begin{longtable}[t]{l>{\raggedright\arraybackslash}p{35em}rr}
\caption{\label{tab:citation_impute_preclipped2-z-hubs-elly-50000-ellz-1e+05-k-5}Z (outgoing citation) factor hubs - k = 5, $\ell_\text{z}$ = 1e+05, $\ell_\text{y}$ = 50000}\\
\toprule
ID & Title & Cited by & Cites\\
\midrule
\endfirsthead
\caption[]{Z (outgoing citation) factor hubs - k = 5, $\ell_\text{z}$ = 1e+05, $\ell_\text{y}$ = 50000 \textit{(continued)}}\\
\toprule
ID & Title & Cited by & Cites\\
\midrule
\endhead

\endfoot
\bottomrule
\endlastfoot
z1 & A Survey Of  $L$ $_{1}$ Regression & 10 & 63\\
z1 & A Selective Overview Of Variable Selection In High Dimensional Feature Space & 293 & 68\\
z1 & A Majorization-Minimization Approach To Variable Selection Using Spike And Slab Priors & 8 & 36\\
\addlinespace
z2 & Asymmetric Clusters And Outliers: Mixtures Of Multivariate Contaminated Shifted Asymmetric Laplace Distributions & 11 & 66\\
z2 & Robust Model-Based Clustering Via Mixtures Of Skew- $T$  Distributions With Missing Information & 7 & 43\\
z2 & Flexible Mixture Modeling Via The Multivariate T Distribution With The Box-Cox Transformation: An Alternative To The Skew-T Distribution & 24 & 36\\
\addlinespace
z3 & $Biometrika$  Highlights From Volume 28 Onwards & 2 & 603\\
z3 & An Overview Of Semiparametric Models In Survival Analysis & 5 & 77\\
\end{longtable}
\endgroup{}

\begingroup\fontsize{8}{10}\selectfont

\begin{longtable}[t]{l>{\raggedright\arraybackslash}p{35em}rr}
\caption{\label{tab:citation_impute_preclipped2-z-hubs-elly-50000-ellz-1e+05-k-10}Z (outgoing citation) factor hubs - k = 10, $\ell_\text{z}$ = 1e+05, $\ell_\text{y}$ = 50000}\\
\toprule
ID & Title & Cited by & Cites\\
\midrule
\endfirsthead
\caption[]{Z (outgoing citation) factor hubs - k = 10, $\ell_\text{z}$ = 1e+05, $\ell_\text{y}$ = 50000 \textit{(continued)}}\\
\toprule
ID & Title & Cited by & Cites\\
\midrule
\endhead

\endfoot
\bottomrule
\endlastfoot
z01 & A Survey Of  $L$ $_{1}$ Regression & 10 & 63\\
z01 & A Selective Overview Of Variable Selection In High Dimensional Feature Space & 293 & 68\\
z01 & A Majorization-Minimization Approach To Variable Selection Using Spike And Slab Priors & 8 & 36\\
\addlinespace
z02 & Approximate Bayesian Inference For Latent Gaussian Models By Using Integrated Nested Laplace Approximations & 812 & 109\\
z02 & Bayesian Statistics In Medicine: A 25 Year Review & 23 & 511\\
z02 & Bayesian Disease Mapping: Past, Present, And Future & 5 & 98\\
\addlinespace
z03 & Estimating Structured High-Dimensional Covariance And Precision Matrices: Optimal Rates And Adaptive Estimation & 63 & 96\\
z03 & Random Matrix Theory In Statistics: A Review & 53 & 148\\
\end{longtable}
\endgroup{}

\begingroup\fontsize{8}{10}\selectfont

\begin{longtable}[t]{l>{\raggedright\arraybackslash}p{35em}rr}
\caption{\label{tab:citation_impute_preclipped2-z-hubs-elly-50000-ellz-1e+05-k-20}Z (outgoing citation) factor hubs - k = 20, $\ell_\text{z}$ = 1e+05, $\ell_\text{y}$ = 50000}\\
\toprule
ID & Title & Cited by & Cites\\
\midrule
\endfirsthead
\caption[]{Z (outgoing citation) factor hubs - k = 20, $\ell_\text{z}$ = 1e+05, $\ell_\text{y}$ = 50000 \textit{(continued)}}\\
\toprule
ID & Title & Cited by & Cites\\
\midrule
\endhead

\endfoot
\bottomrule
\endlastfoot
z01 & Sparsistent And Constansistent Estimation Of The Varying-Coefficient Model With A Diverging Number Of Predictors & 3 & 42\\
z01 & Scad-Penalized Regression In Additive Partially Linear Proportional Hazards Models With An Ultra-High-Dimensional Linear Part & 3 & 37\\
z01 & Partially Linear Structure Selection In Cox Models With Varying Coefficients & 11 & 43\\
\addlinespace
z02 & Adaptive Seamless Designs: Selection And Prospective Testing Of Hypotheses & 31 & 61\\
z02 & Group Sequential And Adaptive Designs - A Review Of Basic Concepts And Points Of Discussion & 12 & 76\\
z02 & Methodological Issues With Adaptation Of Clinical Trial Design & 12 & 41\\
\addlinespace
z03 & Bayesian Statistics In Medicine: A 25 Year Review & 23 & 511\\
z03 & Bayesian Disease Mapping: Past, Present, And Future & 5 & 98\\
\end{longtable}
\endgroup{}

\begingroup\fontsize{8}{10}\selectfont

\begin{longtable}[t]{l>{\raggedright\arraybackslash}p{35em}rr}
\caption{\label{tab:citation_impute_preclipped2-z-hubs-elly-50000-ellz-1e+05-k-40}Z (outgoing citation) factor hubs - k = 40, $\ell_\text{z}$ = 1e+05, $\ell_\text{y}$ = 50000}\\
\toprule
ID & Title & Cited by & Cites\\
\midrule
\endfirsthead
\caption[]{Z (outgoing citation) factor hubs - k = 40, $\ell_\text{z}$ = 1e+05, $\ell_\text{y}$ = 50000 \textit{(continued)}}\\
\toprule
ID & Title & Cited by & Cites\\
\midrule
\endhead

\endfoot
\bottomrule
\endlastfoot
z01 & A Survey Of  $L$ $_{1}$ Regression & 10 & 63\\
z01 & A Selective Overview Of Variable Selection In High Dimensional Feature Space & 293 & 68\\
z01 & Sparse Regression With Exact Clustering & 36 & 18\\
\addlinespace
z02 & Bayesian Statistics In Medicine: A 25 Year Review & 23 & 511\\
z02 & $E$ ($\chi^2$)-Optimal Mixed-Level Supersaturated Designs & 18 & 32\\
z02 & Optimal Mixed-Level Supersaturated Designs And A New Class Of Combinatorial Designs & 11 & 27\\
\addlinespace
z03 & Group Sequential And Adaptive Designs - A Review Of Basic Concepts And Points Of Discussion & 12 & 76\\
z03 & Methodological Issues With Adaptation Of Clinical Trial Design & 12 & 41\\
\end{longtable}
\endgroup{}

\section{Sensitivity to clipping parameters}

This section contains supplementary results for estimates using varying clipping factors $\ell_\text{z}$ and $\ell_\text{y}$, while holding $k = 30$ constant.

\subsection{Y keywords as clipping parameters vary}

\begingroup\fontsize{8}{10}\selectfont

\begin{longtable}[t]{l>{\raggedright\arraybackslash}p{34em}l}
\caption{\label{tab:citation_impute_preclipped-y-keywords-elly-1-ellz-1-k-30}Keywords for Y (incoming citation) factors - k = 30, $\ell_\text{z}$ = 1, $\ell_\text{y}$ = 1}\\
\toprule
Factor Name & Top words & ID\\
\midrule
y01 & series, garch, autoregressive, time, change, extreme & y01\\
y02 & lasso, regularization, coordinate, longitudinal, descent, selection & y02\\
y03 & lasso, selection, regression, shrinkage, via, variable & y03\\
y04 & selection, variable, penalized, oracle, nonconcave, lasso & y04\\
y05 & graphical, covariance, estimation, sparse, lasso, high & y05\\
\addlinespace
y06 & functional, regression, data, principal, linear, longitudinal & y06\\
\bottomrule
\end{longtable}
\endgroup{}

\begingroup\fontsize{8}{10}\selectfont

\begin{longtable}[t]{l>{\raggedright\arraybackslash}p{34em}l}
\caption{\label{tab:citation_impute_preclipped-y-keywords-elly-25000-ellz-25000-k-30}Keywords for Y (incoming citation) factors - k = 30, $\ell_\text{z}$ = 25000, $\ell_\text{y}$ = 25000}\\
\toprule
Factor Name & Top words & ID\\
\midrule
y01 & selection, variable, penalized, oracle, nonconcave, lasso & y01\\
y02 & screening, dimensional, ultrahigh, feature, independence, high & y02\\
y03 & graphical, covariance, lasso, sparse, estimation, selection & y03\\
y04 & high, dimensional, lasso, regression, density, dantzig & y04\\
y05 & treatment, regimes, learning, individualized, optimal, dynamic & y05\\
\addlinespace
y06 & functional, longitudinal, data, regression, linear, principal & y06\\
\bottomrule
\end{longtable}
\endgroup{}

\begingroup\fontsize{8}{10}\selectfont

\begin{longtable}[t]{l>{\raggedright\arraybackslash}p{34em}l}
\caption{\label{tab:citation_impute_preclipped-y-keywords-elly-50000-ellz-50000-k-30}Keywords for Y (incoming citation) factors - k = 30, $\ell_\text{z}$ = 50000, $\ell_\text{y}$ = 50000}\\
\toprule
Factor Name & Top words & ID\\
\midrule
y01 & selection, variable, penalized, oracle, nonconcave, lasso & y01\\
y02 & models, spatial, mixed, bayesian, splines, generalized & y02\\
y03 & graphical, covariance, sparse, estimation, lasso, high & y03\\
y04 & lasso, selection, regression, shrinkage, via, density & y04\\
y05 & high, dimensional, lasso, confidence, regression, selection & y05\\
\addlinespace
y06 & treatment, regimes, learning, individualized, optimal, estimating & y06\\
\bottomrule
\end{longtable}
\endgroup{}

\begingroup\fontsize{8}{10}\selectfont

\begin{longtable}[t]{l>{\raggedright\arraybackslash}p{34em}l}
\caption{\label{tab:citation_impute_preclipped-y-keywords-elly-70000-ellz-70000-k-30}Keywords for Y (incoming citation) factors - k = 30, $\ell_\text{z}$ = 70000, $\ell_\text{y}$ = 70000}\\
\toprule
Factor Name & Top words & ID\\
\midrule
y01 & selection, variable, penalized, oracle, nonconcave, lasso & y01\\
y02 & screening, dimensional, feature, independence, ultrahigh, high & y02\\
y03 & lasso, high, dimensional, regression, selection, dantzig & y03\\
y04 & treatment, regimes, censored, survival, individualized, estimating & y04\\
y05 & spatial, gaussian, bayesian, covariance, datasets, large & y05\\
\addlinespace
y06 & bayesian, complexity, models, disease, fit, mapping & y06\\
\bottomrule
\end{longtable}
\endgroup{}

\subsection{Z keywords as clipping parameters vary}

\begin{table}
\centering
\caption{\label{tab:citation_impute_preclipped-z-keywords-elly-1-ellz-1-k-30}Keywords for Z (outgoing citation) factors - k = 30, $\ell_\text{z}$ = 1, $\ell_\text{y}$ = 1}
\centering
\resizebox{\ifdim\width>\linewidth\linewidth\else\width\fi}{!}{
\fontsize{8}{10}\selectfont
\begin{tabular}[t]{l>{\raggedright\arraybackslash}p{34em}l}
\toprule
Factor Name & Top words & ID\\
\midrule
z01 & skew, density, garch, bibliography, tests, cure & z01\\
z02 & sequential, trials, clinical, group, adaptive, interim & z02\\
z03 & mixed, longitudinal, variance, bayesian, effects, components & z03\\
z04 & selection, high, variable, dimensional, lasso, regression & z04\\
z05 & wavelet, graphical, covariance, high, bayes, regression & z05\\
\addlinespace
z06 & bootstrap, bootstrapping, confidence, jackknife, functional, resampling & z06\\
z07 & survival, hazards, censored, proportional, cox, data & z07\\
\bottomrule
\end{tabular}}
\end{table}

\begin{table}
\centering
\caption{\label{tab:citation_impute_preclipped-z-keywords-elly-25000-ellz-25000-k-30}Keywords for Z (outgoing citation) factors - k = 30, $\ell_\text{z}$ = 25000, $\ell_\text{y}$ = 25000}
\centering
\resizebox{\ifdim\width>\linewidth\linewidth\else\width\fi}{!}{
\fontsize{8}{10}\selectfont
\begin{tabular}[t]{l>{\raggedright\arraybackslash}p{34em}l}
\toprule
Factor Name & Top words & ID\\
\midrule
z01 & selection, variable, high, dimensional, lasso, regression & z01\\
z02 & mixed, models, effects, linear, generalized, data & z02\\
z03 & graphical, covariance, wavelet, high, skew, dimensional & z03\\
z04 & density, kernel, bandwidth, estimation, nonparametric, estimators & z04\\
z05 & unit, root, cointegration, models, roots, tests & z05\\
\addlinespace
z06 & longitudinal, functional, bayesian, data, growth, repeated & z06\\
z07 & regression, transformation, skew, transformations, selection, data & z07\\
\bottomrule
\end{tabular}}
\end{table}

\begin{table}
\centering
\caption{\label{tab:citation_impute_preclipped-z-keywords-elly-50000-ellz-50000-k-30}Keywords for Z (outgoing citation) factors - k = 30, $\ell_\text{z}$ = 50000, $\ell_\text{y}$ = 50000}
\centering
\resizebox{\ifdim\width>\linewidth\linewidth\else\width\fi}{!}{
\fontsize{8}{10}\selectfont
\begin{tabular}[t]{l>{\raggedright\arraybackslash}p{34em}l}
\toprule
Factor Name & Top words & ID\\
\midrule
z01 & selection, variable, lasso, high, dimensional, sparse & z01\\
z02 & mixed, models, effects, generalized, linear, data & z02\\
z03 & wavelet, graphical, high, covariance, estimation, dimensional & z03\\
z04 & density, kernel, bandwidth, regression, nonparametric, estimation & z04\\
z05 & bayesian, models, gibbs, high, monte, carlo & z05\\
\addlinespace
z06 & unit, root, cointegration, tests, roots, series & z06\\
z07 & bayesian, spatial, models, disease, mapping, hierarchical & z07\\
\bottomrule
\end{tabular}}
\end{table}

\begin{table}
\centering
\caption{\label{tab:citation_impute_preclipped-z-keywords-elly-70000-ellz-70000-k-30}Keywords for Z (outgoing citation) factors - k = 30, $\ell_\text{z}$ = 70000, $\ell_\text{y}$ = 70000}
\centering
\resizebox{\ifdim\width>\linewidth\linewidth\else\width\fi}{!}{
\fontsize{8}{10}\selectfont
\begin{tabular}[t]{l>{\raggedright\arraybackslash}p{34em}l}
\toprule
Factor Name & Top words & ID\\
\midrule
z01 & high, wavelet, dimensional, selection, sparse, regression & z01\\
z02 & mixed, models, screening, effects, generalized, dimensional & z02\\
z03 & graphical, high, bayesian, biometrika, spatial, models & z03\\
z04 & survival, frailty, censored, data, treatment, recurrent & z04\\
z05 & bayesian, spatial, models, disease, spatio, temporal & z05\\
\addlinespace
z06 & functional, bayesian, regression, high, robust, depth & z06\\
z07 & longitudinal, mixed, models, data, joint, effects & z07\\
\bottomrule
\end{tabular}}
\end{table}

\subsection{Mixing matrix as clipping parameters vary}

\begin{figure}[H]
  \centering
  \includegraphics[width=0.8\textwidth]{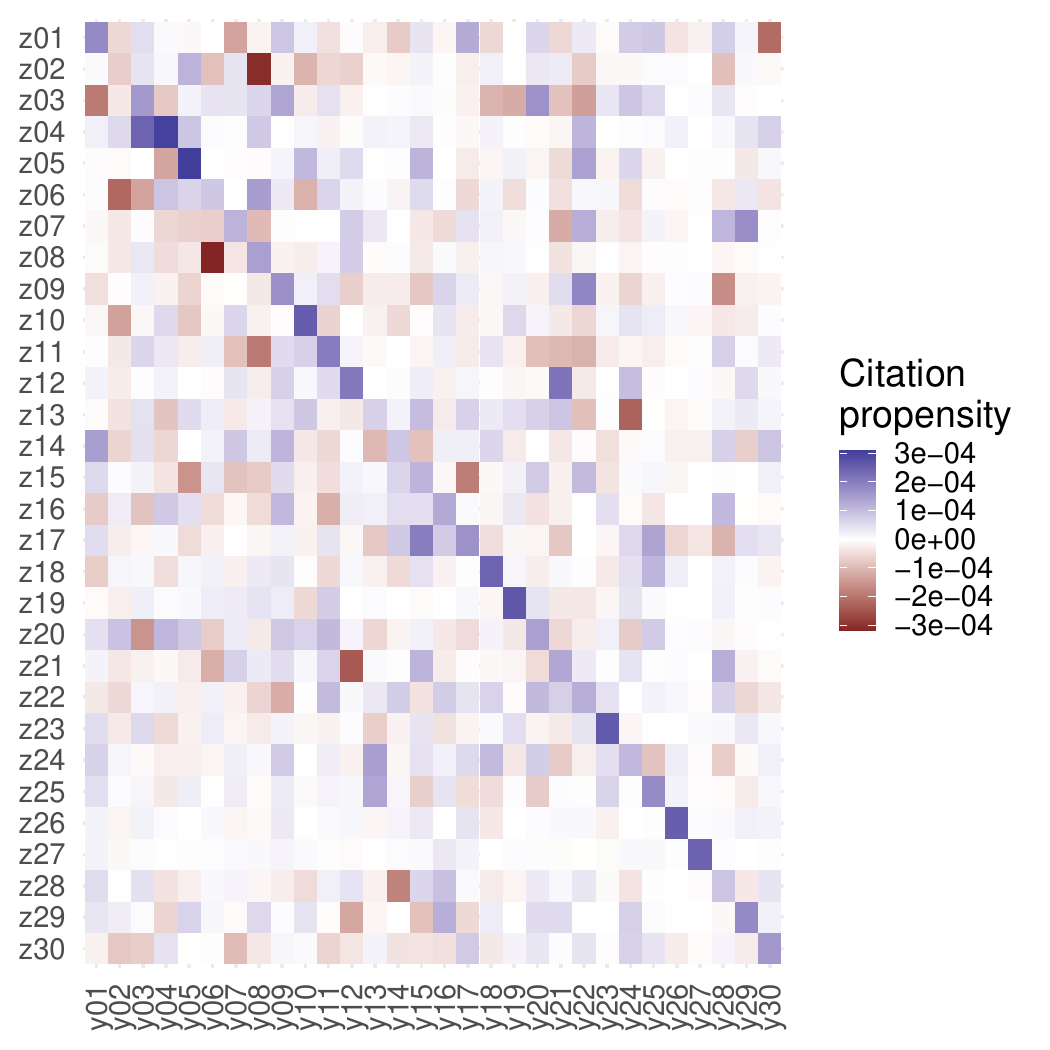}
  \caption{$\widehat B$ when $\ell_\text{z} = 1$ and $\ell_\text{y} = 1$}
\end{figure}

\begin{figure}[H]
  \centering
  \includegraphics[width=0.8\textwidth]{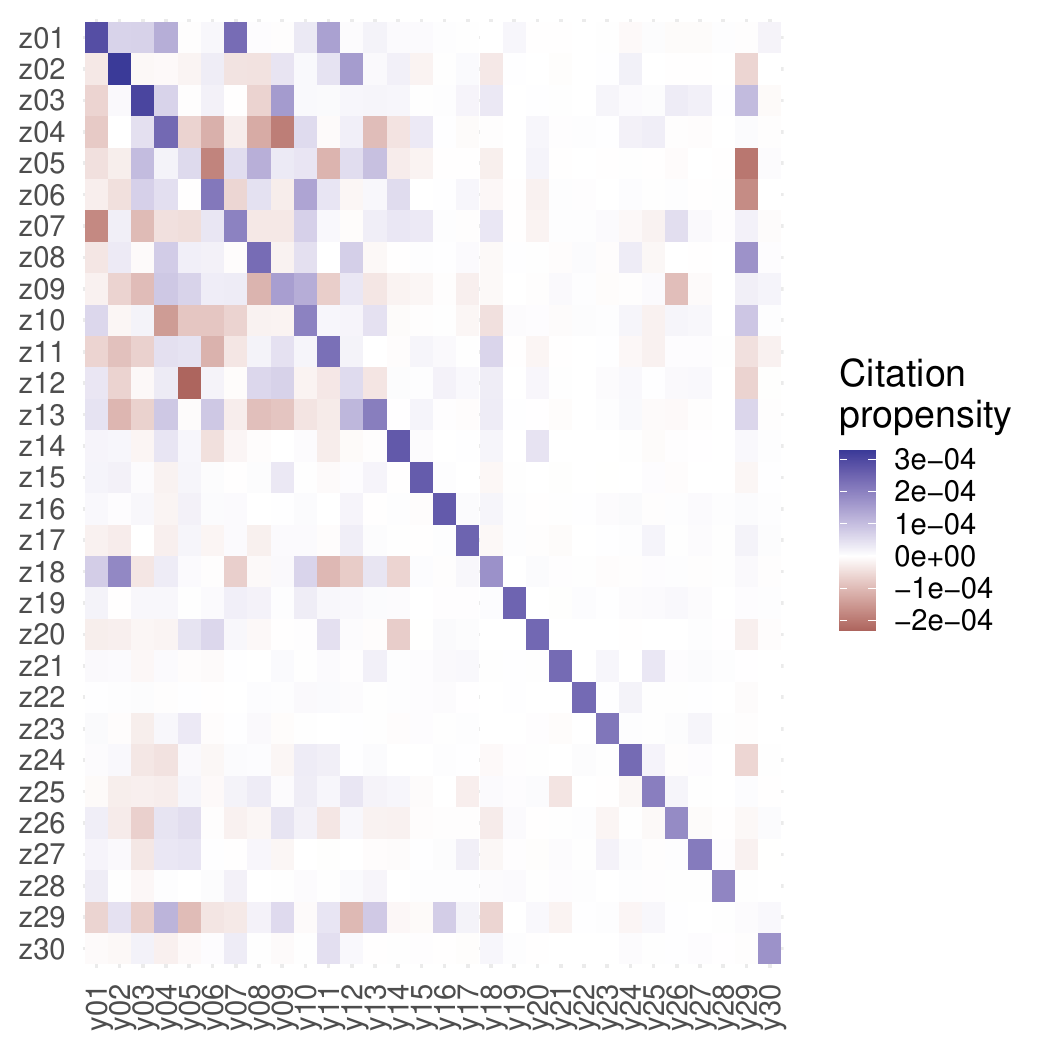}
  \caption{$\widehat B$ when $\ell_\text{z} = 25,000$ and $\ell_\text{y} = 25,000$}
\end{figure}

\begin{figure}[H]
  \centering
  \includegraphics[width=0.8\textwidth]{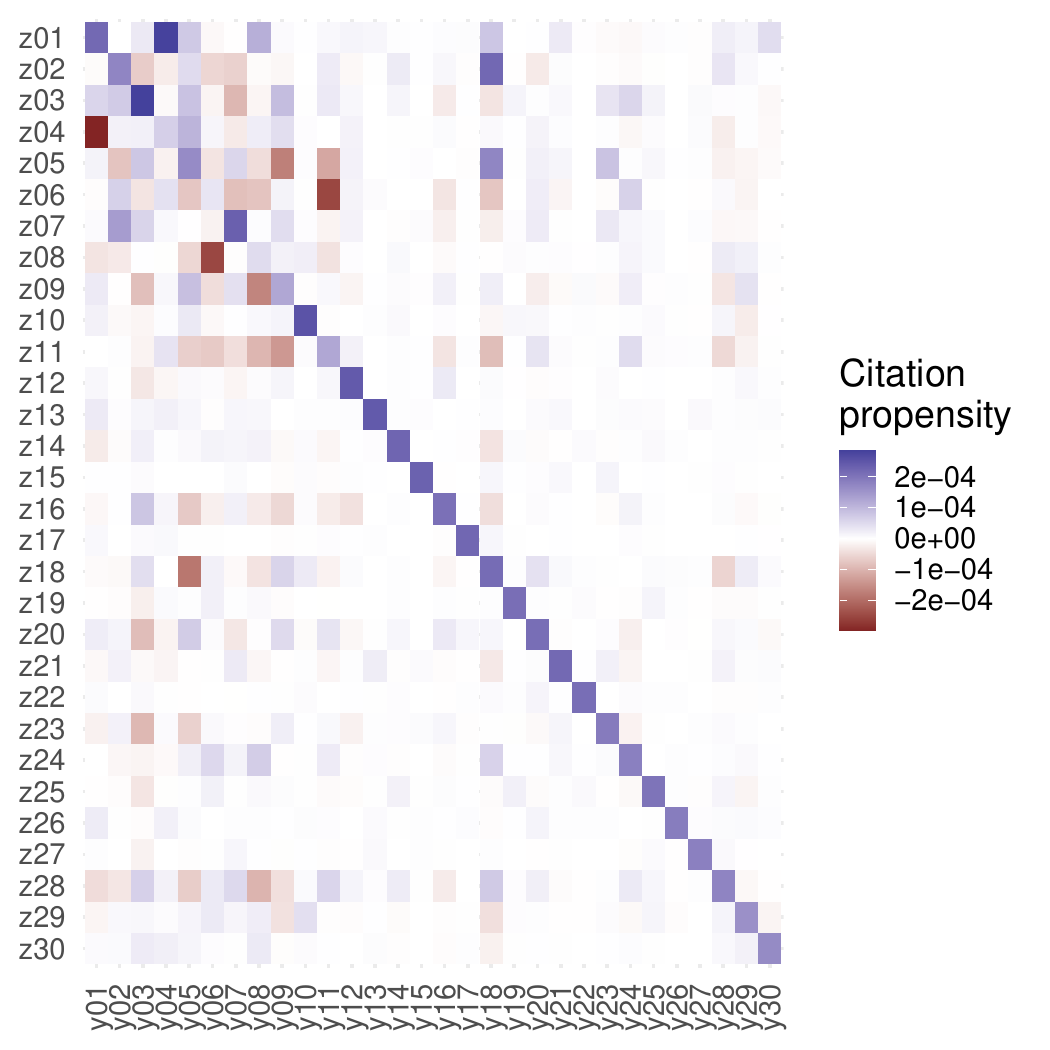}
  \caption{$\widehat B$ when $\ell_\text{z} = 50,000$ and $\ell_\text{y} = 50,000$}
\end{figure}

\begin{figure}[H]
  \centering
  \includegraphics[width=0.8\textwidth]{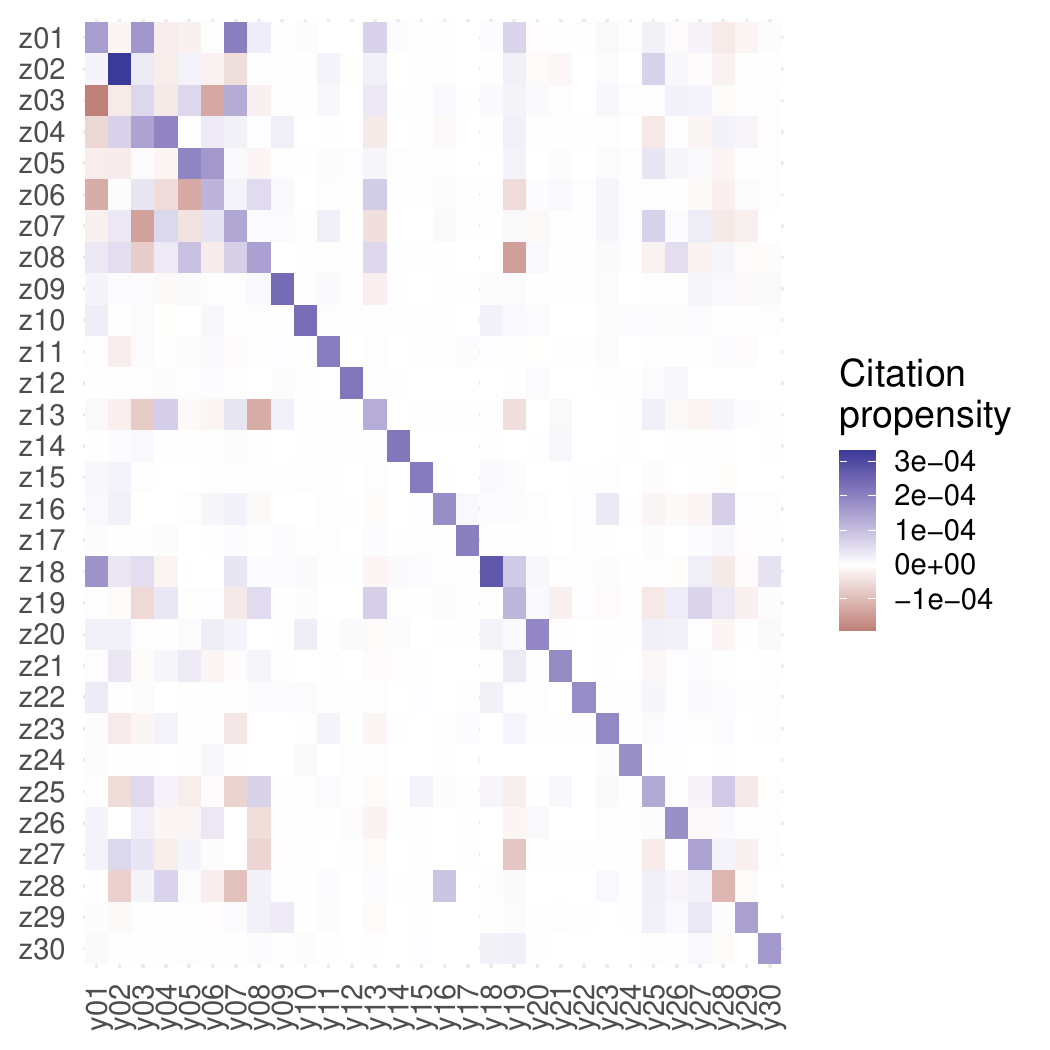}
  \caption{$\widehat B$ when $\ell_\text{z} = 70,000$ and $\ell_\text{y} = 70,000$}
\end{figure}

\subsection{Y hubs as clipping parameters vary}

\begingroup\fontsize{8}{10}\selectfont

\begin{longtable}[t]{l>{\raggedright\arraybackslash}p{35em}rr}
\caption{\label{tab:citation_impute_preclipped2-y-hubs-elly-1-ellz-1-k-30}Y (incoming citation) factor hubs - k = 30, $\ell_\text{z}$ = 1, $\ell_\text{y}$ = 1}\\
\toprule
ID & Title & Cited by & Cites\\
\midrule
\endfirsthead
\caption[]{Y (incoming citation) factor hubs - k = 30, $\ell_\text{z}$ = 1, $\ell_\text{y}$ = 1 \textit{(continued)}}\\
\toprule
ID & Title & Cited by & Cites\\
\midrule
\endhead

\endfoot
\bottomrule
\endlastfoot
y01 & Variable Selection Via Nonconcave Penalized Likelihood And Its Oracle Properties & 2804 & 14\\
y01 & The Adaptive Lasso And Its Oracle Properties & 2052 & 14\\
y01 & Nearly Unbiased Variable Selection Under Minimax Concave Penalty & 987 & 36\\
\addlinespace
y02 & Sure Independence Screening For Ultrahigh Dimensional Feature Space & 905 & 33\\
y02 & Feature Screening Via Distance Correlation Learning & 327 & 22\\
y02 & Sure Independence Screening In Generalized Linear Models With Np-Dimensionality & 305 & 24\\
\addlinespace
y03 & Bayesian Measures Of Model Complexity And Fit & 2107 & 38\\
y03 & Prior Distributions For Variance Parameters In Hierarchical Models(Comment On An Article By Browne And Draper) & 906 & 28\\
\end{longtable}
\endgroup{}

\begingroup\fontsize{8}{10}\selectfont

\begin{longtable}[t]{l>{\raggedright\arraybackslash}p{35em}rr}
\caption{\label{tab:citation_impute_preclipped2-y-hubs-elly-25000-ellz-25000-k-30}Y (incoming citation) factor hubs - k = 30, $\ell_\text{z}$ = 25000, $\ell_\text{y}$ = 25000}\\
\toprule
ID & Title & Cited by & Cites\\
\midrule
\endfirsthead
\caption[]{Y (incoming citation) factor hubs - k = 30, $\ell_\text{z}$ = 25000, $\ell_\text{y}$ = 25000 \textit{(continued)}}\\
\toprule
ID & Title & Cited by & Cites\\
\midrule
\endhead

\endfoot
\bottomrule
\endlastfoot
y01 & Variable Selection Via Nonconcave Penalized Likelihood And Its Oracle Properties & 2804 & 14\\
y01 & The Adaptive Lasso And Its Oracle Properties & 2052 & 14\\
y01 & Nearly Unbiased Variable Selection Under Minimax Concave Penalty & 987 & 36\\
\addlinespace
y02 & Sure Independence Screening For Ultrahigh Dimensional Feature Space & 905 & 33\\
y02 & Feature Screening Via Distance Correlation Learning & 327 & 22\\
y02 & Sure Independence Screening In Generalized Linear Models With Np-Dimensionality & 305 & 24\\
\addlinespace
y03 & Bayesian Measures Of Model Complexity And Fit & 2107 & 38\\
y03 & Prior Distributions For Variance Parameters In Hierarchical Models(Comment On An Article By Browne And Draper) & 906 & 28\\
\end{longtable}
\endgroup{}

\begingroup\fontsize{8}{10}\selectfont

\begin{longtable}[t]{l>{\raggedright\arraybackslash}p{35em}rr}
\caption{\label{tab:citation_impute_preclipped2-y-hubs-elly-50000-ellz-50000-k-30}Y (incoming citation) factor hubs - k = 30, $\ell_\text{z}$ = 50000, $\ell_\text{y}$ = 50000}\\
\toprule
ID & Title & Cited by & Cites\\
\midrule
\endfirsthead
\caption[]{Y (incoming citation) factor hubs - k = 30, $\ell_\text{z}$ = 50000, $\ell_\text{y}$ = 50000 \textit{(continued)}}\\
\toprule
ID & Title & Cited by & Cites\\
\midrule
\endhead

\endfoot
\bottomrule
\endlastfoot
y01 & Variable Selection Via Nonconcave Penalized Likelihood And Its Oracle Properties & 2804 & 14\\
y01 & The Adaptive Lasso And Its Oracle Properties & 2052 & 14\\
y01 & Nearly Unbiased Variable Selection Under Minimax Concave Penalty & 987 & 36\\
\addlinespace
y02 & Sure Independence Screening For Ultrahigh Dimensional Feature Space & 905 & 33\\
y02 & Feature Screening Via Distance Correlation Learning & 327 & 22\\
y02 & Sure Independence Screening In Generalized Linear Models With Np-Dimensionality & 305 & 24\\
\addlinespace
y03 & Bayesian Measures Of Model Complexity And Fit & 2107 & 38\\
y03 & Prior Distributions For Variance Parameters In Hierarchical Models(Comment On An Article By Browne And Draper) & 906 & 28\\
\end{longtable}
\endgroup{}

\begingroup\fontsize{8}{10}\selectfont

\begin{longtable}[t]{l>{\raggedright\arraybackslash}p{35em}rr}
\caption{\label{tab:citation_impute_preclipped2-y-hubs-elly-70000-ellz-70000-k-30}Y (incoming citation) factor hubs - k = 30, $\ell_\text{z}$ = 70000, $\ell_\text{y}$ = 70000}\\
\toprule
ID & Title & Cited by & Cites\\
\midrule
\endfirsthead
\caption[]{Y (incoming citation) factor hubs - k = 30, $\ell_\text{z}$ = 70000, $\ell_\text{y}$ = 70000 \textit{(continued)}}\\
\toprule
ID & Title & Cited by & Cites\\
\midrule
\endhead

\endfoot
\bottomrule
\endlastfoot
y01 & Variable Selection Via Nonconcave Penalized Likelihood And Its Oracle Properties & 2804 & 14\\
y01 & The Adaptive Lasso And Its Oracle Properties & 2052 & 14\\
y01 & Nearly Unbiased Variable Selection Under Minimax Concave Penalty & 987 & 36\\
\addlinespace
y02 & Sure Independence Screening For Ultrahigh Dimensional Feature Space & 905 & 33\\
y02 & Feature Screening Via Distance Correlation Learning & 327 & 22\\
y02 & Sure Independence Screening In Generalized Linear Models With Np-Dimensionality & 305 & 24\\
\addlinespace
y03 & Bayesian Measures Of Model Complexity And Fit & 2107 & 38\\
y03 & Prior Distributions For Variance Parameters In Hierarchical Models(Comment On An Article By Browne And Draper) & 906 & 28\\
\end{longtable}
\endgroup{}

\subsection{Z hubs as clipping parameters vary}

\begingroup\fontsize{8}{10}\selectfont

\begin{longtable}[t]{l>{\raggedright\arraybackslash}p{35em}rr}
\caption{\label{tab:citation_impute_preclipped2-z-hubs-elly-1-ellz-1-k-30}Z (outgoing citation) factor hubs - k = 30, $\ell_\text{z}$ = 1, $\ell_\text{y}$ = 1}\\
\toprule
ID & Title & Cited by & Cites\\
\midrule
\endfirsthead
\caption[]{Z (outgoing citation) factor hubs - k = 30, $\ell_\text{z}$ = 1, $\ell_\text{y}$ = 1 \textit{(continued)}}\\
\toprule
ID & Title & Cited by & Cites\\
\midrule
\endhead

\endfoot
\bottomrule
\endlastfoot
z01 & Scad-Penalized Regression In Additive Partially Linear Proportional Hazards Models With An Ultra-High-Dimensional Linear Part & 3 & 37\\
z01 & Scad-Penalized Least Absolute Deviation Regression In High-Dimensional Models & 5 & 29\\
z01 & Regularization Parameter Selections Via Generalized Information Criterion & 102 & 27\\
\addlinespace
z02 & Bayesian Statistics In Medicine: A 25 Year Review & 23 & 511\\
z02 & Construction Of Optimal Multi-Level Supersaturated Designs & 47 & 32\\
z02 & Methodological Issues With Adaptation Of Clinical Trial Design & 12 & 41\\
\addlinespace
z03 & Are Nonprofit Antipoverty Organizations Located Where They Are Needed? A Spatial Analysis Of The Greater Hartford Region & 1 & 21\\
z03 & Space-Time Interactions In Bayesian Disease Mapping With Recent Tools: Making Things Easier For Practitioners & 0 & 25\\
\end{longtable}
\endgroup{}

\begingroup\fontsize{8}{10}\selectfont

\begin{longtable}[t]{l>{\raggedright\arraybackslash}p{35em}rr}
\caption{\label{tab:citation_impute_preclipped2-z-hubs-elly-25000-ellz-25000-k-30}Z (outgoing citation) factor hubs - k = 30, $\ell_\text{z}$ = 25000, $\ell_\text{y}$ = 25000}\\
\toprule
ID & Title & Cited by & Cites\\
\midrule
\endfirsthead
\caption[]{Z (outgoing citation) factor hubs - k = 30, $\ell_\text{z}$ = 25000, $\ell_\text{y}$ = 25000 \textit{(continued)}}\\
\toprule
ID & Title & Cited by & Cites\\
\midrule
\endhead

\endfoot
\bottomrule
\endlastfoot
z01 & Scad-Penalized Regression In Additive Partially Linear Proportional Hazards Models With An Ultra-High-Dimensional Linear Part & 3 & 37\\
z01 & Scad-Penalized Least Absolute Deviation Regression In High-Dimensional Models & 5 & 29\\
z01 & Regularization Parameter Selections Via Generalized Information Criterion & 102 & 27\\
\addlinespace
z02 & Bayesian Statistics In Medicine: A 25 Year Review & 23 & 511\\
z02 & Construction Of Optimal Multi-Level Supersaturated Designs & 47 & 32\\
z02 & Methodological Issues With Adaptation Of Clinical Trial Design & 12 & 41\\
\addlinespace
z03 & Are Nonprofit Antipoverty Organizations Located Where They Are Needed? A Spatial Analysis Of The Greater Hartford Region & 1 & 21\\
z03 & Space-Time Interactions In Bayesian Disease Mapping With Recent Tools: Making Things Easier For Practitioners & 0 & 25\\
\end{longtable}
\endgroup{}

\begingroup\fontsize{8}{10}\selectfont

\begin{longtable}[t]{l>{\raggedright\arraybackslash}p{35em}rr}
\caption{\label{tab:citation_impute_preclipped2-z-hubs-elly-50000-ellz-50000-k-30}Z (outgoing citation) factor hubs - k = 30, $\ell_\text{z}$ = 50000, $\ell_\text{y}$ = 50000}\\
\toprule
ID & Title & Cited by & Cites\\
\midrule
\endfirsthead
\caption[]{Z (outgoing citation) factor hubs - k = 30, $\ell_\text{z}$ = 50000, $\ell_\text{y}$ = 50000 \textit{(continued)}}\\
\toprule
ID & Title & Cited by & Cites\\
\midrule
\endhead

\endfoot
\bottomrule
\endlastfoot
z01 & Scad-Penalized Regression In Additive Partially Linear Proportional Hazards Models With An Ultra-High-Dimensional Linear Part & 3 & 37\\
z01 & Scad-Penalized Least Absolute Deviation Regression In High-Dimensional Models & 5 & 29\\
z01 & Regularization Parameter Selections Via Generalized Information Criterion & 102 & 27\\
\addlinespace
z02 & Bayesian Statistics In Medicine: A 25 Year Review & 23 & 511\\
z02 & Construction Of Optimal Multi-Level Supersaturated Designs & 47 & 32\\
z02 & Methodological Issues With Adaptation Of Clinical Trial Design & 12 & 41\\
\addlinespace
z03 & Are Nonprofit Antipoverty Organizations Located Where They Are Needed? A Spatial Analysis Of The Greater Hartford Region & 1 & 21\\
z03 & Space-Time Interactions In Bayesian Disease Mapping With Recent Tools: Making Things Easier For Practitioners & 0 & 25\\
\end{longtable}
\endgroup{}

\begingroup\fontsize{8}{10}\selectfont

\begin{longtable}[t]{l>{\raggedright\arraybackslash}p{35em}rr}
\caption{\label{tab:citation_impute_preclipped2-z-hubs-elly-70000-ellz-70000-k-30}Z (outgoing citation) factor hubs - k = 30, $\ell_\text{z}$ = 70000, $\ell_\text{y}$ = 70000}\\
\toprule
ID & Title & Cited by & Cites\\
\midrule
\endfirsthead
\caption[]{Z (outgoing citation) factor hubs - k = 30, $\ell_\text{z}$ = 70000, $\ell_\text{y}$ = 70000 \textit{(continued)}}\\
\toprule
ID & Title & Cited by & Cites\\
\midrule
\endhead

\endfoot
\bottomrule
\endlastfoot
z01 & Scad-Penalized Regression In Additive Partially Linear Proportional Hazards Models With An Ultra-High-Dimensional Linear Part & 3 & 37\\
z01 & Scad-Penalized Least Absolute Deviation Regression In High-Dimensional Models & 5 & 29\\
z01 & Regularization Parameter Selections Via Generalized Information Criterion & 102 & 27\\
\addlinespace
z02 & Bayesian Statistics In Medicine: A 25 Year Review & 23 & 511\\
z02 & Construction Of Optimal Multi-Level Supersaturated Designs & 47 & 32\\
z02 & Methodological Issues With Adaptation Of Clinical Trial Design & 12 & 41\\
\addlinespace
z03 & Are Nonprofit Antipoverty Organizations Located Where They Are Needed? A Spatial Analysis Of The Greater Hartford Region & 1 & 21\\
z03 & Space-Time Interactions In Bayesian Disease Mapping With Recent Tools: Making Things Easier For Practitioners & 0 & 25\\
\end{longtable}
\endgroup{}

\end{document}